\documentclass[aps,prd,twocolumn,notitlepage,nofootinbib,,nobibnotes,superscriptaddress,amsmath,amssymb,preprintnumbers]{revtex4-1}

\usepackage{graphicx} 
\usepackage{dcolumn}
\usepackage{bm}        % for math
\usepackage{amssymb}   % for math
\usepackage{adjustbox}  
\usepackage{amsmath,array}
\usepackage{comment}
\usepackage{natbib}
\usepackage{MnSymbol}
\usepackage[hidelinks]{hyperref}
\usepackage{graphicx}
\usepackage{subfig}
\usepackage{color}
\usepackage{braket}
\usepackage     [utf8]                  {inputenc}
\usepackage     [T1]                    {fontenc}
\usepackage     [english]               {babel}
\usepackage{epigraph}
\usepackage{etoolbox}
\usepackage{ dsfont }
\usepackage{physics}
\usepackage{xcolor}
\usepackage{siunitx}
\makeatletter
\patchcmd{\epigraph}{\@epitext{#1}}{\itshape\@epitext{#1}}{}{}
\newcommand*\eqsize{%
\@setfontsize\mysize{9.0}{9.0}%
    }
    \usepackage{multirow}
\makeatother

\usepackage{xfrac}
\usepackage{amsmath,amssymb}
\usepackage{esdiff} % derivatives
\usepackage{commath} % math macros
\usepackage{bbm} % blackboard style symbols
\usepackage{braket}
\usepackage{slashed}
\def\x{\mathbf{x}}

\newcommand{\trento}{T\raisebox{-.5ex}{R}ENTo}

\newcommand{\ktt}{|\mathbf{k_T}|}

\newcommand{\vp}{\vec{p}}
\newcommand{\vk}{\vec{k}}

\newcommand{\TBg}{\overline{T}}
\newcommand{\TId}{T_\text{id.}}

\newcommand{\tauekt}{\tau_{\scriptscriptstyle \text{EKT}}}
\newcommand{\tauhydro}{\tau_\text{hydro}}
\newcommand{\OmegaId}{\tilde{\omega}_{\rm Id}}

\newcommand{\xT}{\mathbf{x}}

\newcommand{\sT}{\mathbf{s}}
\newcommand{\pT}{\mathbf{p}}
\newcommand{\qT}{\mathbf{q}}
\newcommand{\kT}{\mathbf{k}}
\newcommand{\mbk}{\mathbf{k}}

\newcommand{\aS}{\alpha_S}

\newcommand{\Dipper}{\textsc{McDipper}}

\newcommand{\kompost}{K{\o}MP{\o}ST}

\definecolor{goethe-blau}{cmyk}{1.0,0.2,0.0,0.4}
\definecolor{hellgrau}{cmyk}{0.04,0.04,0.05,0.02}
\definecolor{sandgrau}{cmyk}{0.12,0.09,0.13,0.0}
\definecolor{dunkelgrau}{cmyk}{0.25,0.25,0.30,0.75}
\definecolor{emo-rot}{cmyk}{0.04,1.0,0.8,0.07}
\definecolor{purple}{cmyk}{0.08,1.0,0.3,0.36}
\definecolor{senfgelb}{cmyk}{0.01,0.25,1.0,0.05}
\definecolor{gruen}{cmyk}{0.62,0.4,0.87,0.09}
\definecolor{magenta}{cmyk}{0.08,0.86,0.12,0.12}
\definecolor{orange}{cmyk}{0.0,0.7,1.0,0.04}
\definecolor{sonnengelb}{cmyk}{0.0,0.12,0.95,0.0}
\definecolor{helles-gruen}{cmyk}{0.4,0.17,0.81,0.07}
\definecolor{lichtblau}{cmyk}{0.8,0.0,0.06,0.04}

\definecolor{navyblue}{RGB}{0, 0, 160}
\definecolor{navyred}{RGB}{200, 0, 0}
\definecolor{navygreen}{RGB}{0, 128,0}

\begin{document}

\title{(3+1)D event-by-event pre-equilibrium dynamics in heavy-ion collisions}
\date{\today}

\author{Xiaojian Du} 
%\email{xiaojiandu@outlook.com}
\affiliation{Wilczek Quantum Center, Shanghai Institute for Advanced Studies,\\
University of Science and Technology of China, Shanghai 201315, China}
\affiliation{Shanghai Research Center for Quantum Sciences, Shanghai 201315, China}
\affiliation{Hefei National Laboratory, Hefei 230088, China}
\affiliation{Department of Physics, University of Jyväskylä, P.O. Box 35, 40014 Jyväskylä, Finland}
\affiliation{Helsinki Institute of Physics, P.O. Box 64, 00014 University of Helsinki, Finland}
%\affiliation{Instituto Galego de Física de Altas Enerxías (IGFAE), Universidade de Santiago de Compostela, E-15782 Galicia, Spain}
%\affiliation{Fakult\"at f\"ur Physik, Universit\"at Bielefeld, D-33615 Bielefeld, Germany}

\author{S\"oren Schlichting}
%\email{sschlichting@physik.uni-bielefeld.de}
\affiliation{Fakult\"at f\"ur Physik, Universit\"at Bielefeld, D-33615 Bielefeld, Germany}

\author{Jie Zhu}
\email{Corresponding author: jzhu@physik.uni-bielefeld.de}
\affiliation{Fakult\"at f\"ur Physik, Universit\"at Bielefeld, D-33615 Bielefeld, Germany}
\affiliation{Institute of Particle Physics and Key Laboratory of Quark and Lepton Physics (MOE), Central China Normal University, Wuhan, 430079, China}

\begin{abstract}
So far a major source of uncertainty in the study of heavy-ion collisions arises from the early time dynamics which includes initial state and pre-equilibrium dynamics. The state-of-the-art framework, \kompost~\cite{Kurkela:2018vqr,Kurkela:2018wud}, employs non-equilibrium Green's functions to propagate the initial energy-momentum tensor to the hydrodynamic phase, yet currently only treats transverse plane dynamics under boost-invariant conditions.
In this work, we extend \kompost\ to include non-boost-invariant responses to initial conditions, essential for accurately capturing the longitudinal structures observed in heavy-ion collisions.
Non-boost-invariant fluctuations on top of a homogeneous background are evolved using (3+1)D response functions calculated in kinetic theory. 
To assess kinetic theory's transition towards hydrodynamic evolution, we systematically compare the out-of-equilibrium shear-stress tensor from \kompost-3D with estimates based on Navier-Stokes hydrodynamics.
Subsequently, a comprehensive (3+1)D framework, \Dipper+\kompost-3D+CLVisc+SMASH, is utilized to simulate the complete spacetime evolution of heavy-ion collisions. 
The sensitivity of key observables, including longitudinal structure of anisotropic flow, to variations in the hydrodynamic initialization time is thoroughly investigated.
\end{abstract}

\maketitle
\section{Introduction}\label{sec:intro}
A multi-stage framework consisting of early-time dynamics, hydrodynamic expansion, and a hadronic afterburner has been widely utilized in the study of relativistic heavy-ion collisions~\cite{Muller:2012zq,Muller:2013dea}. 
Recently, most models on the market have been gradually evolving into full three-dimensional formulations in order to describe the complete space-time dynamics at each stage. 
For the initial state modeling, the transition is exemplified by the developments from 2D frameworks such as \trento~\cite{Moreland:2018gsh} to \trento-3D~\cite{Ke:2016jrd, Soeder:2023vdn}, and from CGC-based models like IP-Glasma~\cite{Schenke:2012wb, Schenke:2012hg, Mantysaari:2017cni} to 3D-Glasma~\cite{McDonald:2023qwc, Ipp:2024ykh}, as well as by newer implementations such as EKRT~\cite{Kuha:2024kmq, Hirvonen:2024zne}.
For the hydrodynamic stage, the evolution follows a similar path, from earlier 2D viscous hydrodynamics (e.g., VISHNU~\cite{Shen:2014vra}) to modern (3+1)D frameworks such as MUSIC~\cite{Schenke:2010nt, Schenke:2010rr} and Trajectum~\cite{Nijs:2020roc}.
Finally, for the hadronic afterburner stage, publicly available models such as UrQMD~\cite{Bass:1998ca, Bleicher:1999xi} and SMASH~\cite{wergieluk_2024_10707746, SMASH:2016zqf, Sjostrand:2006za, Sjostrand:2007gs} are already formulated in full (3+1)D.
One of the remaining challenges in this multi-stage picture lies in modeling the pre-equilibrium dynamics, which bridges the gap between the initial conditions and the onset of the hydrodynamic expansion.

Starting from the initial state, parametric models, such as the Glauber model~\cite{Miller:2007ri,Moreland:2014oya,Blaizot:2014nia,Gronqvist:2016hym} or \trento~\cite{Moreland:2018gsh} model, only provide the initial energy or entropy density but miss the remaining hydrodynamic fields, such as the flow velocity and the out-of-equilibrium contributions. Microscopic models, such as IP-Glasma and \Dipper, provide the full information of the initial energy-momentum tensor. 
However, a theoretical description of the thermalization process is still needed to evolve the initial energy-momentum tensor to the point in time where a viscous hydrodynamic description becomes applicable.

In order to describe this pre-equilibrium stage, simple models such as free streaming~\cite{Broniowski:2008qk, Liu:2015nwa} and the gradient expansion~\cite{Vredevoogd:2008id, vanderSchee:2013pia, Romatschke:2015gxa} have been employed to capture some physics during the pre-equilibrium stage. 
However, developing a more realistic model remains an urgent task.
The pre-equilibrium framework, \kompost ~\cite{Kurkela:2018vqr,Kurkela:2018wud}, provides a practical way to propagate the energy-momentum tensor in a far-from-equilibrium initial state to a time when viscous hydrodynamics becomes applicable and some recent works~\cite{Wu:2024pba,Garcia-Montero:2023lrd,Garcia-Montero:2024lbl,Schenke:2020uqq,NunesdaSilva:2020bfs} have already employed \kompost\ in various contexts.
The power and advantage of \kompost\ comes from the fact that the non-equilibrium kinetic evolution can be efficiently described using non-equilibrium Green's functions calculated in kinetic theory, which consequently enables fast event-by-event pre-equilibrium evolutions.

Considering the late time hydrodynamic behavior is fully characterized by the energy-momentum tensor, the central object of \kompost\ is to 
propagate the non-equilibrium energy-momentum tensor $T^{\mu\nu}$  from an initial time $\tau_{\text{EKT}}$ to a time $\tau_{\text{hydro}}$ when hydrodynamics becomes applicable, i.e.
\begin{equation}
\left.T^{\mu \nu}\left({\tau_{\text{EKT}}}, \mathbf{x}\right)\right|_{\text {out-of-equilibrium }} \longrightarrow T^{\mu \nu}\left({\tau_{\text {hydro }}}, \mathbf{x}\right).   
\end{equation}
The energy-momentum tensor $T^{\mu\nu}$ at $(\tau_{\text{hydro}}, \mathbf{x}$) is determined by the initial conditions at an earlier time $\tau_{\text {EKT }}$ in a causal neighborhood of point $\mathbf{x}$, which constraints their position $\mathbf{x}'$ to satisfy
\begin{equation}
\left|\mathbf{x}^{\prime}-\mathbf{x}\right|<c\left(\tau_{\text {hydro }}-\tau_{\text {EKT}}\right).
\end{equation}
Since for collisions of large nuclei, such as Pb and Au, at RHIC and LHC energies,  the time duration of pre-equilibrium evolution $\Delta\tau=\tau_{\rm hydro}-\tau_{\rm EKT}$ is much smaller than the typical size ($R\sim 5$ fm) of collision region, long wavelength fluctuations within the causal circle are small, whereas short wavelength fluctuations are typically washed out as the Quark Gluon Plasma (QGP) equilibrates and becomes a hydrodynamically expanding fluid.

Based on this separation of scales, the evolution of energy-momentum tensor can be split into the expansion of a longitudinally boost-invariant and locally transverse homogeneous background and the propagation of respective perturbations around the background~\cite{Kurkela:2018vqr,Kurkela:2018wud},
\begin{equation}
T^{\mu \nu}\left(\tau, \mathbf{x}^\prime\right)
=
\TBg_{\mathbf{x}}^{\mu \nu}\left(\tau\right)
+
\delta T_{\mathbf{x}}^{\mu \nu}\left(\tau, \mathbf{x}^\prime\right).    
\end{equation}
By virtue of the scaling behavior found in kinetic theory~\cite{Kurkela:2018vqr, Kurkela:2015qoa}, the non-equilibrium evolution of the background energy-momentum tensor ($\TBg^{\mu\nu}$)  can be described by a universal attractor curve~\cite{Giacalone:2019ldn},
if the time variable ($\tau$) is scaled by the equilibrium relaxation time $\tau_R(\tau)={4\pi \eta_v/s \over T}$.
While the small perturbations around the background can be described by linear response theory,
\begin{equation}
\frac{\delta 
T^{\mu\nu}(\tau,\mathbf{x})}{\TBg^{\tau\tau}_\mathbf{x}(\tau)}=
\int 
d^2\mathbf{x}_0 \,G^{\mu\nu}_{\alpha\beta}\left(\mathbf{x},\mathbf{x}_0,\tau,\tau_0\right) 
\frac{\delta T^{\alpha\beta}_\x(\tau_0,\mathbf{x}_0)}{\TBg^{\tau\tau}_\x (\tau_0)}.
\end{equation}
where the non-equilibrium Green's functions $G^{\mu\nu}_{\alpha\beta}$ need to be determined from the underlying microscopic theory, using e.g. QCD kinetic theory.

However, the current implementation of \kompost\ is limited to describe the dynamics in the transverse plane, and therefore only applies at mid-rapidity. A full three-dimensional description becomes essential in the forward and backward rapidity regions of high-energy nucleus–nucleus collisions, especially at lower collision energies, where the longitudinal structure of the final observables plays an important role.
In this work, we extend the state-of-the-art \kompost\ framework, to construct a new framework for the three-dimensional pre-equilibrium dynamics. \kompost-3D provides a smooth transition from 3D initial conditions to (3+1)D viscous hydrodynamics, thereby reducing the dependence of the hydrodynamic model on the initialization
time $\tauhydro$~\cite{vanderSchee:2013pia,Kurkela:2016vts}.

This paper is organized as follows. In Sec.~\ref{sec:response-RTA}, we briefly introduce the solution to the kinetic theory under the relaxation time approximation, and present the corresponding response functions in both Fourier and coordinate space, as well as their implementation in the \kompost-3D framework.
In Sec.~\ref{sec:3D-framework}, we introduce the (3+1)D framework employed to simulate event-by-event heavy-ion collisions.
This framework consists of the {\Dipper} initial state model, the \kompost-3D pre-equilibrium dynamics, the CLVisc viscous hydrodynamics, and the hadronic afterburner SMASH. 
In Sec.~\ref{sec:results}, we first visualize the input and output of the \kompost-3D framework. Then we examine the smooth transition from kinetic theory pre-equilibrium dynamics to hydrodynamic expansion by 
analyzing the evolution of hydrodynamic fields, including energy density, radial flow, and momentum azimuthal ellipticity, 
and the hydrodynamization of the out-of-equilibrium shear stress tensor as well as the scaled time variable.
Furthermore, we examine the sensitivity of final state observables, including longitudinal structure of the multiplicity, mean transverse momentum and anisotropic flow, to the hydro starting time ($\tau_{\rm hydro}$). 
Finally, we summarize the findings and point out possible directions for future work in Sec.~\ref{sec:summary}.

\section{Response functions}\label{sec:response-RTA}
Building on the foundational work of \kompost~\cite{Kurkela:2018wud,Kurkela:2018vqr}, in this work, we continue to assume that the background is locally transverse homogeneous and longitudinally boost-invariant. Hence, the non-equilibrium evolution of the background energy-momentum tensor $\bar{T}^{\mu\nu}$ remains identical to the original \kompost\ framework,  while the non-equilibrium response functions evolve the full three-dimensional structure of the initial energy-momentum tensor. Since the linearization is performed on a boost-invariant and homogeneous background, the 3D response functions depend only on the differences $(\mathbf{x-x_0}, \eta-\eta_0)$, such that
\begin{equation}
    \begin{aligned}
    \frac{
    {\delta T^{\mu \nu}(\tau, \mathbf{x}, \eta)}
    }{\overline{T}_{\mathbf{x},\eta}^{\tau \tau}(\tau)}
    =\int d^2 \mathbf{x}_0\ d\eta_0\ 
    &{G_{\alpha \beta}^{\mu \nu}\left(\mathbf{x}, \eta,  \mathbf{x}_0, \eta_0, \tau, \tau_{0}\right)}\\
    &\times \frac{
    {\delta T_{\mathbf{x},\eta}^{\alpha \beta}\left(\tau_{0}, \mathbf{x}_0, \eta_0\right)}
    }{\overline{T}_{\mathbf{x},\eta}^{\tau \tau}\left(\tau_{0}\right)},
    \end{aligned}
\end{equation}
where the response functions, ${G_{\alpha \beta}^{\mu \nu}\left(\mathbf{x}, \eta,  \mathbf{x}_0, \eta_0, \tau, \tau_{0}\right)}$, can be more easily solved in momentum space,
\begin{equation}
\begin{aligned}
&G_{\alpha \beta}^{\mu \nu}
\left(\mathbf{x}-\mathbf{x}_0, \eta-\eta_0, \tau, \tau_0, \overline{T}_{\mathbf{x}}^{\tau \tau}\left(\tau_0\right)\right)=
\int \frac{d^2 \mathbf{k_T} dk_\eta}{(2 \pi)^3}\\
&
\tilde{G}_{\alpha \beta}^{\mu \nu}\left(\mathbf{k_T}, k_\eta, \tau, \tau_0, \overline{T}_{\mathbf{x}}^{\tau \tau}\left(\tau_0\right)\right) e^{i\{ \mathbf{k_T} \cdot\left(\mathbf{x}-\mathbf{x}_0 \right) + k_\eta (\eta-\eta_0)\} }.
\end{aligned}
\end{equation}
where the longitudinal momentum is defined as $k_\eta\equiv{|\mathbf{k_T}| \tau }\sinh(y-\eta)$. 
So in momentum space, we have
\begin{equation}
\frac{
{\delta T_k^{\mu \nu}(\tau)}
}{e(\tau)}=
\frac{1}{2} 
{G_{\alpha \beta}^{\mu \nu}\left(\mathbf{k}_{\mathbf{T}}, k_\eta, \tau, \tau_0\right)}
\frac{
{\delta T_k^{\alpha \beta}\left(\tau_0\right)}
}{e\left(\tau_0\right)}.
\end{equation}
We note that the symmetries of the background constrain the possible structure of the response functions, and it is therefore advantageous to perform a decomposition into different tensor structures.

\subsection{Decomposition of response functions}
Based on rotational symmetry in the transverse plane, the response functions in momentum space for (scalar) energy perturbations can be decomposed as follows.
\begin{align}
    & \tilde{G}_{\tau \tau}^{\tau \tau}(\mathbf{k_T}, k_\eta)=\tilde{G}_s^s(|\mathbf{k_T}|, |k_\eta|)\nonumber\\
    &\tilde{G}_{\tau \tau}^{\tau i}(\mathbf{k_T}, k_\eta)=-i \tilde{G}_s^v(|\mathbf{k_T}|, |k_\eta|) \frac{\mathbf{k_T}^i}{|\mathbf{k_T}|} \nonumber\\
    & \tilde{G}_{\tau \tau}^{i j}(\mathbf{k_T}, k_\eta)=\tilde{G}_s^{t, \delta}(|\mathbf{k_T}|, |k_\eta|) \delta^{i j}+\tilde{G}_s^{t, k}(|\mathbf{k_T}|, |k_\eta|) \frac{\mathbf{k_T}^i \mathbf{k_T}^j}{|\mathbf{k_T}|^2}\nonumber\\
    &\tilde{G}_{\tau \tau}^{\eta \eta}\left(\mathbf{k}_{\mathbf{T}}, k_\eta\right) =\tilde{G}_s^\eta\left(|\mathbf{k_T}|, |k_\eta|\right)\\ 
    &\tilde{G}_{\tau \tau}^{\tau \eta}\left(\mathbf{k}_{\mathbf{T}}, k_\eta\right)  =-i \frac{k_\eta}{\left|k_\eta\right|} \tilde{G}_s^{s, \eta}\left(|\mathbf{k_T}|, |k_\eta|\right) \nonumber\\
    &\tilde{G}_{\tau \tau}^{i \eta}\left(\mathbf{k}_{\mathbf{T}}, k_\eta\right)  =\frac{k_T^i k_\eta}{\left|\mathbf{k}_{\mathbf{T}}\right|\left|k_\eta\right|} \tilde{G}_s^{v, \eta}\left(|\mathbf{k_T}|, |k_\eta|\right)\nonumber
\end{align}
Similarly, in position space, the decomposition for energy perturbations takes the following form.
\begin{align}
    \begin{aligned}
    G_{\tau \tau}^{\tau \tau}\left(\mathbf{r}, \eta\right) & =G_s^s\left(|\mathbf{r}|, |\eta|\right) \\
    G_{\tau \tau}^{\tau i}\left(\mathbf{r}, \eta\right) & =\frac{\mathbf{r}^i}{|\mathbf{r}|} G_s^v\left(|\mathbf{r}|, |\eta|\right) \\
    G_{\tau \tau}^{i j}\left(\mathbf{r}, \eta\right) & =G_s^{t, \delta}\left(|\mathbf{r}|, |\eta|\right) \delta^{i j}+G_s^{t, r}\left(|\mathbf{r}|, |\eta|\right) \frac{\mathbf{r}^i \mathbf{r}^j}{|\mathbf{r}|^2}\\
    G_{\tau\tau}^{\eta \eta}\left(\mathbf{r}, \eta\right) & =G_s^\eta\left(|\mathbf{r}|, |\eta|\right) \\
    G_{\tau\tau}^{\tau \eta}\left(\mathbf{r}, \eta\right) & =\frac{\eta}{|\eta|} G_s^{s,\eta}\left(|\mathbf{r}|, |\eta|\right) \\
    G_{\tau\tau}^{i \eta}\left(\mathbf{r}, \eta\right) & = \frac{\mathbf{r}^i \eta}{|\mathbf{r}||\eta|} G_s^{v,\eta}(|\mathbf{r}|, |\eta|) 
    \end{aligned}
\end{align}
The coordinate space Green's functions are related to their momentum space counterparts according to the Fourier-Hankel transformation as follows.
\begin{widetext}
    \begin{equation}\label{eq:rev-fourier}
    \begin{aligned}
    &G_s^s\left(|\mathbf{r}|, |\eta|\right) =\frac{1}{2 \pi^2} \int_0^\infty d|\mathbf{k_T}||\mathbf{k_T}| {dk_\eta \cos(k_\eta\cdot \eta)} J_0(|\mathbf{k_T}||\mathbf{r}|) \tilde{G}_s^s\left(|\mathbf{k_T}|, |k_\eta| \right) \\
    &G_s^v\left(|\mathbf{r}|, |\eta|\right) =\frac{1}{2 \pi^2} \int_0^\infty d|\mathbf{k_T}||\mathbf{k_T}| {dk_\eta \cos( k_\eta\cdot \eta)} J_1(|\mathbf{k_T}||\mathbf{r}|) \tilde{G}_s^v\left(|\mathbf{k_T}|, |k_\eta|\right) \\
    &G_s^{t,\delta}\left(|\mathbf{r}|,|\eta|\right)=\frac{1}{2 \pi^2} \int_0^\infty d|\mathbf{k_T}||\mathbf{k_T}|{dk_\eta \cos( k_\eta\cdot \eta)} \left[J_0(|\mathbf{k_T}||\mathbf{r}|)\tilde{G}_s^{t,\delta}\left(|\mathbf{k_T}|,|k_\eta|\right)+\frac{J_1(|\mathbf{k_T}||\mathbf{r}|)}{|\mathbf{k_T}||\mathbf{r}|} \tilde{G}_s^{t,k}\left(|\mathbf{k_T}|,|k_\eta|\right)\right]\\
    &G_s^{t, r}\left(|\mathbf{r}|, |\eta|\right) =\frac{-1}{2 \pi^2} \int_0^\infty d|\mathbf{k_T}||\mathbf{k_T}| {dk_\eta \cos(k_\eta\cdot \eta)} J_2(|\mathbf{k_T} \| \mathbf{r}|) \tilde{G}_s^{t, k}\left(|\mathbf{k_T}|, |k_\eta|\right)\\
    &G_s^\eta(|\mathbf{r}|, |\eta|)=\frac{1}{2 \pi^2} \int_0^\infty d|\mathbf{k_T}||\mathbf{k_T}| {dk_\eta \cos( k_\eta\cdot \eta)} J_0(|\mathbf{k_T}||\mathbf{r}|) \tilde{G}_s^\eta\left(|\mathbf{k_T}|, |k_\eta|\right)\\
    &G_s^{s, \eta}(|\mathbf{r}|, |\eta|)=\frac{\eta}{2 \pi^2|\eta|} \int_0^\infty d|\mathbf{k_T}||\mathbf{k_T}| {dk_\eta \sin( k_\eta\cdot \eta)} J_0(|\mathbf{k_T}||\mathbf{r}|) \tilde{G}_s^{s, \eta}\left(|\mathbf{k_T}|, |k_\eta|\right)\\
    &G_s^{v, \eta}(|\mathbf{r}|, |\eta|)=\frac{-\eta}{2 \pi^2 |\eta|} \int_0^\infty d|\mathbf{k_T}||\mathbf{k_T}| {dk_\eta \sin( k_\eta\cdot \eta)} J_1(|\mathbf{k_T}||\mathbf{r}|) \tilde{G}_s^{v, \eta}\left(|\mathbf{k_T}|, |k_\eta|\right)
    \end{aligned}
    \end{equation}
\end{widetext}
In this work, the independent components of macroscopic response functions $G_{\alpha \beta}^{\mu\nu}$ will be calculated using kinetic theory in the relaxation time approximation, which will be explained in next section.

\subsection{Response functions from kinetic theory in the relaxation time approximation (RTA)}
We will employ the Boltzmann equation in the relaxation time approximation, 
\begin{equation}
\begin{aligned}
\left[p^\tau \partial_\tau+p^i \partial_i-\frac{p_\eta}{\tau^2} \partial_\eta\right]
&f(x, p)=-\frac{p_\mu u^\mu(x)}{\tau_R}\\
&\times\left[f(x, p)-f_{e q}\left(p_\mu \beta^\mu(x)\right)\right].
\label{eq:BoltzmannRTAEq}
\end{aligned}
\end{equation}
for a conformal system of ultra-relativistic massless bosons, to calculate the response functions, as exemplified in ~\cite{Kamata:2020mka}, and in selected cases also compare the RTA results to calculations in QCD kinetic theory. We employ Milne coordinates 
\begin{equation}
x^{\mu}=(\tau,\mathbf{x}_T,\eta)
\end{equation}
with metric $g_{\mu\nu}=\text{diag}(1,-1,-1,-1/\tau^2)$, such that $p^\mu$ in Eq.~(\ref{eq:BoltzmannRTAEq}) denotes the four momentum 
\begin{equation}
    p^\mu \equiv (p^\tau, \mathbf{p_T}, p^\eta)
\end{equation}
which can be parametrized as
\begin{equation}
    p^\mu =(p_T\cosh(y-\eta), \mathbf{p_T}, {p_T \over \tau }\sinh(y-\eta)),
\end{equation}
In equilibrium, the distribution is assumed to be the Bose-Einstein distribution,
$$
f_{\rm eq}(x)=\frac{1}{e^x-1}.
$$
such that the equation of state is given by $e_{\rm eq}(T)=\nu_{\rm eff} \frac{\pi^2}{30}T^4$,
where $\nu_{\rm eff}$ is the effective number of degrees of freedom. The relaxation time is determined by $T\tau_R=5 \eta_v/s$, such that for a conformal system with constant ratio of shear-viscosity to entropy density $\eta_v/s$,
the relaxation time is inversely proportional to the temperature. 

In principle, the distribution function can be decomposed into a background ($f_{BG}(\tau, p_T, |p_\eta|)$) that is longitudinally boost invariant and locally homogeneous in the transverse plane, together with small perturbations (${\delta f(x,p)}$) around it
    \begin{equation}
    {f(x,p)}={f_{BG}(\tau, p_T, |p_\eta|)}+{\delta f(x,p)}.
    \end{equation}
Therefore, we obtain the Boltzmann equations for the background and the perturbations, respectively, 
    \begin{equation}\label{eq:boltz-dist}
    \begin{aligned}
    \tau \partial_\tau 
    f_{B G}
    &=-\frac{\tau}{\tau_R}\left[f_{B G}\left(\tau, p_T,\left|p_\eta\right|\right)-f_{e q}\left(\frac{p^\tau}{T(\tau)}\right)\right],\\
    p^\mu \partial_\mu 
    \delta f
    &= -\frac{p^\tau}{\tau_R} \delta f+\frac{p_\mu \delta u^\mu}{\tau_R}\left[\left(f_{e q}-f_{B G}\right)+\frac{p^\tau}{T} f_{e q}^{\prime}\right]\\
    &\phantom{=}-\frac{p^\tau}{\tau_R} \delta T\left[\frac{1}{\tau_R} \frac{\partial \tau_R}{\partial T}\left(f_{e q}-f_{B G}\right)+\frac{p^\tau}{T^2} f_{e q}^{\prime}\right].
    \end{aligned}
    \end{equation}
where only the linear-order terms are retained for the latter.
Since the response functions are first derived in Fourier space, we also rewrite the Boltzmann equation for the perturbations into Fourier space to eliminate the explicit spatial dependence, as follows:

\begin{equation}
\begin{aligned}
\delta f\left(\tau, \mathbf{x}, \eta, \mathbf{p}_{\mathbf{T}}, p_\eta\right) =
&\int \frac{d^2 \mathbf{k}_{\mathbf{T}}}{(2 \pi)^2} \int \frac{d k_\eta}{2 \pi} \\
&\delta f_k\left(\tau, \mathbf{p}_{\mathbf{T}}, p_\eta\right) e^{i\left(\mathbf{k}_{\mathbf{T}} \cdot \mathbf{x}+k_\eta \eta\right)} .
\end{aligned}
\end{equation}

The Boltzmann equations in Eq.~(\ref{eq:boltz-dist}) are then reformulated using the method of moments, which transforms the partial differential equations into a set of coupled ordinary differential equations.
The quantities to be solved are the moments $E_l^m$ and $\delta E_{l,k}^m$ of the corresponding distribution function, defined as
\begin{equation}
\begin{aligned}
{E_l^m(\tau)}=&\nu_{e f f} 
\int \frac{d p_\eta}{2 \pi} \int \frac{d^2 \mathbf{p}_{\mathbf{T}}}{(2 \pi)^2} 
\tau^{1 / 3} p^\tau \\
&{\times Y_l^m\left(\phi_{\mathbf{p}_{\mathbf{T}}}, \theta_\eta\right) f_{B G}\left(\tau, p_T,\left|p_\eta\right|\right)}.\\
{\delta E_{l, k}^m(\tau)}=
&\int \frac{d p_\eta}{2 \pi} \int \frac{d^2 \mathbf{p}_{\mathbf{T}}}{(2 \pi)^2} \tau^{1 / 3} p^\tau \\
&\times Y_l^m\left(\phi_{\mathbf{p}_{\mathbf{T}} \mathbf{k}_{\mathbf{T}}}, \theta_\eta\right) 
{\times \delta f_k\left(\tau, \mathbf{p}_{\mathbf{T}}, p_\eta\right)}.
\end{aligned}
\end{equation}
Following~\cite{Kamata:2020mka}, the resulting coupled ordinary differential equations are numerically solved using the fourth-order Runge-Kutta method.

Since large longitudinal momenta cannot be maintained at early times, and the non-equilibrium attractor of kinetic theory suggests that all initial conditions eventually converge to the same evolution, we employ an initial condition in which the longitudinal momentum distribution is taken as a Dirac delta function.

\begin{equation}
    \begin{aligned}
    f_{B G}\left(\tau_0, p_T, p_\eta\right)
    &=\frac{(2 \pi)^3}{\nu_{\text {eff }}} \delta\left(p_\eta\right) \frac{\mathrm{d} N_0}{\mathrm{~d} \eta \mathrm{~d}^2 \mathbf{p}_{\mathbf{T}} \mathrm{d}^2 \mathbf{x}}, 
        \end{aligned}
\end{equation}  
where, due to the simplicity of the kinetic theory in relaxation time approximation, the concrete form of the initial (transverse) momentum distribution ${\mathrm{d} N_0}/{\mathrm{~d} \eta \mathrm{~d}^2 \mathbf{p}_{\mathbf{T}} \mathrm{d}^2 \mathbf{x}}$ is irrelevant.

By following the logic of \kompost~\cite{Kurkela:2018wud,Kurkela:2018vqr,Kamata:2020mka}, the initial conditions for the  linearized perturbations are taken to correspond to a small shift of the typical transverse momentum $p_T$, free-streaming up to $\tau_0$, such that
\begin{equation}
    \begin{aligned}
    \delta f_k\left(\tau_0, \mathbf{p}_{\mathbf{T}}, p_\eta\right)
    &=-\left(\frac{\left|\mathbf{p}_{\mathbf{T}}\right|}{3} \partial_{\left|\mathbf{p}_{\mathbf{T}}\right|} f_{B G}^{(0)}\left(\left|\mathbf{p}_{\mathbf{T}}\right|, p_\eta\right)\right)\\
    &\times e^{-i\left(\frac{\mathbf{p_T} \cdot \mathbf{k_T}}{\left|\mathbf{p_T}\right|} \tau_0-k_\eta \operatorname{arcsch}\left(\frac{\mathbf{p_T} \tau_0}{p_\eta}\right)\right)}.
    \end{aligned}
\end{equation}    

We note that due to the conformal scaling properties of the Boltzmann equation in the relaxation time approximation, the dependence on the relaxation time or coupling constant can be completely eliminated by considering $\tilde{\omega}=T(\tau) \tau / (4\pi \eta_v/s)$ as the relevant time variable~\cite{Kamata:2020mka}, thus making the Green's functions universal functions of the variables $\tilde{\omega},|\mathbf{k_T}| \Delta \tau$ and $k_\eta$ in Fourier space, and respectively  $\tilde{\omega},|\mathbf{r}|/\Delta\tau$ and  $\eta$ in coordinate space.

\subsubsection{Green's functions in Fourier space}
\begin{figure*}
    \centering
    \includegraphics[width=0.9\linewidth]{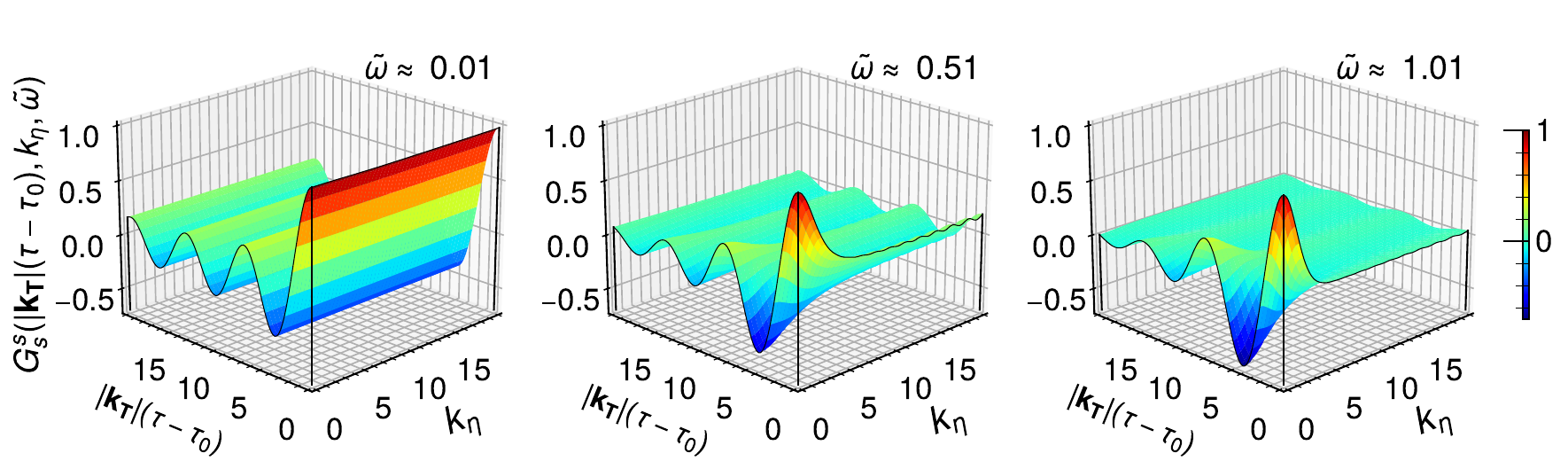}\\
    \includegraphics[width=0.9\linewidth]{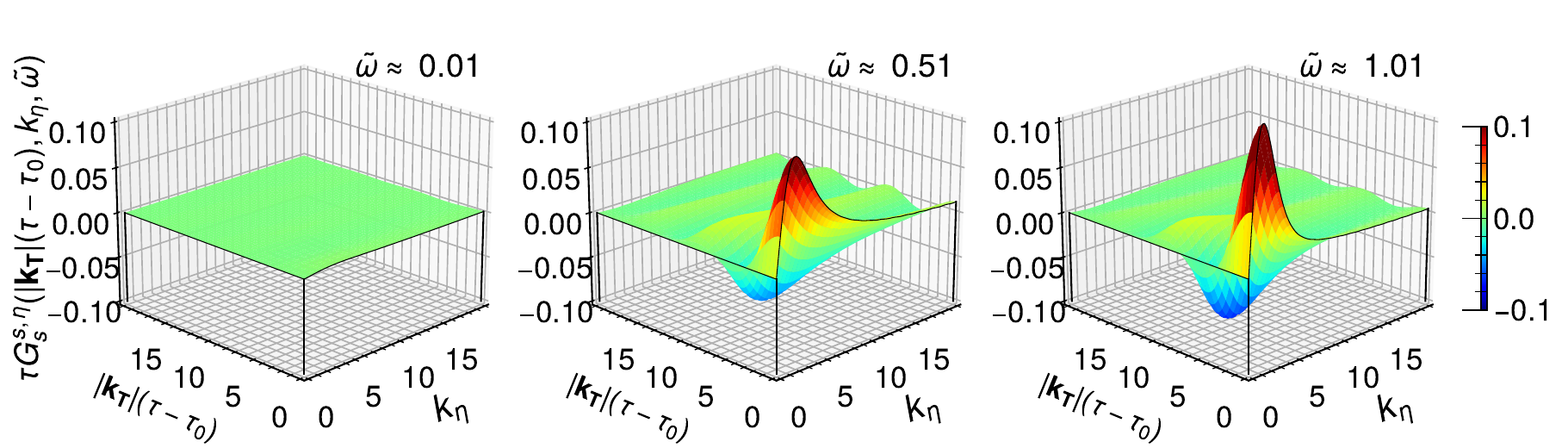}
    \caption{Evolution of the energy response ($G_s^s(|\mathbf{k_T}|\Delta\tau, k_\eta)$) and longitudinal momentum response ($G_s^{s,\eta}(|\mathbf{k_T}|\Delta\tau, k_\eta)$) to an initial energy perturbation as a function of $|\mathbf{k_T}|\Delta\tau$ and $k_\eta$. Different panels correspond to different points in time. The color indicates the magnitude of respective response functions.}
    \label{fig:Gs-k-RTA}
\end{figure*}

In Fig.~\ref{fig:Gs-k-RTA}, we show the evolution of the energy response $G_s^s(|\mathbf{k_T}|\Delta\tau\equiv |\mathbf{k_T}|(\tau-\tau_0), k_\eta)$ (upper panels) and the longitudinal momentum response $G_s^{s,\eta}(|\mathbf{k_T}|\Delta\tau, k_\eta)$ (lower panels) to an initial energy perturbation as functions of transverse ($|\mathbf{k_T}|\Delta\tau$) and longitudinal ($k_\eta$) wavenumber. Different columns in Fig.~\ref{fig:Gs-k-RTA} correspond to different evolution times $\tilde{\omega}=T(\tau) \tau / (4\pi \eta_v/s)$.
For the energy response, at very early times, a free-streaming behavior is observed (see App.~\ref{free-streaming} for more details), where the oscillations extend to large $|\mathbf{k_T}|\Delta\tau$ values and exhibit an almost uniform pattern along the $k_\eta$ direction. As time evolves, the response gradually decays, and only the low-wavenumber modes eventually survive.
In contrast, a longitudinal momentum response to an initial energy perturbation is absent at early times, as the momentum response needs to build up dynamically. As the system evolves, however, the response is gradually generated through the energy gradients and increases over time, while remaining mainly concentrated in the low-wavenumber region.
The additional response functions $G_s^v(|\mathbf{k_T}|\Delta\tau, k_\eta)$, $G_s^{t,\delta}(|\mathbf{k_T}|\Delta\tau, k_\eta)$, $G_s^{t,k}(|\mathbf{k_T}|\Delta\tau, k_\eta)$, $G_s^{\eta}(|\mathbf{k_T}|\Delta\tau, k_\eta)$ and $G_s^{v,\eta}(|\mathbf{k_T}|\Delta\tau, k_\eta)$ are presented in Fig.~\ref{fig:more-Gs-k-RTA} of Appendix~\ref{more-res}.
\subsubsection{Green's functions in coordinate space}
\begin{figure*}
    \centering
    \includegraphics[width=0.9\linewidth]{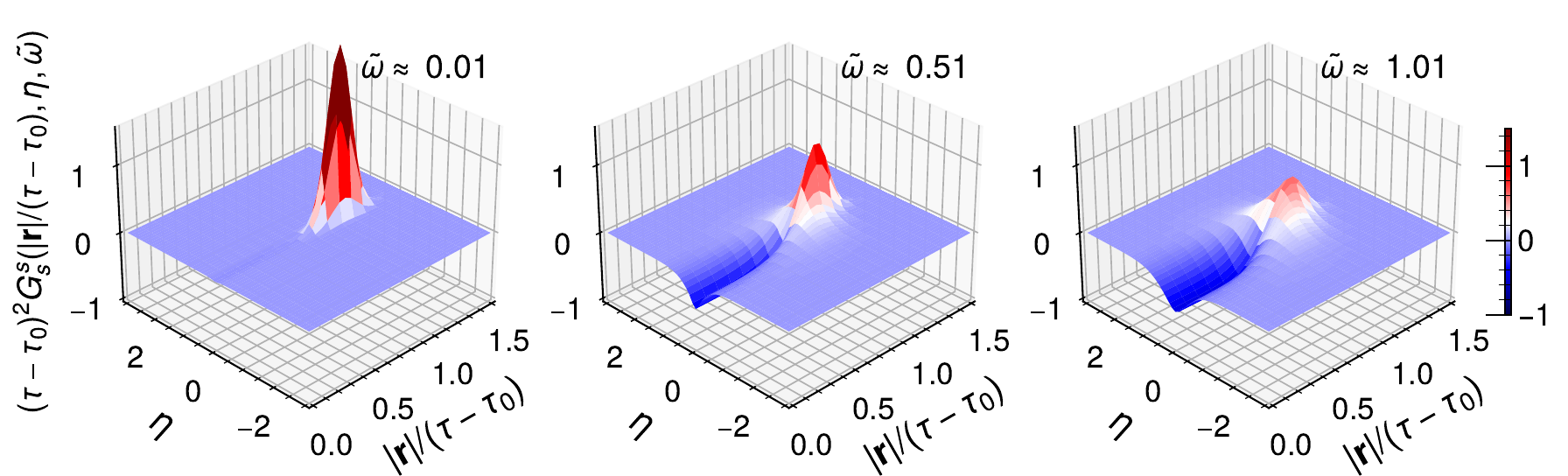}\\
    \includegraphics[width=0.9\linewidth]{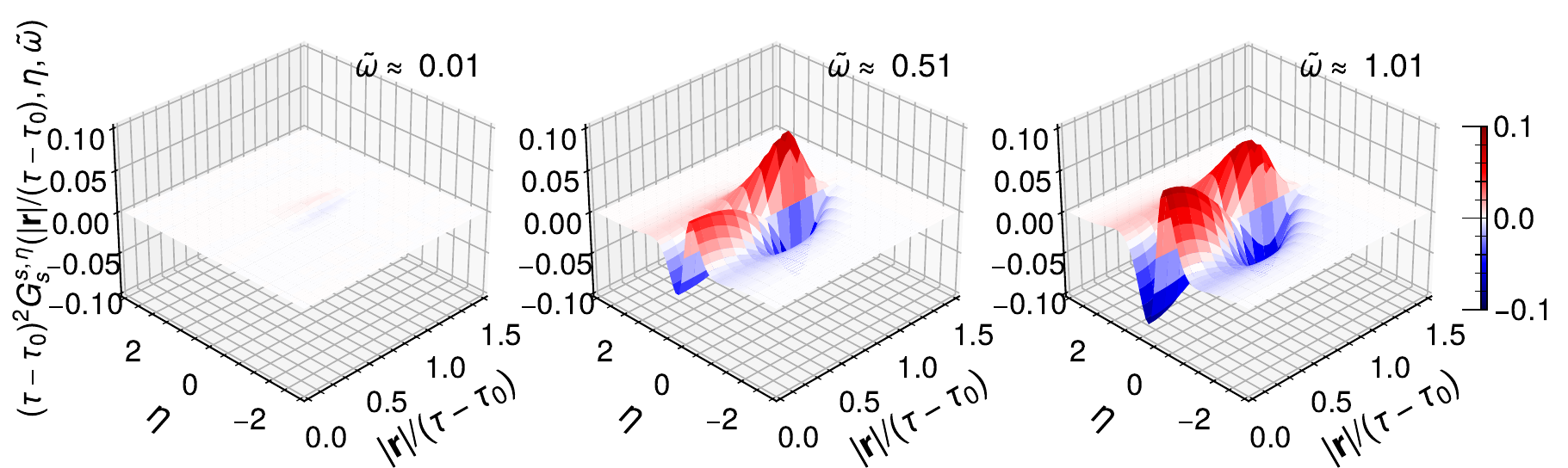}
    \caption{Evolution of the energy response ($G_s^s(|\mathbf{r}|/\Delta\tau, \eta)$) and longitudinal momentum response ($G_s^{s,\eta}(|\mathbf{r}|/\Delta\tau, \eta)$) to an initial energy perturbation as a function of $|\mathbf{r}|/\Delta\tau$ and $\eta$. Different panels correspond to different points in time. The color indicates the magnitude of respective response functions.}
    \label{fig:Gs-r-RTA}
\end{figure*}
Based on the results in Fourier space, the response functions in coordinate space are obtained through the reverse Fourier transformation as defined in Eq.~(\ref{eq:rev-fourier}).
In practice, since only a finite set of discrete points in Fourier space ($|\mathbf{k_T}|\Delta\tau, k_\eta$) are calculated within a limited region of the kinetic theory evolution, 
interpolation between these points and extrapolation beyond the covered region (using the free-streaming approximation) are performed to obtain a smooth Fourier transform as in Refs.~\cite{Kurkela:2018vqr,Keegan:2016cpi}.
Furthermore, Gaussian smearing kernels, ${{\exp(-\sigma^2 |\mathbf{k_T}|^2\Delta \tau^2/2)}}$ and ${{\exp(-\sigma_\eta^2 k_\eta^2/2)}}$, are applied to regulate the large wavenumber tails, at early times where viscous damping of large $k_T$ and $k_\eta$ modes is not yet present.
If not state otherwise, in this work, we use $\sigma=0.1, \sigma_\eta=0.1$. %{\color{blue} $\sigma$ is not dimensionless, should be $\sigma=0.1 Q_s^{-1}$?, SS: probably $\Delta \tau$ -- fixed this  }

In Fig.~\ref{fig:Gs-r-RTA}, we show the evolution of energy response $G_s^s(|\mathbf{r}|/\Delta\tau, \eta)$ (upper panels) and longitudinal momentum response $G_s^{s,\eta}(|\mathbf{r}|/\Delta\tau, \eta)$ (lower panels) to an initial energy perturbation as functions of $|\mathbf{r}|/\Delta\tau$ and $\eta$.
At very early times, the energy response exhibits free-streaming behavior resulting in a localized peak around $|\mathbf{r}|/\Delta\tau=1$ and $\eta=0$ as
in the transverse plane the perturbation propagates at the speed of light ($|\mathbf{r}|/\Delta\tau\approx 1$), while along the longitudinal direction, it remains nearly a delta function ($\delta(\eta)$), indicating negligible longitudinal diffusion.
As the system evolves,  a finite longitudinal width of the response appears due to the underlying diffusive dynamics, such that by the time $\tilde{\omega}\approx 1$ when the system can be considered sufficiently close to equilibrium for hydrodynamics to apply, initial perturbations have spread around one unit in rapidity.  By spreading out, transversely and longitudinally, the overall magnitude of the response gradually decreases, where at later times the Green's function features an outer peak signaling the outward transport of the initial energy perturbation as well as a central dip characterizing the wake. While the longitudinal momentum response, depicted in the lower panels vanishes at early times, as the system evolves, the longitudinal momentum is gradually generated by the energy density gradients in longitudinal direction and increases over time. 
The flow pattern shows inward motion around the dip and outward motion around the peak.
The remaining response functions, namely $G_s^v(|\mathbf{r}|/\Delta\tau, \eta)$, $G_s^{t,\delta}(|\mathbf{r}|/\Delta\tau, \eta)$, $G_s^{t,k}(|\mathbf{r}|/\Delta\tau, \eta)$, $G_s^{\eta}(|\mathbf{r}|/\Delta\tau, \eta)$, $G_s^{v,\eta}(|\mathbf{r}|/\Delta\tau, \eta)$ are presented in Fig.~\ref{fig:more-Gs-r-RTA} of the Appendix.

\subsubsection{Hydrodynamic constitutive relations}
\begin{figure*}
    \centering
    \includegraphics[width=0.33\linewidth]{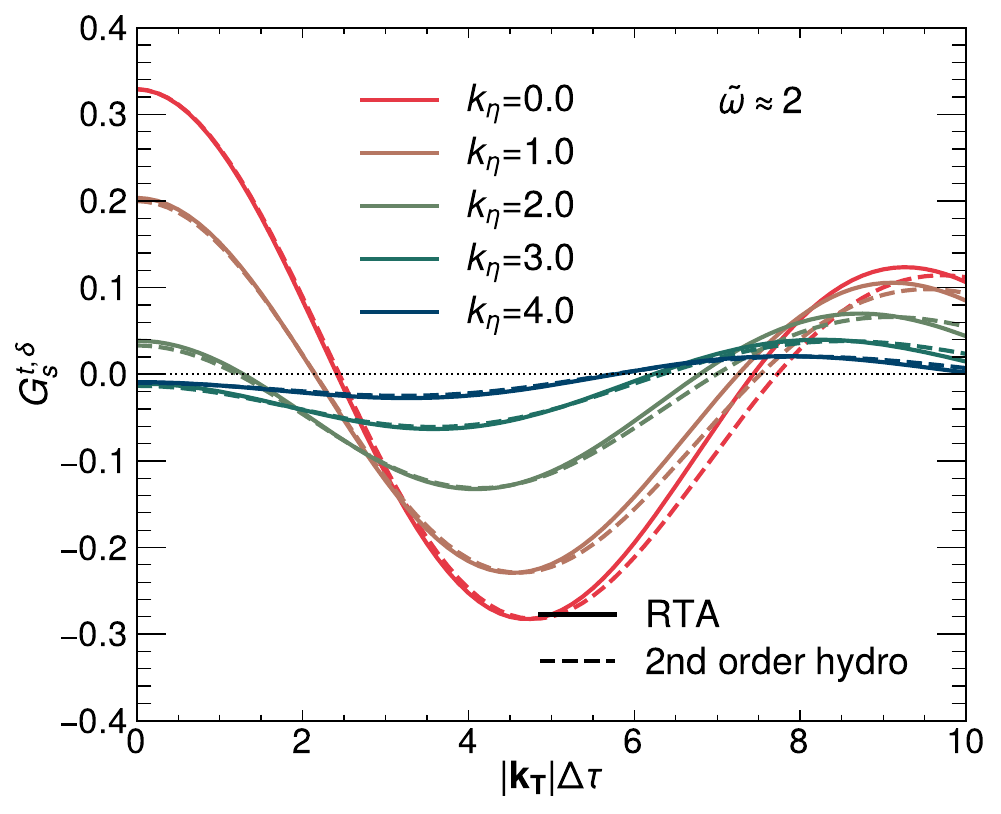}
    \includegraphics[width=0.64\linewidth]{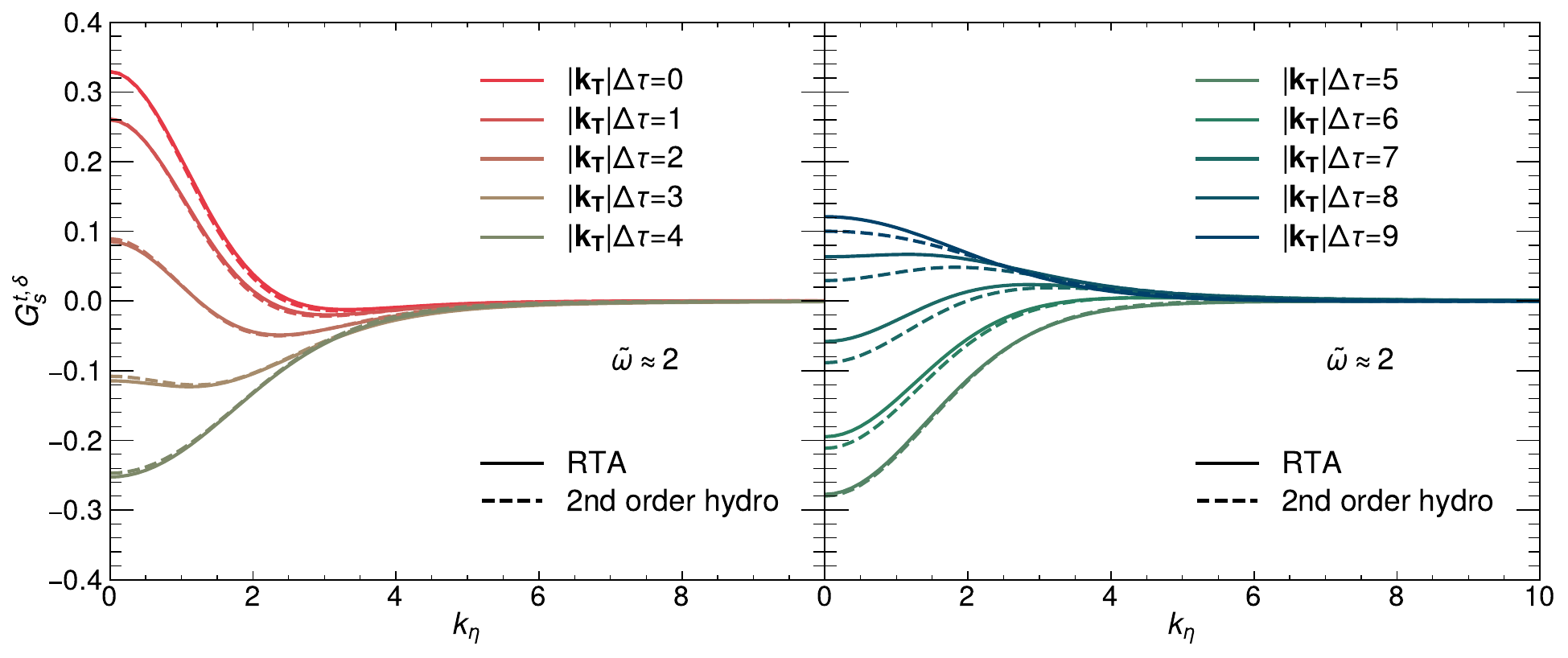}
    \includegraphics[width=0.32\linewidth]{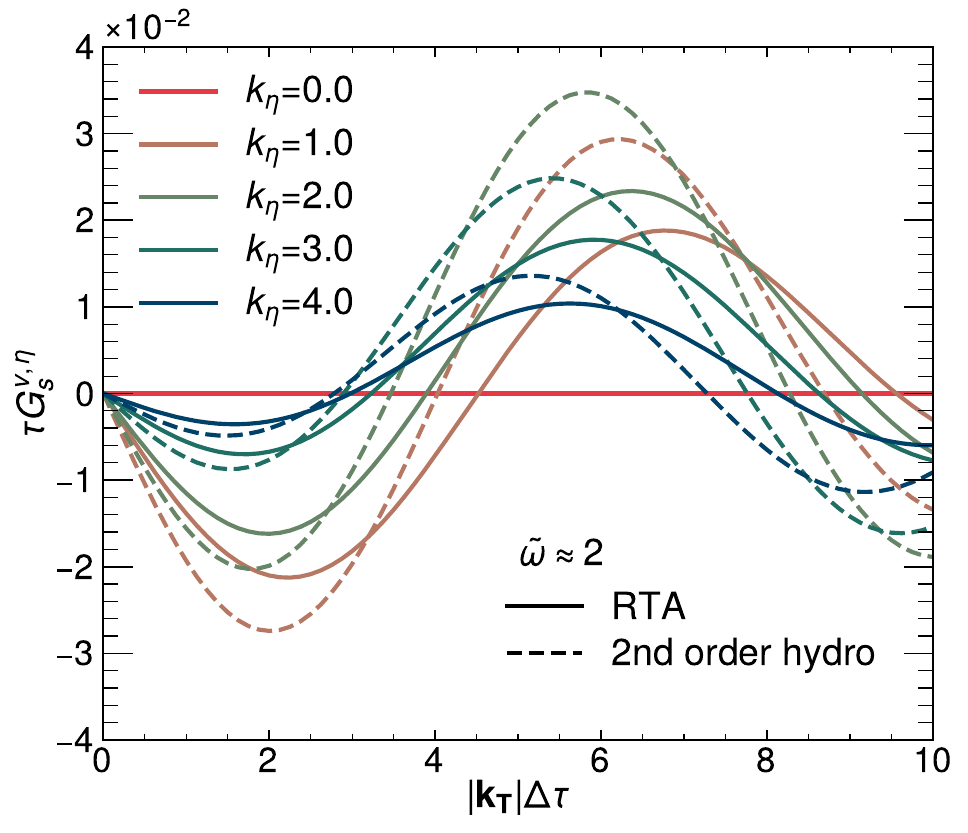}
    \includegraphics[width=0.32\linewidth]{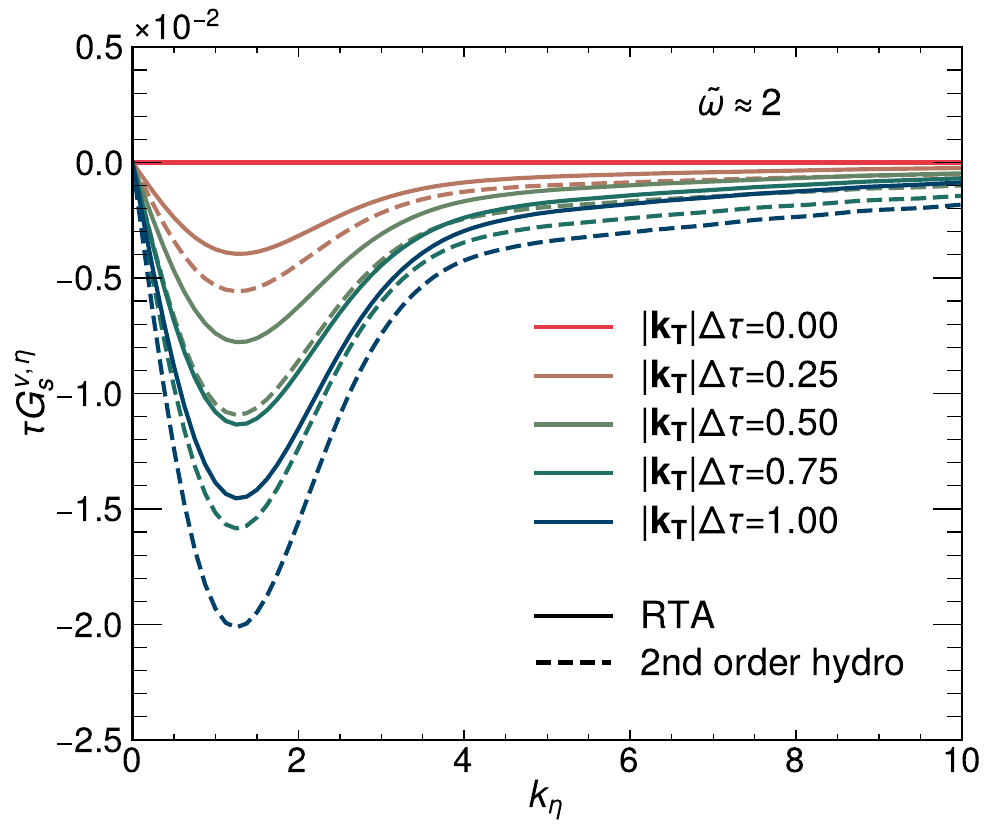}
    \includegraphics[width=0.32\linewidth]{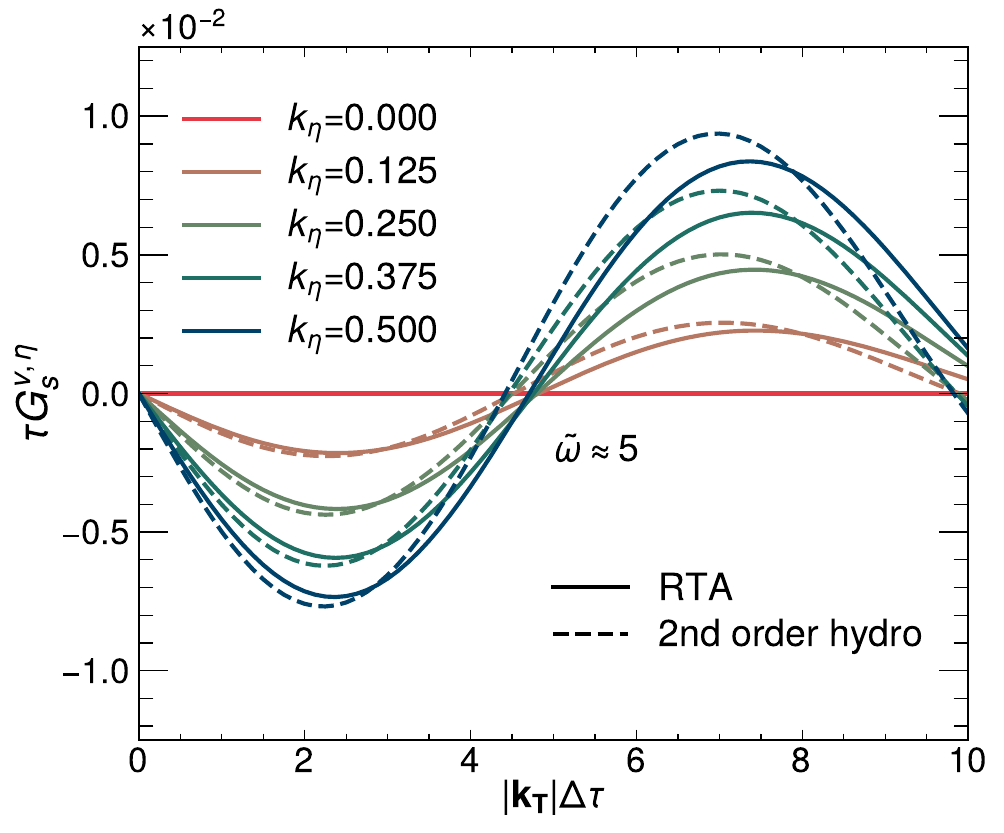}
    \caption{Examination of the constitutive relations, Eqs.(\ref{eq:const1}) and (\ref{eq:const2}), at scaled times $\tilde{w}\approx 2$ and $5$.
    The response functions $\tilde{G}_s^{t,\delta}$ (upper panels) and $\tilde{G}_s^{v,\eta}$ (lower panels) from RTA dynamics (denoted as RTA) are compared with those reconstructed from $\tilde{G}_s^s$, $\tilde{G}_s^v$ and $\tilde{G}_s^{s,\eta}$, $\tilde{G}_s^v$ according to the constitutive relations at various $|\mathbf{k_T}|\Delta\tau$ or $k_\eta$.}
    \label{fig:consti}
\end{figure*}
Since the non-equilibrium system approaches equilibrium over the course of its evolution, at late times, the response functions from kinetic theory are expected to approach the hydrodynamic regime, in which the different components are interrelated by the hydrodynamic constitutive relations.
By following the strategy of Ref.~\cite{Kurkela:2018vqr}, we can thus identify relations between different components of energy-momentum tensor, which emerge in second order hydrodynamics for small wavenumber perturbations. Extending the analysis of Ref.~\cite{Kurkela:2018vqr} to include longitudinal dynamics $(k_\eta\neq 0)$,
we obtain the following relations between independent  components of the response functions for an initial energy perturbations 
\begin{align}\label{eq:const1}
&\tilde G_s^{t,\delta} (\tau,\ktt,k_\eta)=\nonumber\\ 
&{ \left(P+\frac{\eta_v}{2\tau}+\frac{1-c_s^2}{3}\frac{\tau_\pi
	\eta_v-{\lambda_1}}{\tau^2}-\frac{c_s^2 \tau_\pi \eta_v}{2} |\mathbf{k_T}|^2
	\right)\!\frac{\tilde 
	G_s^s 
(\tau,\ktt,k_\eta)}{\TBg^{\tau\tau}}}\nonumber\\
&{\small+\left(\frac{2}{3} 
\eta_v-\eta_v 
\tau_\pi 
\frac{6c_s^2+4}{9\tau}
+\frac{16}{9\tau}\lambda_1 \right)\frac{\ktt \tilde G_s^v(\tau,\ktt,k_\eta)}{\TBg^{\tau\tau}+\frac{1}{2}\TBg^k_k},}
\end{align}
\begin{equation}\label{eq:const2}
    \begin{aligned}
    &\tilde{G}_s^{v,\eta}(\tau,|\mathbf{k_T}|,|k^\eta|)= 
    \left(
    \eta_v \tau_\pi {c_s^2+1/3 \over \tau} - \eta_v - 
    {4\lambda_1\over 3\tau}
    \right) \times \\
    &\left(
    {|\mathbf{k_T}| \tilde{G}_s^{s,\eta}(\tau,|\mathbf{k_T}|,|k^\eta|) \over \TBg^{\tau\tau}+\TBg^{\eta}_\eta}
    +{|k^\eta| \tilde{G}_s^v(\tau,|\mathbf{k_T}|,|k^\eta|) \over \TBg^{\tau\tau}+{1\over 2}\TBg^k_k}
    \right).
    \end{aligned}
\end{equation}
where, in order to match the underlying microscopic kinetic description, we have 
$c_s^2=1/3
$ and the transport coefficients are taken as~\cite{York:2008rr, Dash:2023ppc, Jaiswal:2013npa}
\begin{equation}
\begin{aligned}
    &\tau_\pi=\frac{5\eta_v}{\epsilon+P},\\
    &\lambda_1= \frac{25 \eta_v^2}{7(\epsilon+P)}.
\end{aligned}
\end{equation}

In Fig.~\ref{fig:consti}, we examine the constitutive relations, Eqs.~(\ref{eq:const1}) and (\ref{eq:const2}) at scaled time $\tilde{w}\approx 2 \text{ or } 5$. 
The response functions $\tilde{G}_s^{t, \delta}$ and $\tilde{G}_s^{v,\eta}$ obtained from RTA dynamics are compared with those (denoted as 2nd order hydro) reconstructed from $\tilde{G}_s^s$, $\tilde{G}_s^v$ and $\tilde{G}_s^{s,\eta}$, $\tilde{G}_s^v$ according to the constitutive relations at various $|\mathbf{k_T}|\Delta\tau$ or $k_\eta$.
Strikingly, as shown in the upper panels, the constitutive relation between the pressure, energy and momentum responses, Eq.~(\ref{eq:const1}),  is well satisfied  by $\tilde{\omega}=2$ for even for relatively large wavenumbers $|\mathbf{k_T}|\Delta\tau$ and $k_\eta$.
Conversely, as shown in the lower panels, the hydrodynamic constitutive relation between the longitudinal component of the shear stress and momentum response, is
much harder to fulfill, except at very late times ($\tilde{w}\approx 5$) and for rather small wavenumbers $|\mathbf{k_T}|\Delta\tau, k_\eta \lesssim 1$, indicating that hydrodynamization is more difficult to achieve for the longitudinal components of the energy-momentum tensor than for the transverse ones. 

\subsubsection{From RTA to QCD: Comparison of energy response functions at $k_T=0$ and $k_\eta=0$}
So far we have presented the calculation of (3+1)D non-equilibrium response functions in kinetic theory in the relaxation time approximation. One step further is to calculate these response functions in QCD kinetic theory by formulating the collision integrals for background and perturbation contributions as 
\begin{equation}
    \begin{aligned}
    \partial_\tau 
    f_{a,\rm BG}
    &=-C_a[f]\\
    \frac{p^\mu}{p^{\tau}} \partial_{\mu} \delta f_{\vk,a}
    &= -\delta C_a[f,\delta f_{\vk}].
    \end{aligned}
\end{equation}
where the index $a$ runs through all QCD species $a=g,q,\bar{q}$ and the wavenumber vector is $\vk=(\mbk_T,k_{\eta})$. The background collision integrals for QCD include both $2\leftrightarrow 2$ elastic and $1\leftrightarrow 2$ inelastic processes. The elastic collision process is calculated 
\begin{align}
\label{eq:ElasticKern}
&C^{{2\leftrightarrow2}}_a[f](\vp_1)
=\frac{1}{2 \nu_{a}}\frac{1}{2 E_{p_1}}\\
\nonumber
&{\sum_{b,c,d \in \{g,q_f,\bar{q}_f\}}}
\int d\Pi_{2\leftrightarrow2}|\mathcal{M}_{cd}^{ab}(\vp_1,\vp_2|\vp_3,\vp_4)|^2F_{cd}^{ab}(\vp_1,\vp_2|\vp_3,\vp_4)
\end{align}
up to leading order scattering in matrix amplitude square $|\mathcal{M}_{cd}^{ab}(\vp_1,\vp_2|\vp_3,\vp_4)|^2$ for all possible processes $a,b\leftrightarrow c,d$ with $a,b,c,d=g,q,\bar{q}$ in QCD, and $F_{cd}^{ab}(\vp_1,\vp_2|\vp_3,\vp_4)$ is the quantum statistical factor.
The inelastic radiation process is calculated  as
\begin{align}
\label{eq:inelasticKern}
\nonumber
&C^{{1\leftrightarrow2}}_a[f](p)=\frac{1}{2\nu_{a}} \sum_{b,c \in \{g,q_f,\bar{q}_f\}}\int_0^1dz\;, \\ &\bigg[
\frac{d\Gamma_{bc}^{a}}{dz}(p,z)\nu_{a}F_{bc}^{a}(p|zp,\bar{z}p)\\
\nonumber
&-\frac{1}{z^3}\frac{d\Gamma_{ab}^{c}}{dz}(\frac{p}{z},z)\nu_{c}F_{ab}^{c}(\frac{p}{z}|p,\frac{\bar{z}}{z}p)  \\
\nonumber
&-
\frac{1}{\bar{z}^3}\frac{d\Gamma_{ab}^{c}}{dz}(\frac{p}{\bar{z}},\bar{z})\nu_{c}F_{ab}^{c}(\frac{p}{\bar{z}}|p,\frac{z}{\bar{z}}p)
\bigg]\\
\end{align}
with effective inelastic rates $d\Gamma_{bc}^{a}(p,z)/dz$ for all possible process $a\leftrightarrow b,c$ with $a,b,c=g,q,\bar{q}$ in QCD in the AMY framework~\cite{Arnold:2002zm}, and $F_{ab}^{c}(\frac{p}{z}|p,\frac{\bar{z}}{z}p)$ is the quantum statistical factor.
The detailed implementations, including numerical algorithms, can be found in our previous papers~\cite{Du:2020dvp,Du:2020zqg}.
The perturbation components receive contributions from variations of the scattering matrix elements $\delta |\mathcal{M}^{ab}_{cd}|^2$, radiation rates $\delta (d\Gamma_{bc}^{a}(p,z)/dz)$, as well as the statistical factors $\delta F^{ab}_{cd}$, $\delta F^{a}_{bc}$ and a more detailed discussions of all the above quantities and their numerical implementation will be provided in a forthcoming paper~\cite{Dore:BI2025}.\footnote{Note that linearized perturbations in QCD kinetic theory have already been studied in~\cite{Du:2023bwi} to investigate hydrodynamic/non-hydrodynamic modes in QCD, however, without providing technical details.} 

\begin{figure*}
    \centering
    \includegraphics[width=.8\linewidth]{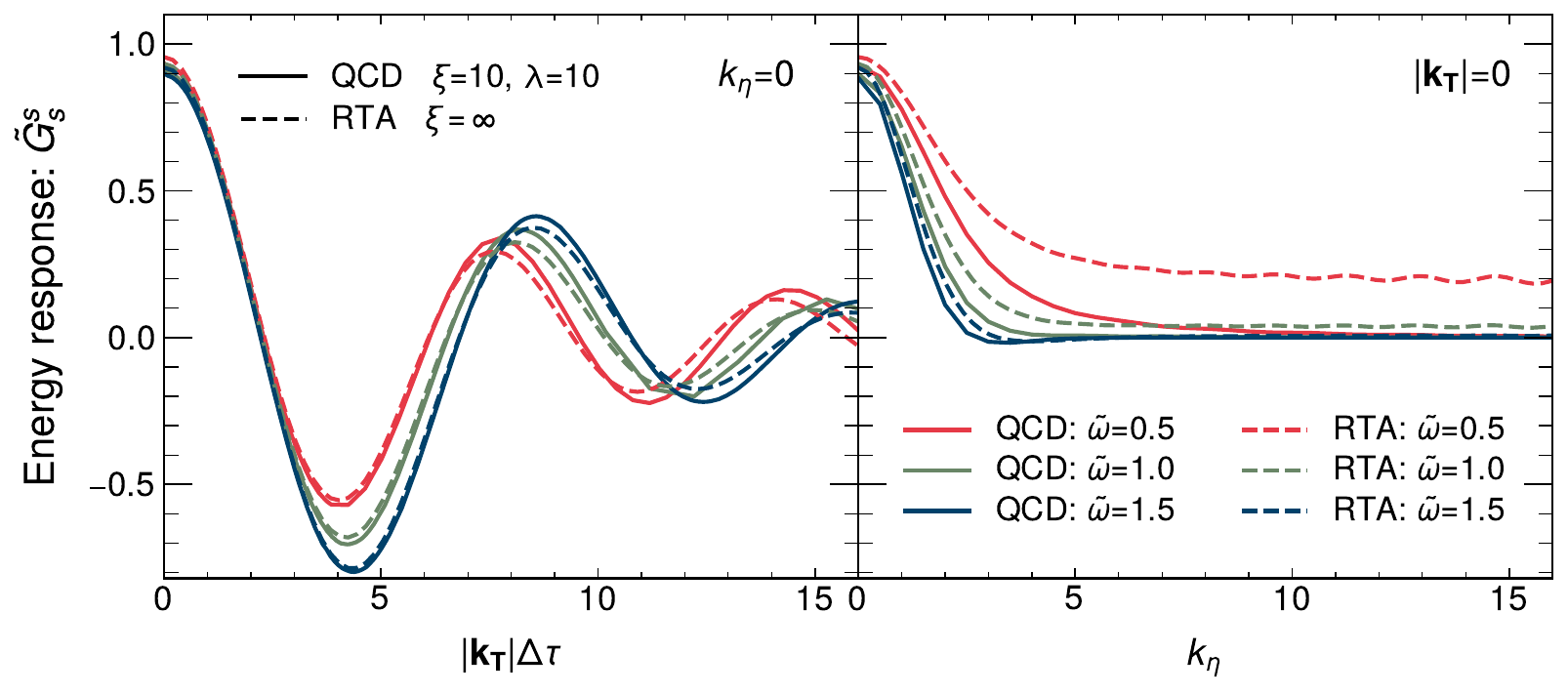}
    \caption{Comparison of the energy response functions in QCD and in RTA for $k_\eta=0$ (left) and $k_T=0$ (right) at three selected times.}
    \label{fig:RTA-QCD}
\end{figure*}

Evidently, the numerical calculations in QCD kinetic theory are significantly more demanding compared to the relaxation time approximation (RTA), and we will therefore only compare the results for the special cases $k_\eta=0$ and $|\mathbf{k_T}|=0$ depicted in the left and right panels of Fig.~\ref{fig:RTA-QCD}. By inspecting the case of $k_\eta=0$, which characterizes the transverse response, we observe no qualitative differences between the behavior in QCD and RTA kinetic theory, which is in line with earlier studies of equilibrium \cite{Du:2023bwi} and non-equilibrium response functions~\cite{Kamata:2020mka} in different kinetic theories.  When considering the longitudinal dynamics of energy perturbations, depicted in the right panel, we again observe the same qualitative behavior between QCD and RTA kinetic theory. 
Differences emerge at early times, where the results in QCD kinetic theory show a stronger damping of the large $k_\eta$ modes, which can be attributed to a difference in the initial conditions, where for numerical convenience in the case of QCD kinetic theory the simulations are initialized with a large but finite momentum anisotropy, as opposed to the infinite momentum anisotropy employed in RTA calculations. Due to this difference longitudinal energy diffusion starts right away in the case of QCD kinetic theory, however at later times, where the system can sustain a larger longitudinal pressure and longitudinal energy diffusion becomes more and more pronounced the QCD and RTA results are again in good agreement.

\subsection{\kompost-3D}
We proceed by embedding the three-dimensional response functions discussed above into the pre-equilibrium evolution model, \kompost, thereby constructing a complete (3+1)D evolution of the pre-hydrodynamic stage, referred to as \kompost-3D.

Following the general formalism of \kompost~\cite{Kurkela:2018wud, Kurkela:2018vqr, Kamata:2020mka} outlined in Sec.~\ref{sec:intro}, we include the non-equilibrium response functions discussed above to extend the description to  (3+1)D. We aim to describe the evolution of the energy-momentum tensor from $\tau_{\rm EKT}$ to $\tau_{\rm hydro}$ at any given point $(\x_0, \eta_0)$.
To this end, at the starting time $\tau_{\rm EKT}$, we first build the background as a Gaussian weighted average around the point of interest, 
\begin{align}\label{eq:bg}
    \TBg_{\mathbf{x}_0, \eta_0}^{\tau \tau}(\tau_{\rm EKT}) \equiv 
    \int d^2 \mathbf{x} d\eta 
    &\frac{1}{2 \pi \sigma^2} e^{\frac{-\left(\mathbf{x}^{\prime}-\mathbf{x}_0\right)^2}{2 \sigma^2}}
    \frac{1}{\sqrt{2 \pi \sigma_{\eta}^2}} e^{\frac{-\left(\mathbf{\eta}-\mathbf{\eta}_0\right)^2}{2 \sigma_{\eta}^2}}\nonumber\\
    &\times\ 
    T^{\tau \tau}\left(\tau_{\rm EKT}, \mathbf{x},\eta\right),
\end{align}
where the arguments $(\x_0, \eta_0)$ on the left-hand side are rewritten as subscripts to emphasize that the background is locally transverse homogeneous and longitudinally boost-invariant.
In this work, the respective Gaussian widths are chosen as 
$\sigma=(\tau_{\rm hydro}-\tau_{\rm EKT}) /2$, $\sigma_{\eta}=3/2$. 
In principle, all components of perturbations
around the homogeneous diagonal background energy-momentum tensor $\TBg^{\mu\nu}_{\mathbf{x}_0, \eta_0}$ 
should contribute to the input of the subsequent hydrodynamic evolution.
As a first step, this work only considers the initial energy perturbation $\delta T^{\tau\tau}_{\mathbf{x}_0, \eta_0}(\tau_{\rm EKT},\mathbf{x},\eta)$ 
around the spatial point $(\x_0, \eta_0)$, 
which is obtained as 
\begin{equation}
    \delta T^{\tau\tau}_{\mathbf{x}_0, \eta_0}(\tau_{\rm EKT},\mathbf{x},\eta)
    = 
    T^{\tau\tau}(\tau_{\rm EKT}, \mathbf{x}, \eta)-\TBg^{\tau\tau}_{\mathbf{x}_0, \eta_0}(\tau_{\rm EKT}).
\end{equation}
It should be noted that the double dependence on spatial coordinates in $\delta T^{\tau\tau}_{\mathbf{x}_0, \eta_0}(\tau_{\rm EKT},\mathbf{x},\eta)$ reflects that the perturbation evaluated at point $(\x, \eta)$ generally varies with the choice of the reference point $(\x_0, \eta_0)$.

With regard to the evolution of the background components of the energy-momentum tensor during the pre-equilibrium stage, the well-known hydrodynamic attractor behavior is used to infer the components of the background energy-momentum tensor at hydrodynamization time as in Ref.~\cite{Kurkela:2018vqr, Kurkela:2018wud}.
We will briefly summarize it here again.
In the study of background evolution, we only consider the diagonal components of the energy-momentum tensor, i.e. $\TBg^{\mu\nu}=\text{diag}\,( e, P_T, P_T, \tfrac{1}{\tau^2}P_L)$, and based on the study of kinetic theory~\cite{Kurkela:2018vqr}, the background evolution  can be parametrized in terms of a single scaling function $\mathcal{E}(\OmegaId)$ for the energy density. 
In practice, the energy density, transverse and longitudinal pressure at the hydrodynamic initialization time $\tauhydro$ can be read off from the following scaling functions
\begin{align}
&e\phantom{L}=\nu_g \frac{\pi^2}{30}\TId^4\,\mathcal{E}
\left(\OmegaId\right)
,\label{eq:para-E}\\
&P_L=\nu_g \frac{\pi^2}{90}\TId^4\,\mathcal{P}(\OmegaId),\label{eq:para-PL}\\
&P_T=\nu_g \frac{\pi^2}{90}\TId^4\,\Big(\frac{3}{2}\mathcal{E}(\OmegaId)-\frac{1}{2}\mathcal{P}(\OmegaId)\Big).\label{eq:para-PT}
\end{align}
where $\TId$ denotes the overall energy scale and $\OmegaId=\frac{\tau}{\tau_{\rm eq}}$ is a scaled time variable, which is determined by the shear viscosity 
and the ideal temperature $\TId(\tau;\Lambda_T)$ as
\begin{equation}
   \tau_{\rm eq}(\tau;\Lambda_T)
   \equiv\frac{4\pi \eta_v/s}{\TId(\tau;\Lambda_T)}.\label{eq:tauR}
\end{equation}
Note that in contrast, to the more frequently used scaling variable $\tilde{\omega}=\frac{T(\tau)\tau}{4\pi \eta/s}$, the scaling variable $\OmegaId$ in \kompost\ is defined in terms of the ideal temperature~\cite{Kurkela:2018vqr}
\begin{equation}
\TId(\tau;\Lambda_T)\equiv\frac{\Lambda_T}{(\tau \Lambda_T)^{1/3}} 
,\label{eq:TId} 
\end{equation}
such that $\TId$ behaves as $\tau^{1/3}$ at all times. The energy scale $\Lambda_T=(T\tau^{1/3})_{\infty}^{3/2} $ is self-consistently determined from the initial background energy density $T^{\tau\tau}(\tau_{\rm EKT})$ and the parametrization in Eq.~(\ref{eq:para-E}) as 
\begin{multline}
e(\tauekt)=\TBg^{\tau\tau}_{\x_0}(\tauekt)
= 
\nu_{g}~\frac{\pi^2}{30}~\TId^4(\tauekt;\Lambda_T) \\ \times \mathcal{E}\left[\OmegaId=\frac{\tauekt
 \TId(\tauekt;\Lambda_T) }{4\pi \eta_v/s}\right].\label{eq:implicit}
\end{multline}
By virtue of Eq.~(\ref{eq:para-E}), the scaling variables are simply related according to
$\tilde{\omega}= \OmegaId \mathcal{E}^{1/4}(\OmegaId)$, and  as in the original version of \kompost~we use $\OmegaId$ in the numerical implementation.

While the energy attractor function $\mathcal{E}(\OmegaId)$ is parametrized as in Ref.~\cite{Kurkela:2018vqr}, the auxiliary function $\mathcal P(\OmegaId)$ in Eq.~(\ref{eq:para-PL}) is readily determined by the energy balance equation for Bjorken expansion, i.e. $\partial_\tau (\tau e)=-P_L$,  such that
\begin{eqnarray}
\mathcal{P}(\OmegaId)=\mathcal{E}(\OmegaId) - 2 \OmegaId  \mathcal{E}'(\OmegaId)\;.
\end{eqnarray}

On top of the background field, the perturbations $\delta T^{\mu \nu}(\x_0, \eta_0)$
at the hydrodynamic starting time $\tau_\text{hydro}$ 
are determined by 
convoluting the pre-calculated  linear kinetic response functions, $G^{\mu\nu}_{\alpha\beta}$, with the initial energy perturbation as follows.
\begin{equation}
    \begin{aligned}
    &\frac{
    {\delta T^{\mu \nu}(\tau_{\rm hydro}, \x_0, \eta_0)}
    }{\overline{T}_{\mathbf{x},\eta}^{\tau \tau}(\tau_{\rm hydro})}
    =\int d^2 \mathbf{x}\ d\eta \\
    &\phantom{aaaaaa}
    {G_{\tau\tau}^{\mu \nu}\left(\x, \eta,  \x_0, \eta_0, \tau_{\rm hydro}, \tau_{\rm EKT}\right)}
    \frac{
        {\delta T_{\x_0, \eta_0}^{\tau\tau}
        \left(\tau_{\rm EKT}, \mathbf{x},\eta \right)}
    }
    {\overline{T}_{\x_0,\eta_0}^{\tau \tau}\left(\tau_{\rm EKT}\right)}.
    \end{aligned}
\end{equation}
where the integration is confined to a causal region. While in the transverse plane this restricts the integration to $|\x -\x_0| \leq \Delta\tau$ in the limit $\tau_0 \to 0$, there is no limit on the rapidity separation $|\eta-\eta_0|$. Since the Green's functions themselves have a limited support in both  $|\x -\x_0|/\Delta \tau$ and $|\eta-\eta_0|$, in practice we limit the integration to $|\x -\x_0|<1.3 \Delta \tau$ and $\Delta \eta < 3$, which is sufficient to account for all contributions.

Eventually, the full energy-momentum tensor is reconstructed from the background and perturbations, as
\begin{equation}
T^{\mu\nu}(\tauhydro,\x_0,\eta_0)=\TBg^{\mu\nu}_{\x_0,\eta_0}(\tauhydro)+\delta 
T^{\mu\nu}(\tauhydro,\x_0,\eta_0).
\end{equation}
In order to examine the dependence of hydrodynamic model on the starting time $\tau_{\rm hydro}$, we will consider different values of the hydrodynamization time,
\begin{equation}
    \tau_{\rm hydro}=0.4,\ 0.6,\ 0.8,\ 1.0,\ 1.2,\ 1.4,\ 1.6\ \rm fm.
\end{equation}
and study the subsequent evolution up to the final state hadronic observables.
During the pre-equilibrium stage, the ratio between shear viscosity and entropy is taken as $\eta_v/s$=0.16.
Once we have the energy-momentum tensor at $\tau_{\rm hydro}$, it is decomposed with energy density, flow velocity and shear tensor in Landau frame which can be directly input into the subsequent hydrodynamic expansion,
\begin{equation}
    T^{\mu\nu}=\left(e+P\right) u^\mu u^\nu - P g^{\mu\nu}+\pi^{\mu\nu}
    .
\end{equation}
We note that the regulator developed in Ref.~\cite{Kurkela:2018vqr}
is also utilized for handling the failure to find a local fluid rest frame, especially at the edges or peaks of the medium where the gradients are particularly steep.

\section{Application of \kompost-3D: (3+1)D full evolution of HICs}\label{sec:3D-framework}
Now that we have established \kompost-3D to perform the pre-equilibrium evolution, we can embed this into a (3+1)D  multistage framework employed for the description of relativistic heavy-ion collisions.
Prior to the pre-equilibrium evolution, a saturation-based three-dimensional initial state model, \Dipper, is utilized to generate the initial conditions for \kompost-3D. 
Subsequently, the event-by-event (3+1)D CLVisc hydrodynamic framework~\cite{Pang:2012he, Pang:2018zzo, Wu:2021fjf}, which incorporates second-order dissipative hydrodynamic evolution and a Cooper-Frye hadronic sampler, is employed to simulate the space-time evolution of the QGP medium and thermal hadron production.
Finally, the microscopic hadronic cascade model SMASH is used to describe the subsequent hadronic rescattering and decays. Before we turn to the numerical results of such simulations, we briefly describe the individual modules.

\subsection{\Dipper\ initial state model}
The {\Dipper} model~\cite{Garcia-Montero:2025bpn, Garcia-Montero:2023gex} provides a framework for the initial energy and charge deposition in heavy-ion collisions, formulated within the $k_T$-factorization limit of the Color Glass Condensate (CGC) effective theory.
In this approach, the initial energy density $(e\tau)_0$ is obtained from the first moment of the gluon and (sea)quark distributions,
\begin{equation}
(e\tau)_0 =\int d^2\pT~|\pT|~\left[K_g \frac{dN_{g}}{d^2\xT d^2\pT dy} + \sum_{f,\bar{f}} \frac{dN_{q_f}}{d^2\xT d^2\pT dy}\right]_{y=\eta_s}\;,
\label{eq:EnergyWithKFactor}
\end{equation}

Here, the phenomenological normalization factor $K_g$ is introduced to account for the normalization uncertainties and higher-order corrections to gluon production in accordance with Ref.~\cite{Garcia-Montero:2023gex}.
In the present implementation, \Dipper\ employs the single parton production formulas computed at leading order (LO) to determine the corresponding gluon and quark distributions. %{\color{blue} specify LO here?}

For gluons, the single inclusive production originates from radiative processes. Consequently, the gluon spectrum, $\frac{dN_{g}}{d^2\xT d^2\pT dy}$, can be expressed as the integration  of the two momentum space dipole functions~\cite{Blaizot:2004wv,Gelis:2001da,Gelis:2001dh} of the projectile and target, in the adjoint representation, together with a hard factor (see Refs.~\cite{Dumitru:2001ux,Lappi:2017skr})
\begin{equation}
\begin{split}
\frac{dN_{g}}{d^2\xT d^2\pT dy}=  &\frac{ (N_c^2-1)}{4\pi^3 g^2 N_c}  \int \frac{d^2\qT}{(2\pi)^2}   \frac{d^2\kT}{(2\pi)^2}~\frac{\qT^2\kT^2}{\pT^2}\\
&\times~D_{1,\rm adj}(x_1,\xT,\qT)~D_{2,\rm adj}(x_2,\xT,\kT)\\
&\times~(2\pi)^2\delta^{(2)}(\qT+\kT-\pT)\,\\
\end{split}
\label{eq:gluon_prod_w_uGDFs}
\end{equation}
where $x_{1/2}=\frac{|\pT|}{\sqrt{s_{NN}}} e^{\pm y}$ denotes the light-cone momentum fraction in the projectile and target, $g$ is the strong coupling constant, and $N_c=3$ denotes the number of colors.

In the high energy limit, single inclusive quark production arises from the stopping of the collinear quarks inside the projectile (target) nucleus via multiple scattering off the color fields of the target (projectile) nucleus. 
The corresponding quark spectra, $\frac{dN_{q_{f}}}{d^2\xT d^2\pT dy}$, can be written as~\cite{Dumitru:2002qt, Dumitru:2005gt}
\begin{equation}
\begin{split}
\frac{dN_{q_{f}}}{d^2\xT d^2\pT dy} &= \frac{x_{1}q^{A}_{f}(x_{1},\pT^2,\xT)~D_{\rm fun}(x_2,\xT,\pT)}{(2\pi)^2} \\
&+ \frac{x_{2}q^{A}_{f}(x_{2},\pT^2,\xT)~D_{\rm fun}(x_1,\xT,\pT)}{(2\pi)^2}\,.
\end{split}
\label{eq:single_quark_density}
\end{equation}
where $q^{A}_{f}(x_{1/2},Q^2,\xT)$ represent the collinear quark distributions in the two nuclei, which in \Dipper\ are modeled as the incoherent sum of the parton distributions of individual nucleons, $q_f^{p/n}$. 
Invoking isospin symmetry, the parton distribution functions of neutrons are obtained from those of protons by interchanging $u\leftrightarrow d$ quark flavors.

In this work, the dipole gluon distributions are modeled using the impact-parameter dependent saturation (IP-Sat) model~\cite{Kowalski:2006hc,Kowalski:2003hm}, where the fundamental dipole is given by
\begin{equation}\label{eq:IPSatDip}
D_{\rm fun}(x,\xT,\sT)=\exp\left[-\frac{\pi^2\sT^2}{2\,Nc}\aS(\mu^2)\, xg(x,\mu^2)\, T(\xT)\right].
\end{equation}
where $g(x,\mu^2)$ denotes the gluon distribution function initialized at a scale $\mu_0$ via the parametrization 
\begin{equation}
xg(x,\mu^2_0)= a_g\,x^{-\lambda_g}\, (1-x)^{5.6}\,.
\end{equation}
and evolved to further scales, $\mu$, via the Dokshitzer-Gribov-Lipatov-Altarelli-Parisi (DGLAP) equation. 
By assuming that local color correlations of charges inside the nucleus are Gaussian, the adjoint dipole can be related to the fundamental one as 
$
D_{\rm adj}(x,\xT,\sT)=[D_{\rm fun}(x,\xT,\sT)]^{C_A/C_F}\;,
$
which is the relation employed throughout this work. 

With the initial energy density specified, the energy-momentum tensor for the subsequent pre-equilibrium evolution is initialized as 
\begin{equation}\label{eq:mcdipper-ini}
T^{\mu\nu}=\mathrm{diag}(e,e/2,e/2,0)
\end{equation}
where only energy perturbations are present and the strong pressure anisotropy of the above energy–momentum tensor indicates that the system at $\tauekt$ is still far from local equilibrium and cannot be properly described by viscous hydrodynamics, motivating the use of \kompost.

Since the \Dipper\ model does not include final state interactions, we always assume the initial time $\tauekt$ to be very small, to account for kinetic final state interactions in \kompost. In practice, we take the starting time of the subsequent pre-equilibrium evolution as $\tau_{\rm EKT}=0.01\ \rm fm/c$, however we expect our results to be rather insensitive to the choice of initial time $\tauekt$. 

We note that in this work, we will only examine the applicability of our (3+1)D framework in the most central 0-5\% 
Pb+Pb collisions at $\sqrt{s_{NN}}=2.76$ TeV. The normalization factor $K_g=2$ in \Dipper\ is chosen to approximately reproduce the charged particle multiplicity at midrapidity.

\subsection{CLVisc hydrodynamics}
Following the pre-equilibrium evolution in \kompost-3D, the subsequent space-time evolution of the QGP fireball will be simulated in the CLVisc hydrodynamic framework~\cite{Pang:2012he, Pang:2018zzo, Wu:2021fjf} which solves the energy-momentum conservation equation,
\begin{equation}
\partial_{\mu} T^{\mu\nu} = 0,
\end{equation}
based on hydrodynamic constitutive equations for the energy-momentum tensor $T^{\mu\nu}$ of the QGP. 
Since the QGP medium remains out of local thermal equilibrium throughout its entire evolution, dissipative corrections to $T^{\mu\nu}$ are evolved according to Israel-Stewart type equations~\cite{Denicol:2018wdp}. Considering the conformal limit assumed in the pre-equilibrium stage, the evolution of bulk pressure in the QGP phase is ignored, while the shear stress tensor is evolved according to
\begin{equation}
\begin{gathered}
\tau_\pi \Delta_{\alpha \beta}^{\mu \nu} D \pi^{\alpha \beta}+\pi^{\mu \nu}=\eta_v \sigma^{\mu \nu}-\delta_{\pi \pi} \pi^{\mu \nu} \theta-\tau_{\pi \pi} \pi^{\lambda\langle\mu} \sigma_\lambda^{\nu\rangle}\\+\varphi_1 \pi_\alpha^{\langle\mu} \pi^{\nu\rangle \alpha}.
\end{gathered}
\end{equation}
Here $D$ is the covariant derivative $u^\mu \partial_\mu$, $\Delta^{\mu\nu}$ is the spatial projector $g^{\mu\nu}-u^\mu u^\nu$, $\nabla^\mu=\Delta^{\mu\nu}\partial_\nu$, $\theta=\partial_{\mu}u^{\mu}$ is the expansion rate and $\sigma^{\mu\nu}$ is the symmetric shear tensor $2\nabla^{<\mu} u^{\nu>}$. 
The traceless tensor $A^{\langle\mu \nu\rangle}=\Delta_{\alpha \beta}^{\mu \nu}A^{\alpha \beta}$ projects out the part that is traceless and transverse to the flow velocity using the double, symmetric, and traceless projection operator $\Delta_{\alpha \beta}^{\mu \nu}=\frac{1}{2}\left(\Delta_\alpha^\mu \Delta_\beta^\nu+\Delta_\alpha^\nu \Delta^\mu{ }_\beta\right)-\frac{1}{3} \Delta^{\mu \nu} \Delta_{\alpha \beta}$. 
$\tau_{\pi}$ is relaxation times of shear tensor. $\delta_{\pi \pi},\tau_{\pi \pi}$ and $\varphi_1$ are second-order transport coefficients. The values of relaxation time and transport coefficients are taken from Ref.~\cite{Denicol:2014vaa}.
In this work, the specific shear viscosity $\eta_v/s$ is taken to be a constant value of 0.16 throughout the hydrodynamic evolution, consistent with the value used in \kompost-3D evolution.
The equation of state used in the hydrodynamic evolution is taken from the HotQCD Collaboration~\cite{HotQCD:2014kol}.
This choice differs from the equation of state employed in the \textsc{\kompost-3D} pre-equilibrium stage, where 
we assume a conformal equation of state $e = 3P$ with an effective number of degrees of freedom $\nu_{\rm eff} = 40$, chosen such that the resulting energy density $e(T)$ approximately matches that of high-temperature QCD~\cite{HotQCD:2014kol,Borsanyi:2016ksw}.
We note that this mismatch between the pre-equilibrium and hydrodynamic equations of state constitutes a systematic approximation of our framework.
Its phenomenological consequences will be reflected in the final observables and are implicitly included in the theoretical uncertainty of our results.

\subsection{Particlization and after-burner}

When the local temperature of the QGP fluid drops down to the freeze-out temperature ($T_{\rm frz}$=0.154 GeV), we use the Cornelius algorithm~\cite{Huovinen:2012is} to infer the hypersurface for the subsequent Cooper-Frye sampler. The momentum of produced hadron is sampled from the following formula, 
\begin{equation}
\frac{d N_i}{d Y p_T d p_T d \phi}=\frac{g_i}{(2 \pi)^3} \int_\Sigma p^\mu d \Sigma_\mu f_{\mathrm{eq}}(1+\delta f_\pi,
\end{equation}
where the thermal equilibrium contributions $f_{\rm {eq}}$ and out-of-equilibrium correction $\delta f_\pi$ are given as,
\begin{equation}
\begin{aligned}
f_{\mathrm{eq}} & =\frac{1}{\exp \left[\left(p\cdot u - B \mu_B\right) / T\right] \pm 1}, \\
\delta f_\pi & =(1 \pm f_{\mathrm{eq}}) \frac{p_\mu p_\nu \pi^{\mu \nu}}{2 T^2(e+P)}. %\\
\end{aligned}
\end{equation}
We note that in this work, we oversampled  each hydrodynamic event 1000 times to suppress non-flow effects and reduce statistical fluctuations in final observables. Once the hadrons are produced, a hadronic afterburner SMASH~\cite{wergieluk_2024_10707746, SMASH:2016zqf, Sjostrand:2006za, Sjostrand:2007gs} is employed to simulate final state interactions between them%until $\cdots$
.

\section{Results and discussions}\label{sec:results}

In this section, we show the results from the (3+1)D framework in which the \Dipper\ initial conditions are evolved with the kinetic theory in \kompost-3D, followed by hydrodynamic evolution in CLVisc and a hadronic afterburner in SMASH.
We first investigate the smoothness of the transition from the early pre-equilibrium stage of a heavy-ion collision to the subsequent hydrodynamic stage, by looking at averages and profiles of the hydrodynamics fields. Subsequently, we will also consider hadronic observables, and investigate their dependence on the hydrodynamic initialization time $\tau_{\textrm{hydro}}$. In order to scrutinize the effects, and not lose any information in statistical averages, in this work, we will only examine  one single random event from \Dipper\ in 0-5\% Pb+Pb collisions at $\sqrt{s_{NN}}$=2.76 TeV.

\subsection{Visualization of pre-equilibrium evolution}
\begin{figure} 
    \centering
    \includegraphics[width=.8\linewidth]{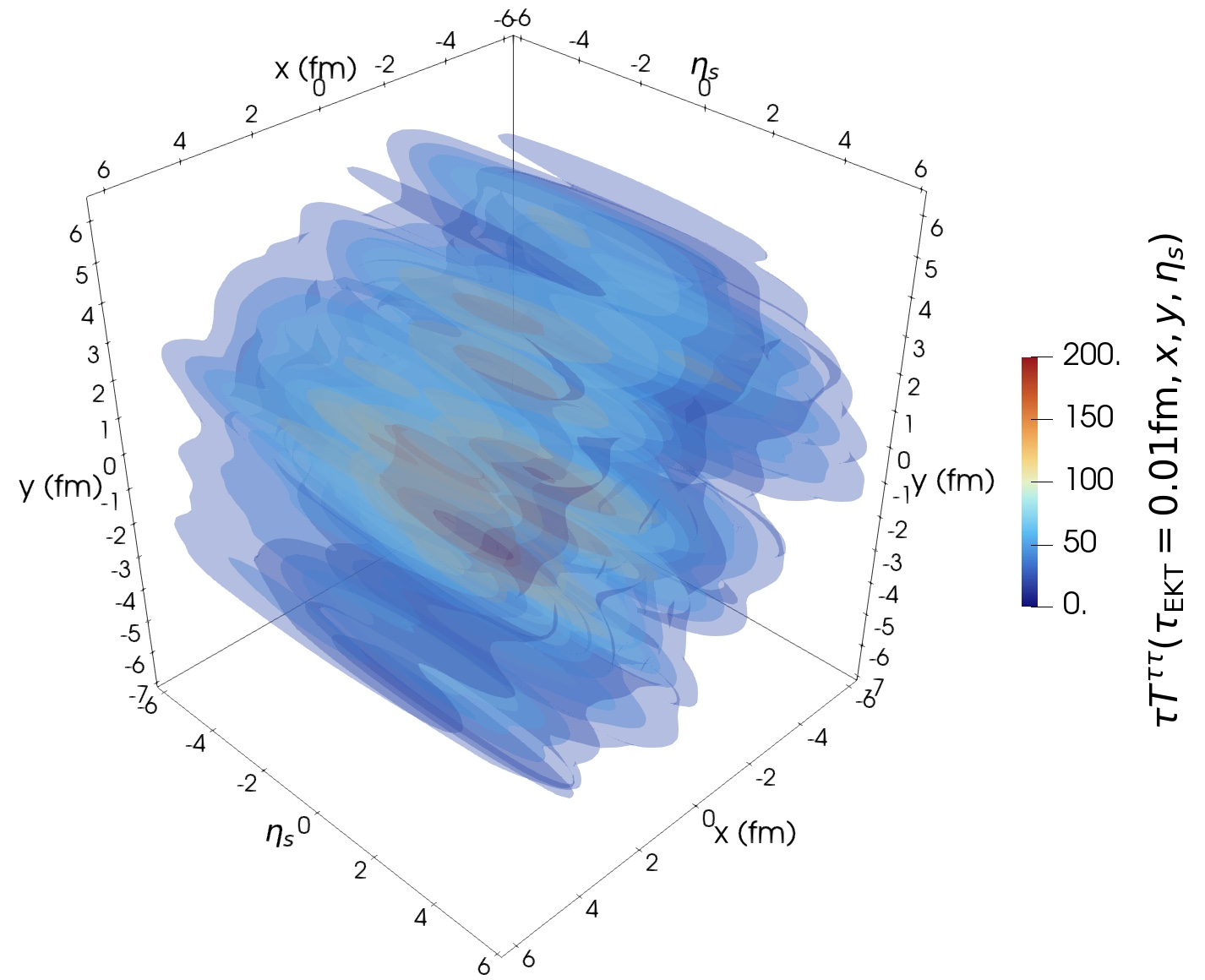}
    \includegraphics[width=.8\linewidth]{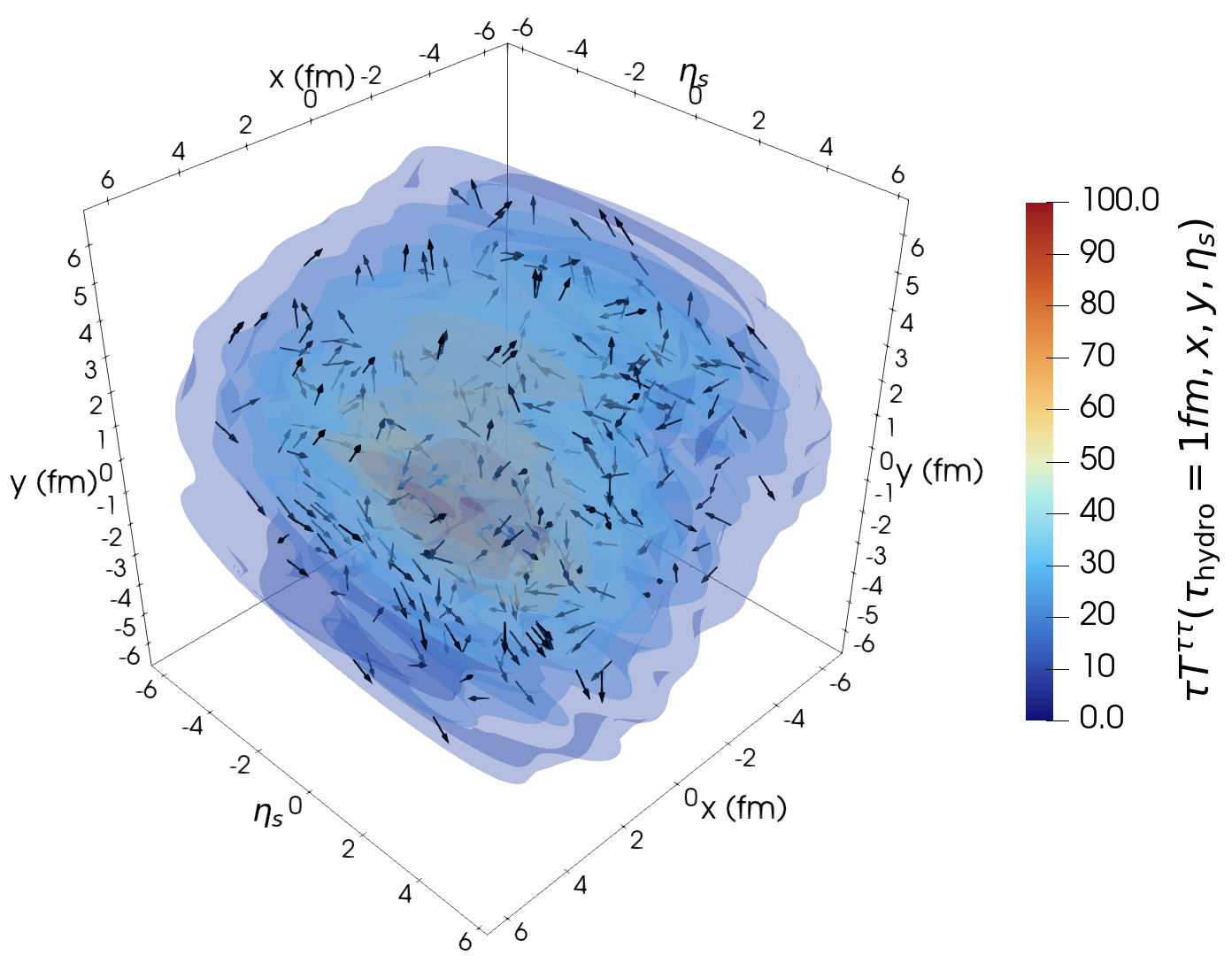}
    \caption{Three-dimensional energy density distribution $\tau T^{\tau\tau}(x,y,\eta_s)$ for the \Dipper\ initial condition, shown before the kinetic-theory pre-equilibrium evolution at $\tau_{\mathrm{EKT}} = 0.01~\mathrm{fm}$ (upper) and at the hydrodynamic initialization time $\tau_{\mathrm{hydro}} = 1.0~\mathrm{fm}$ (lower). Arrows in the lower panel indicate the direction of the energy-flow components $(T^{\tau x}, T^{\tau y}, \tau T^{\tau\eta})$.}
    \label{fig:paraview}
\end{figure}
\begin{figure} 
    \centering
    \includegraphics[width=\linewidth]{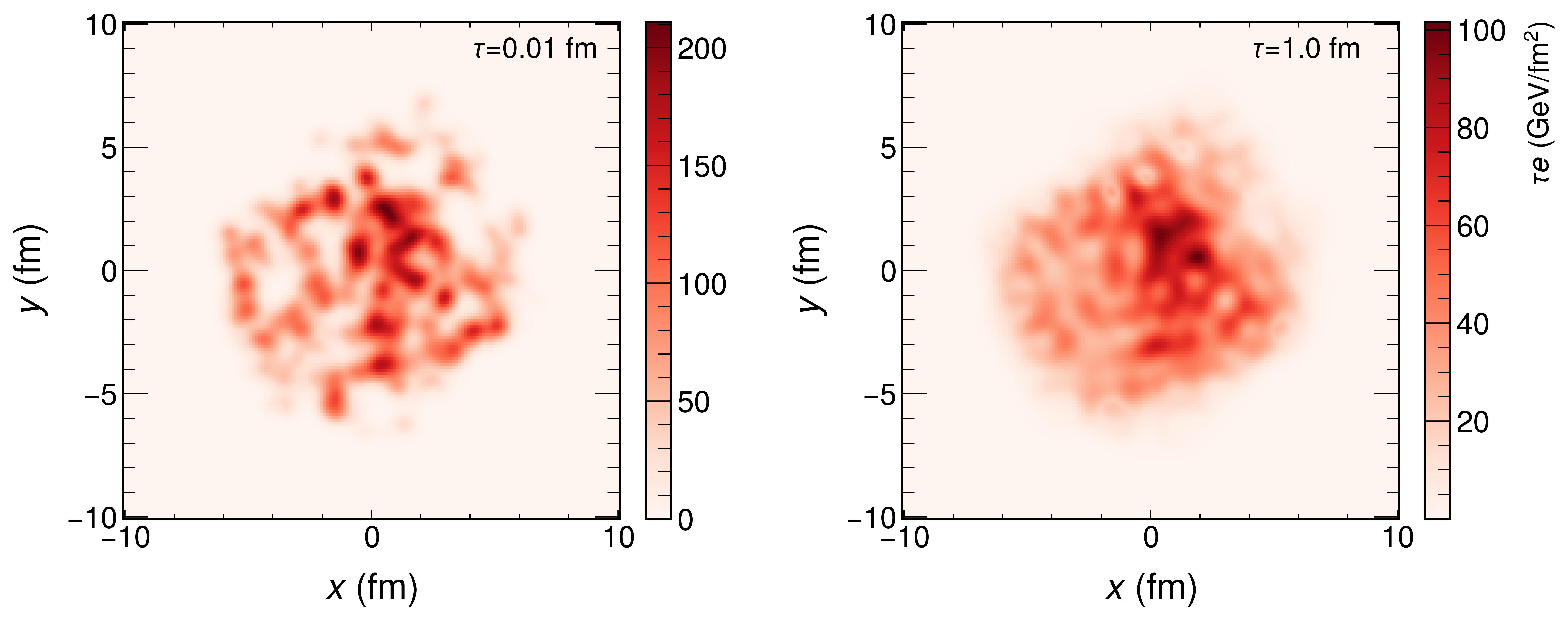}
    \includegraphics[width=\linewidth]{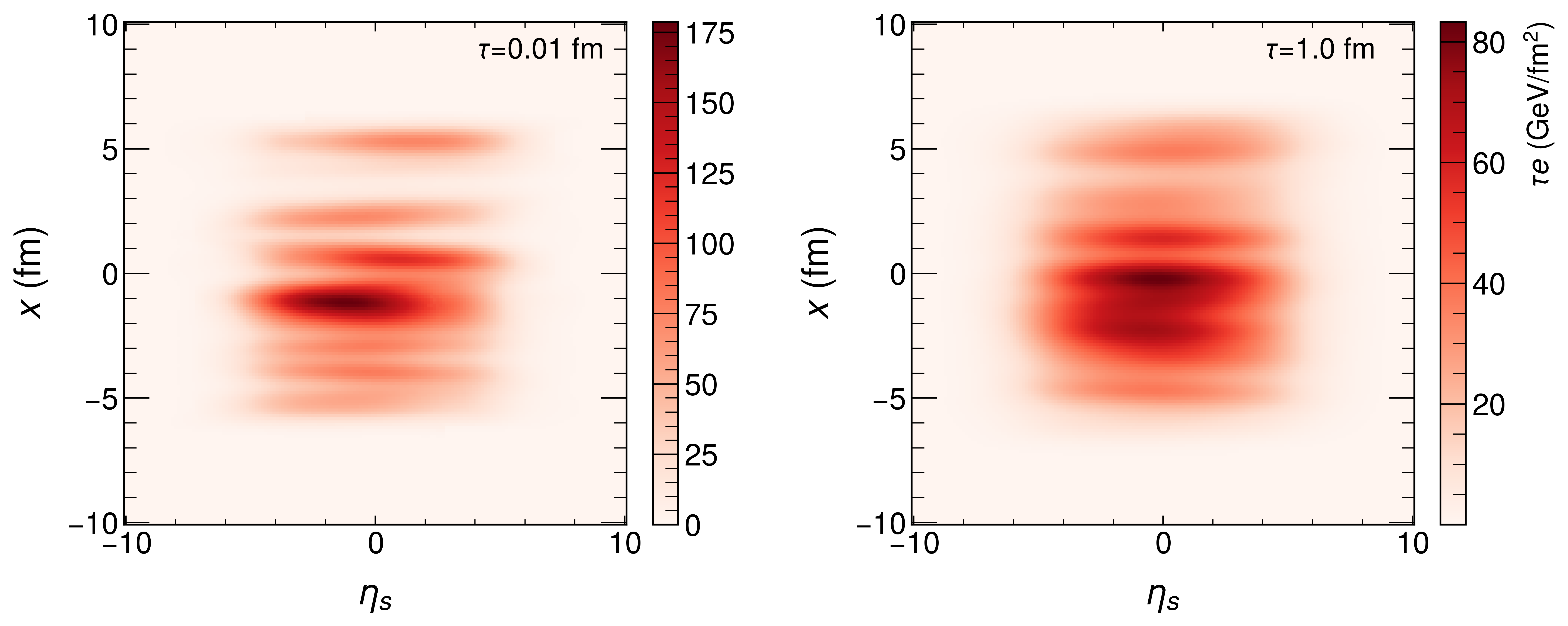}
    \caption{Time evolution of the energy density $\tau e(x,y,\eta_s)$ during the kinetic-theory pre-equilibrium stage, shown as two-dimensional cuts in the transverse plane at $\eta_s = 0$ (top) and in the reaction plane at $y = 0~\mathrm{fm}$ (bottom).}
    \label{fig:taued}
\end{figure}

In order to gain intuition for the pre-equilibrium evolution, we visualize the energy density and flow for the \Dipper\ initial condition, both 
in full three dimensions (Fig.~\ref{fig:paraview})
and in two-dimensional slices (Fig.~\ref{fig:taued}).
The three-dimensional energy density profile $\tau T^{\tau\tau}(x,y,\eta_s)$ from the \Dipper\ initial state model is evolved with the kinetic-theory based pre-equilibrium solver \kompost-3D 
from $\tau_{\mathrm{EKT}} = 0.01~\mathrm{fm}$ to the hydrodynamic initialization time $\tau_{\mathrm{hydro}} = 1.0~\mathrm{fm}$.
In the lower panel of Fig.~\ref{fig:paraview}, we also show the direction of initial energy flow $(T^{\tau x}, T^{\tau y}, \tau T^{\tau\eta})$ with black arrows where the magnitude of the flow is omitted for a clearer illustration.
All in all, we observe both the spreading of the initial fireball energy profile and the generation of initial flow, demonstrating significant expansion during the pre-equilibrium stage.

For a clearer view of the evolution during the pre-equilibrium stage, Fig.~\ref{fig:taued} shows the energy density $\tau e(x,y,\eta_s)$ in the transverse plane at $\eta_s = 0$ (top) and in the reaction plane at $y = 0\ \mathrm{fm}$ (bottom). Again, we can see how the onset of dissipation during the pre-equilibrium phase results in a smearing of the energy density profile both in transverse plane and reaction plane.

\subsection{Hydrodynamic fields}
Next, to examine the transition from the kinetic pre-equilibrium stage to hydrodynamics, we first consider the transversely averaged hydrodynamic fields, namely the energy density, radial flow, and momentum anisotropy, at midrapidity and in the forward rapidity regions.
The averaged energy density and radial flow, are defined as 
\begin{align}
    &\langle \tau e^{3/4}\rangle_{\eta_s} \equiv 
    \frac{\int d^2 \mathbf{x} u^\tau e \cdot \tau e^{3/4} }{\int d^2 \mathbf{x} u^\tau e},
    \\
    &\langle v_\perp\rangle_{\eta_s} \equiv 
    \frac{\int d^2 \mathbf{x} u^\tau e \cdot v_\perp}{\int d^2 \mathbf{x} u^\tau e},
\end{align}
where the radial flow is defined as $v_\perp \equiv \sqrt{v_x^2 + v_y^2}$ and $\int d^2 \mathbf{x}$ denotes an integral over the transverse coordinates.
The azimuthal momentum anisotropy is constructed from the transversely integrated stress tensor as 
\begin{align}
    &[T^{\mu\nu}]_s \equiv \int d^2\mathbf{x} u^\tau T^{\mu\nu}\\
    &\varepsilon_p(\tau) 
    \equiv \frac{\sqrt{\left(\left[T^{x x}\right]_s-\left[T^{y y}\right]_s\right)^2+4\left[T^{x y}\right]_s^2}}{\left[T^{x x}\right]_s+\left[T^{y y}\right]_s}
\end{align}

\begin{figure*}
    \centering
    \includegraphics[width=\linewidth]{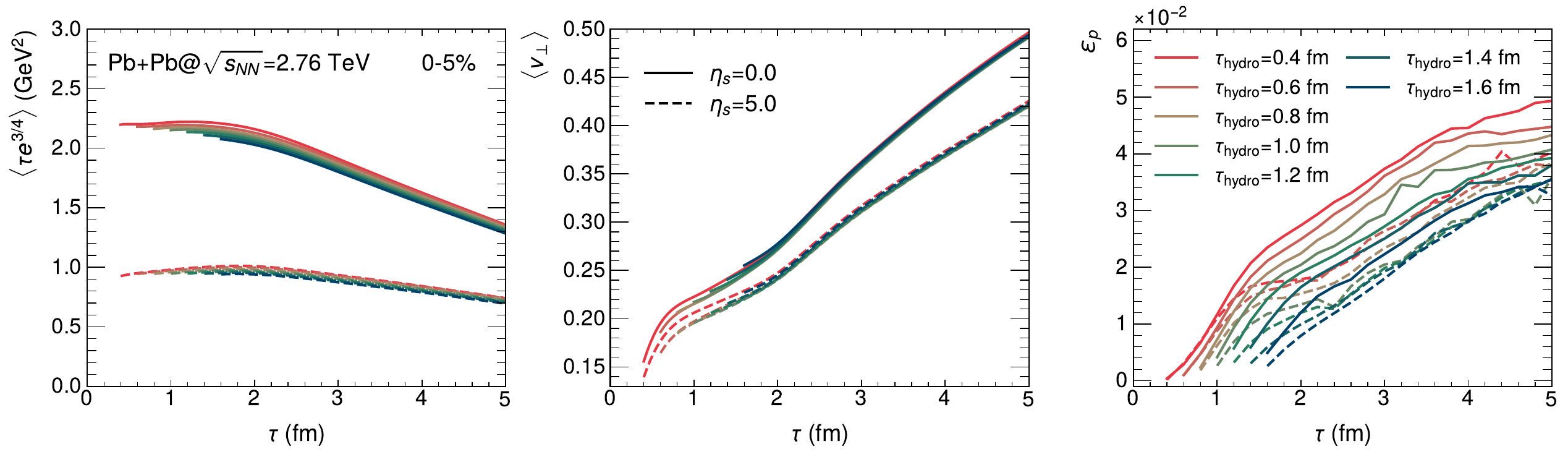}
    \caption{Transverse average of $\tau e^{3/4}$, transverse velocity $v_\perp$ and momentum eccentricity as a function of time $\tau$ in midrapidity ($\eta_s$=0) and forward rapidity ($\eta_s=5$). The three average fields are plotted for different hydrodynamic initialization time $\tau_{\rm hydro}$. }
    \label{fig:ave-field}
\end{figure*}
In Fig.~\ref{fig:ave-field}, we present the averaged transverse profiles of the energy density $\langle \tau e^{3/4}\rangle$, transverse velocity $\langle v_\perp\rangle$ and momentum eccentricity $\varepsilon_p$ as a function of evolution time $\tau$ at midrapidity ($\eta_s$=0) and forward rapidity ($\eta_s=5$) for different hydrodynamization time $\tau_{\rm hydro}$.
We find that the evolution of energy density $\langle \tau e^{3/4}\rangle$ in midrapidity and forward rapidity shows only a very weak dependence on the hydrodynamics initialization time $\tauhydro$, with the tiny differences originating from the switch from a conformal equation of state to a realistic QCD equation of state~\cite{Borghini:2024kll,daSilva:2022xwu}.
On the other hand, we find that the evolution of the radial flow is almost independent of the hydrodynamic initialization time $\tauhydro$, except for the very early initialization at $\tauhydro=0.4$ fm.
This means the kinetic theory captures well the rapid rise in the radial flow at early times which ensures a smooth transition to hydrodynamic regime.
However, when we look at the time evolution of the momentum eccentricity $\varepsilon_p$, an evident dependence on hydro starting time is found for both midrapidity and forward rapidity as in previous (2+1)D pre-equilibrium evolution~\cite{Kurkela:2018vqr}.
This indicates that \kompost~cannot generate a momentum anisotropy $\varepsilon_p$ comparable to that in the hydrodynamic evolution. While the origins of this behavior are still under active investigation, one possible explanation would be that $\varepsilon_p$ could be sensitive to non-linear effects of fluctuations which are not included in the linearized treatment of \kompost~\cite{Kurkela:2018vqr}.
\begin{figure*}
    \centering
    \includegraphics[width=.8\linewidth]{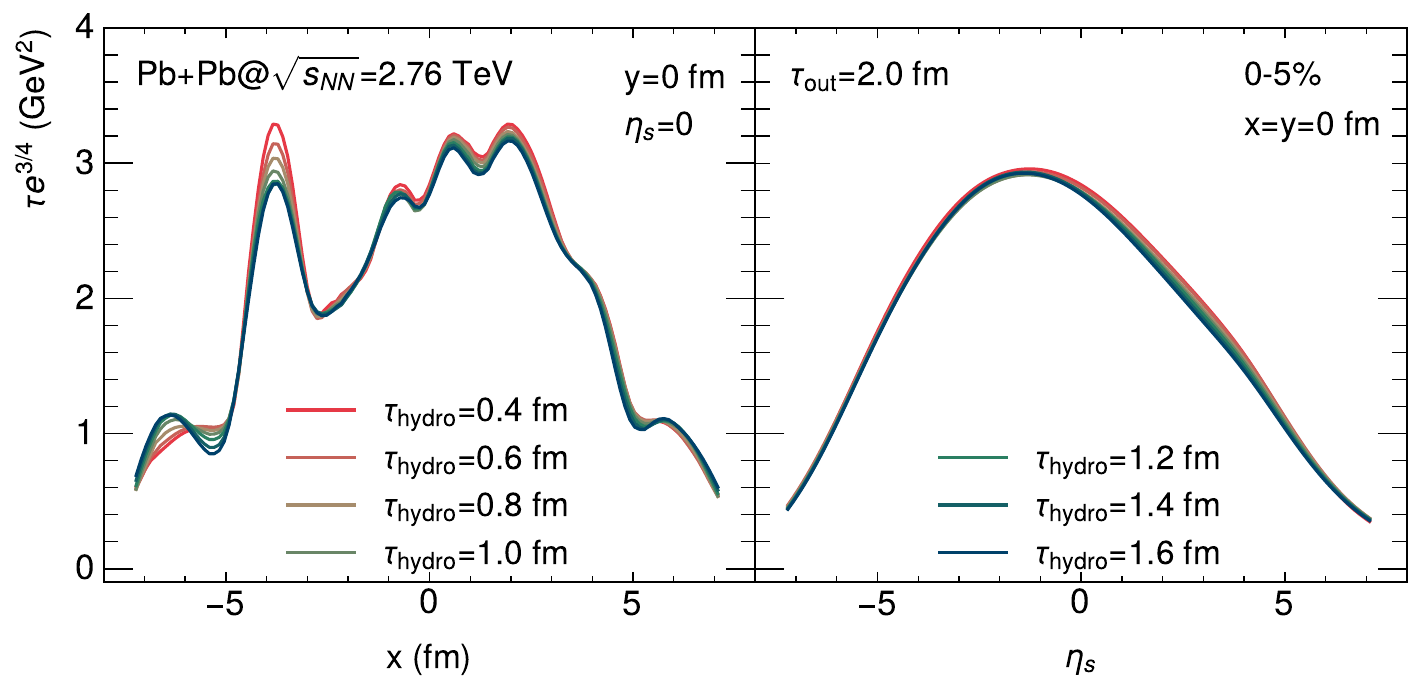}
    \includegraphics[width=.8\linewidth]{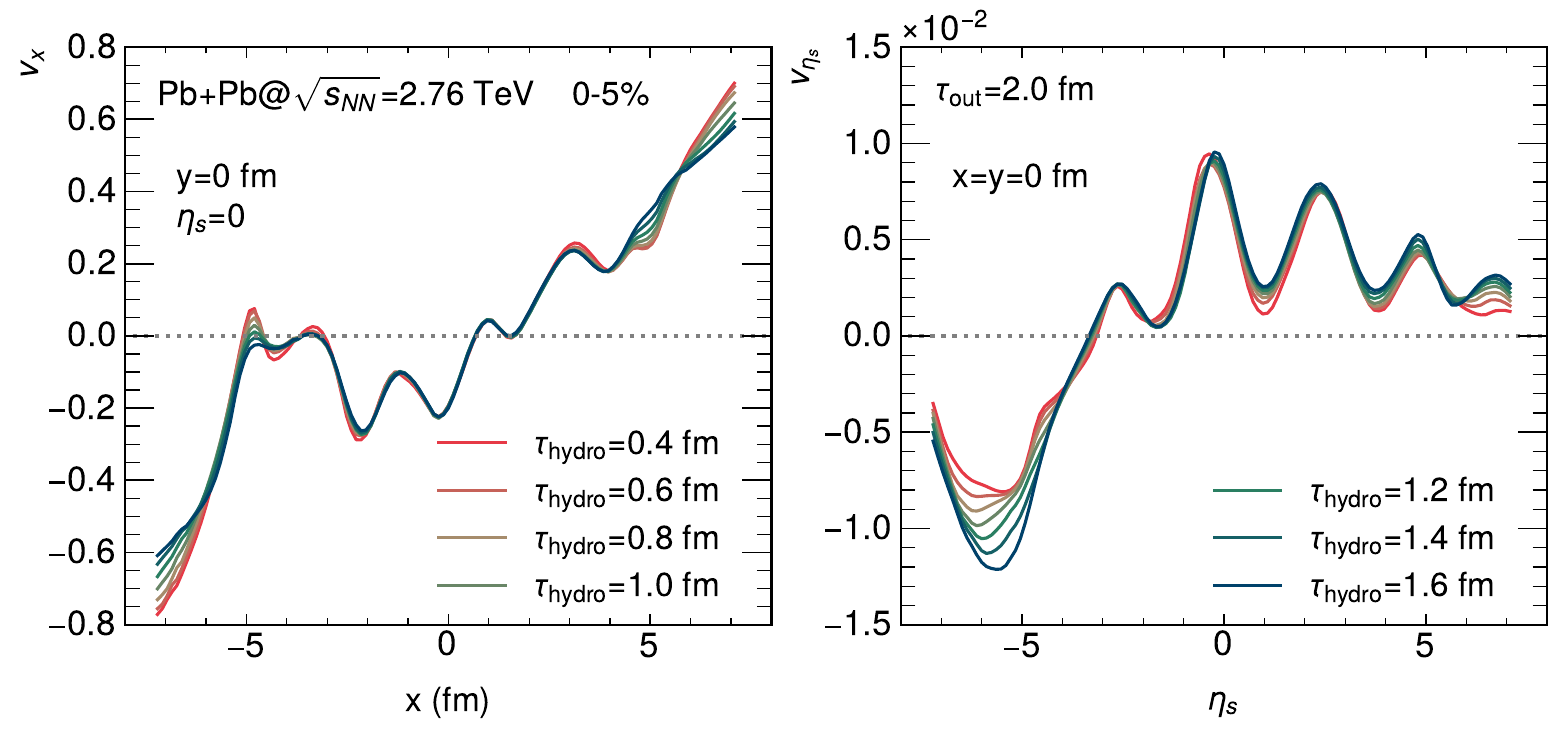}
    \caption{Local profile of energy density $\tau e^{3/4}$ and flow velocity $v_x$ (or $v_{\eta_s}$) for different hydrodynamics transition times $\tau_{\rm hydro}$ in transverse plane (left panels) or longitudinal direction (right panels).}
    \label{fig:local-field}
\end{figure*}
The equivalence between kinetic theory and hydrodynamics at the hydrodynamization time $\tau_{\rm hydro}$ ensures that the above examined averaged hydrodynamic fields are insensitive to the choice of the hydrodynamic starting time, especially for energy density and radial flow. We now proceed to test whether this insensitivity also holds for the local distribution of hydrodynamic fields.
In Fig.~\ref{fig:local-field}, we show the spatial distributions of the energy density $\tau e^{3/4}$ and the flow velocity $v_x$ (or $v_{\eta_s}$) at the time slice $\tau_{\rm out} = 2\ \mathrm{fm}$ in the transverse plane (left panels) and along the longitudinal direction (right panels).
The different curves stand for different hydrodynamic transition times $\tau_{\rm hydro}$.
Again, we find that the local structure of $\tau e^{3/4}$ and the flow velocities exhibit only a very weak dependence on the hydrodynamic initialization time~$\tau_{\rm hydro}$ both in transverse plane and longitudinal direction, indicating a well-controlled matching between the pre-equilibrium dynamics and the subsequent viscous hydrodynamic evolution in full (3+1) dimensions.

\subsection{Hydrodynamization}
As in Ref.~\cite{Kurkela:2018vqr}, the approach of kinetic theory towards hydrodynamic regime can be assessed by comparing the out-of-equilibrium components of the energy-momentum tensor to the corresponding estimates in first order dissipative hydrodynamics.
\begin{figure*}
    \centering
    \includegraphics[width=0.8\linewidth]{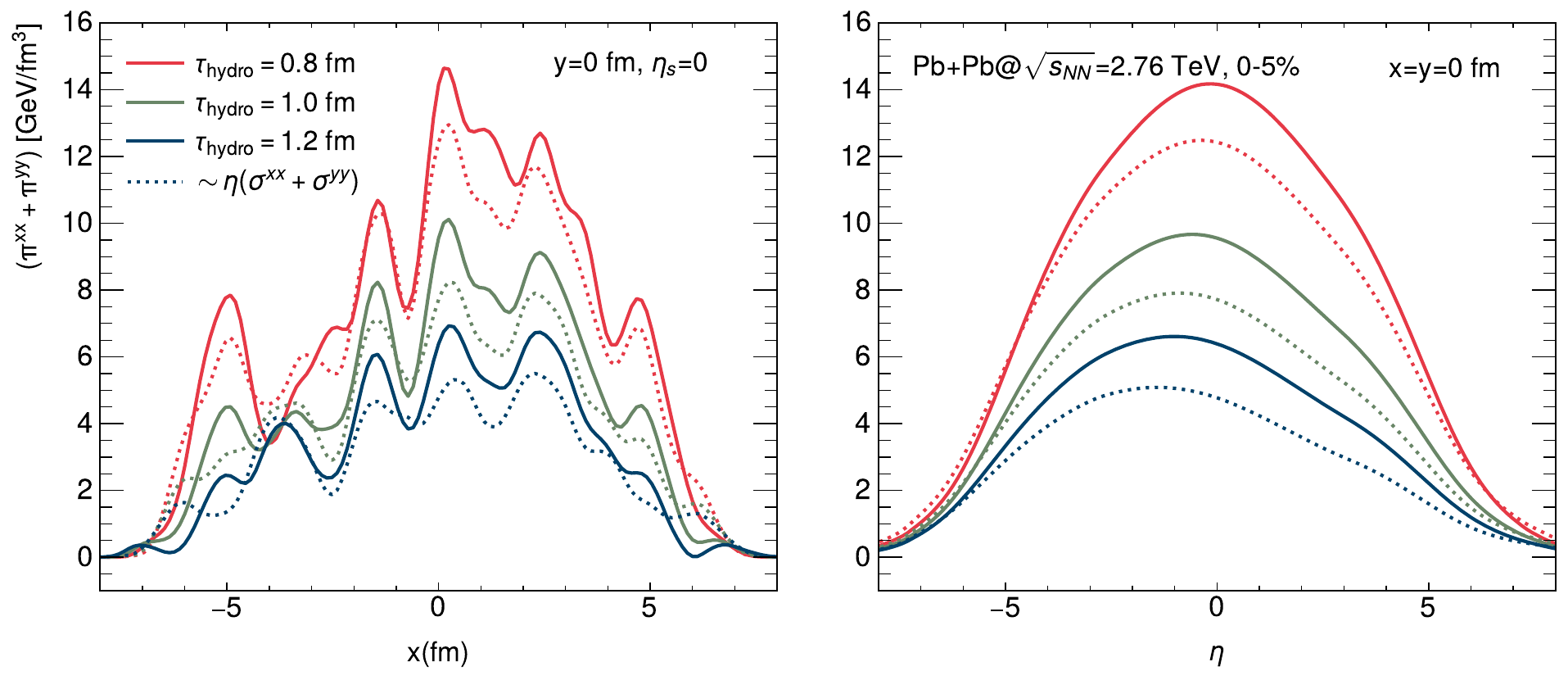}
    \includegraphics[width=0.8\linewidth]{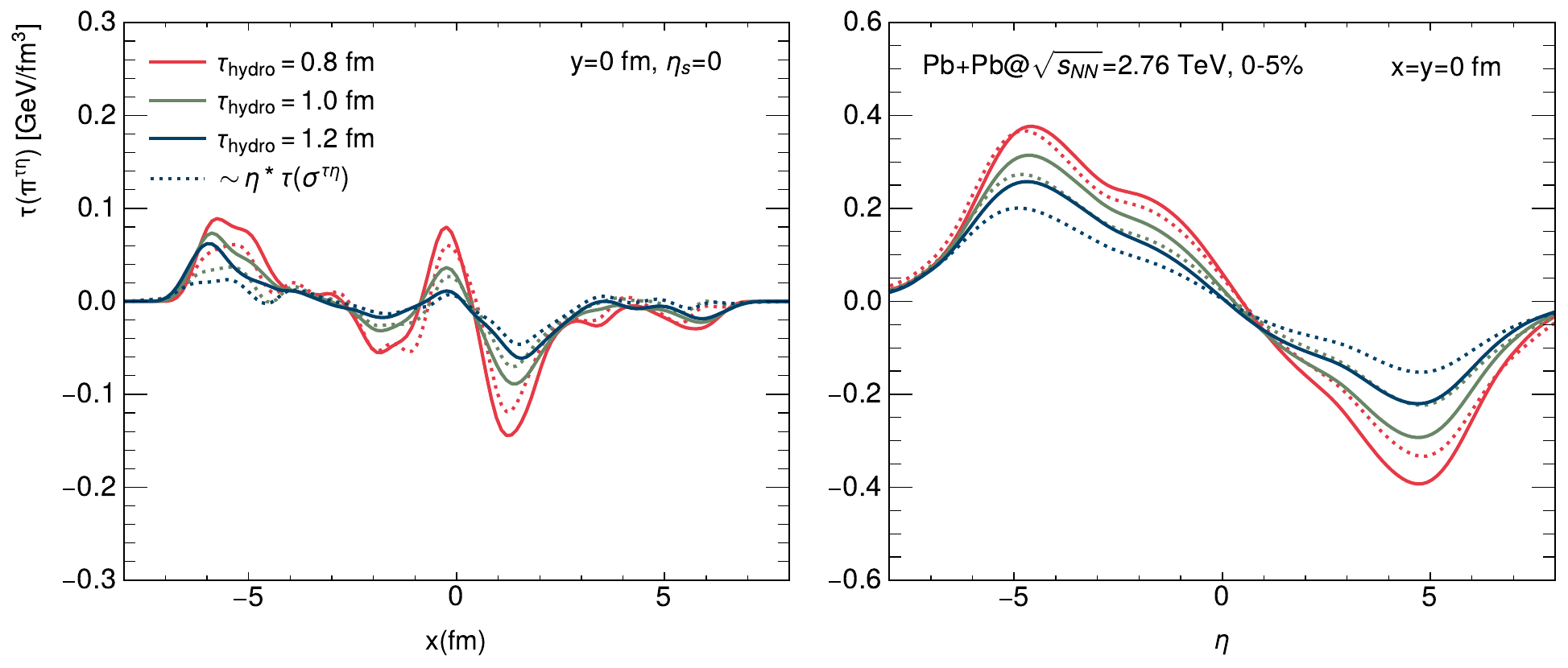}
    \caption{Comparison of the out-of-equilibrium shear stress tensor with the Navier-Stokes estimate at different hydrodynamics initialization times $\tau_{\rm hydro}$ = 0.8, 1.0, 1.2 fm.}
    \label{fig:shear-hydro}
\end{figure*}
In Figs.~\ref{fig:shear-hydro}, we compare the out-of-equilibrium shear stress tensor $\pi^{\mu\nu}$ with the Navier-Stokes estimate at different hydrodynamic initialization  times $\tau_{\rm hydro}$ = 0.8, 1.0, 1.2 fm. 
Here we take the transverse diagonal component and the longitudinal momentum component as examples, while the remaining components of the shear tensor are shown in App.~\ref{more-shear}.
We find that both the transverse and longitudinal components of the shear stress tensor from kinetic theory agree well with the hydrodynamic estimate over most of the collision region, both in the transverse plane and along the longitudinal direction, except near sharp edges where the small gradient assumption breaks down. This indicates that the evolution of the system can now be smoothly matched onto hydrodynamics.

\begin{figure*}
    \centering
    \includegraphics[width=0.8\linewidth]{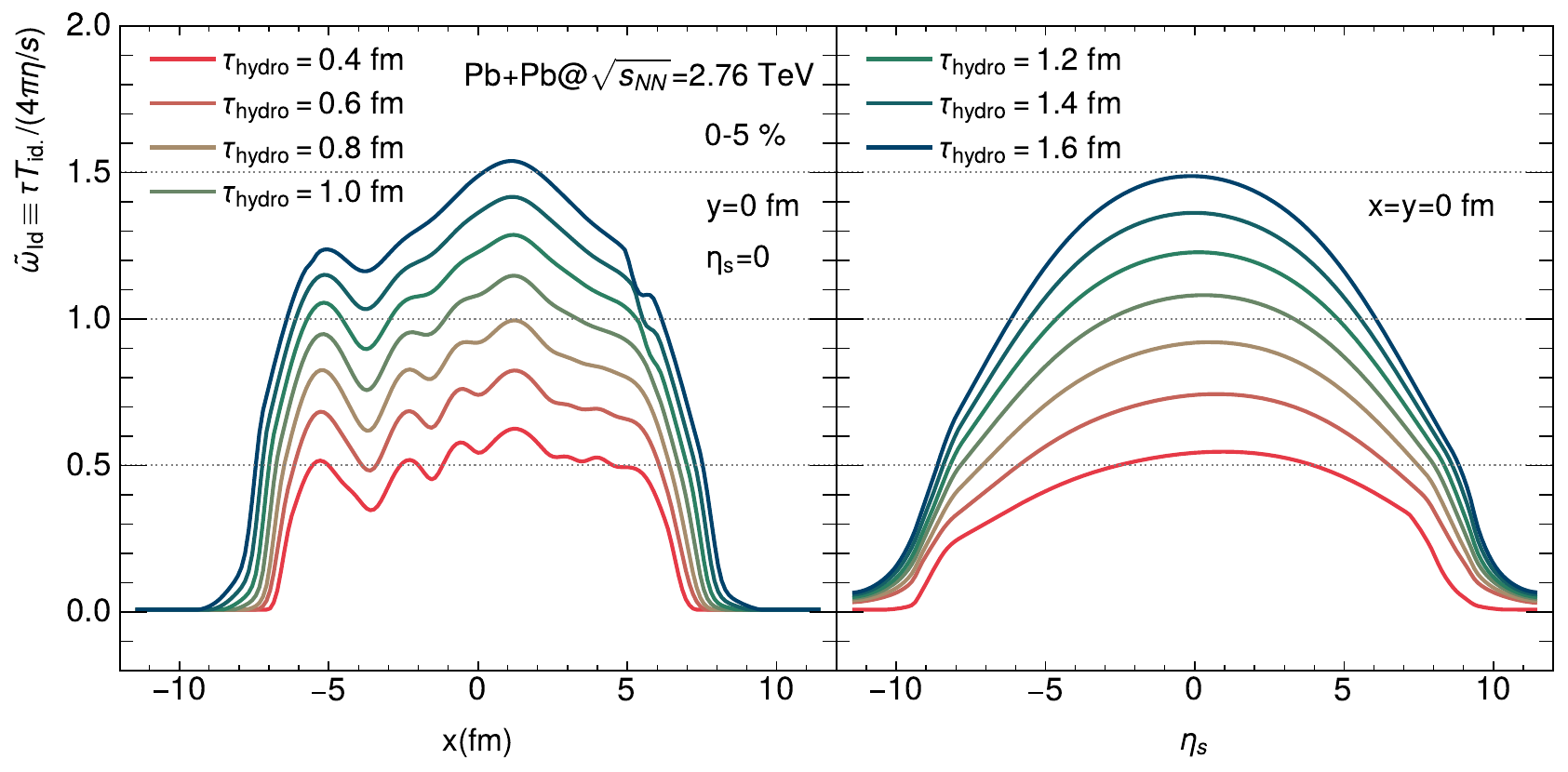}
    \caption{Scaled evolution time variable $\OmegaId$ at different hydro starting times. Values of $\OmegaId>1$ indicate that the system is close enough to local thermal equilibrium for hydrodynamics to become applicable.}
    \label{fig:w}
\end{figure*}

In Fig.\ref{fig:w}, we also present the scaled evolution time variable $\OmegaId\equiv\tau T_{\rm id}/(4\pi\eta_v/s)$ at different hydro starting times.
Since previous studies~\cite{Kurkela:2015qoa,Kurkela:2018vqr,Kurkela:2018wud,Romatschke:2017vte,Strickland:2018ayk} 
have shown that values of $\OmegaId>1$ indicate that the system can be described by hydrodynamics, we find that the central region in the transverse plane and the midrapidity region along the longitudinal direction enter the hydrodynamic regime for sufficiently large times, while deviations still persist at the edges of fireball where the local energy density is significantly smaller.

\subsection{Final observables}
Finally, we move on to investigate the effect of a consistent description of the early time dynamics on the final observables in heavy-ion collisions.
Since we observe a smooth matching from the pre-equilibrium to the hydrodynamic stage, we expect that, for reasonable values, the choice of the matching time $\tauhydro$ does not significantly affect the final state hadronic observables.

\begin{figure}
    \centering
    \includegraphics[width=\linewidth]{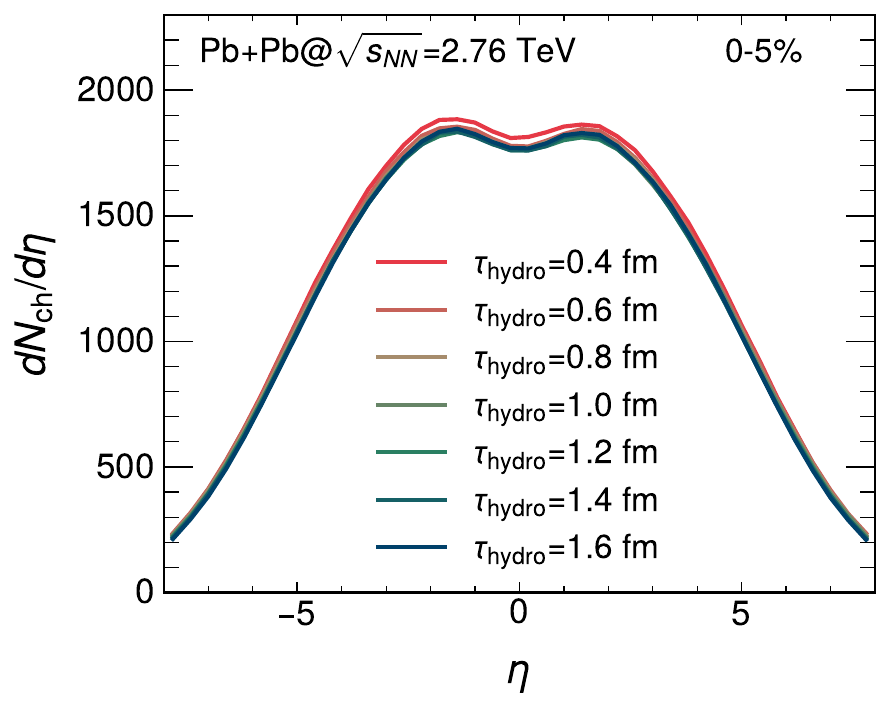}
    \caption{Dependence of charged particle multiplicity on hydrodynamization time $\tau_{\rm hydro}$.}
    \label{fig:dndeta}
\end{figure}

\begin{figure}
    \centering
    \includegraphics[width=\linewidth]
    {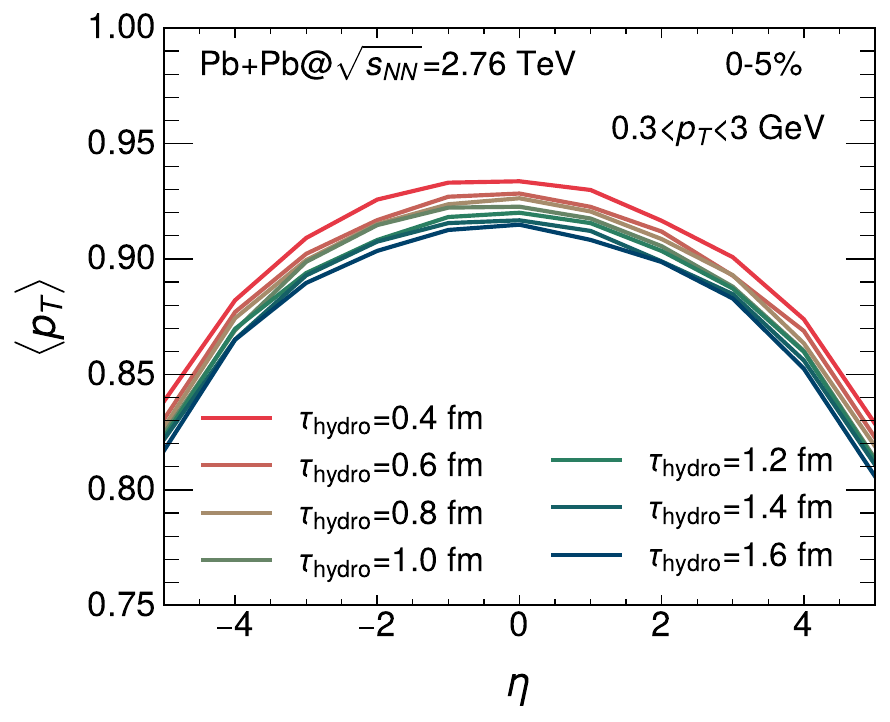}
    \caption{Sensitivity of the mean transverse momentum of charged hadron to hydrodynamization time $\tau_{\rm hydro}$.}
    \label{fig:pT}
\end{figure}

We first demonstrate the effects of hydrodynamization time $\tau_{\rm hydro}$ on the pseudorapidity dependence of the charged hadron multiplicity in Fig.~\ref{fig:dndeta}. 
Overall, the multiplicity shows only a very weak dependence on $\tauhydro$ across rapidity direction when \kompost-3D is used to describe the early stage of the medium evolution.
The slightly decreasing trend in $dN_{\rm ch}/d\eta$ is already anticipated by the corresponding behavior of $\langle \tau e^{3/4}\rangle$, and can be traced back to the mismatch between the equations of state used in the pre-equilibrium stage and in the subsequent hydrodynamic evolution. However, for variations of $\tauhydro$ between 0.8 and 1.6 fm this effect is less than 1 percent.
Similarly, the examination of radial flow, quantified by the mean transverse momentum of charged hadrons in Fig.~\ref{fig:pT}, also shows only marginal sensitivity to the hydrodynamization time $\tauhydro$ over the whole rapidity range, at the level of about 2\%. This tiny difference can be attributed to the slight variation of the multiplicity in each case. If the overall normalization is consistently tuned to reproduce the experimental multiplicity, the residual variation in the mean transverse momentum can be further suppressed.

\begin{figure} 
    \centering
    \includegraphics[width=\linewidth]
    {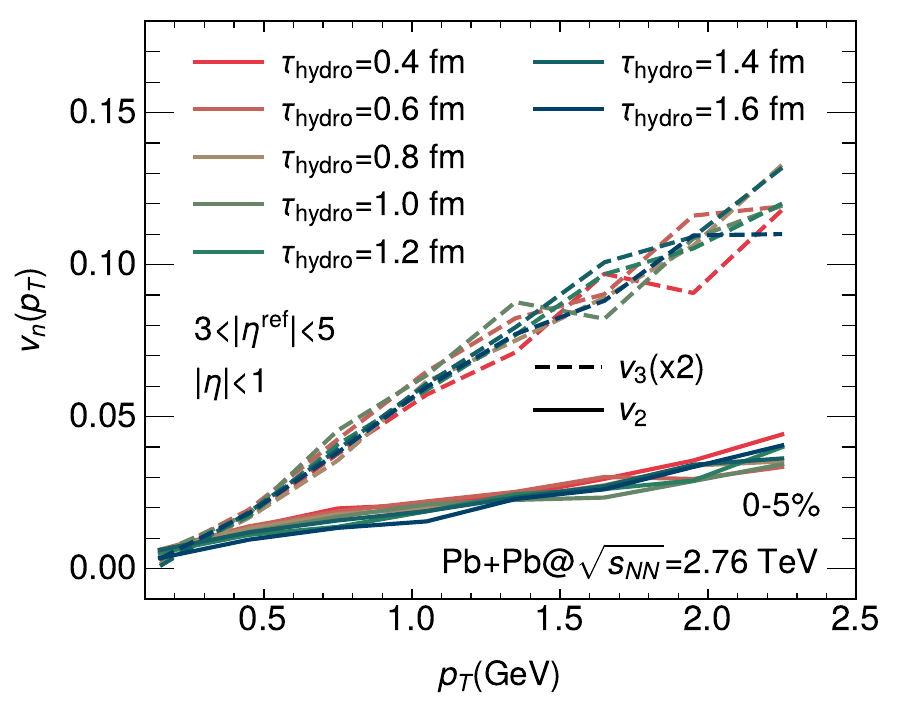}
    \includegraphics[width=.94\linewidth]
    {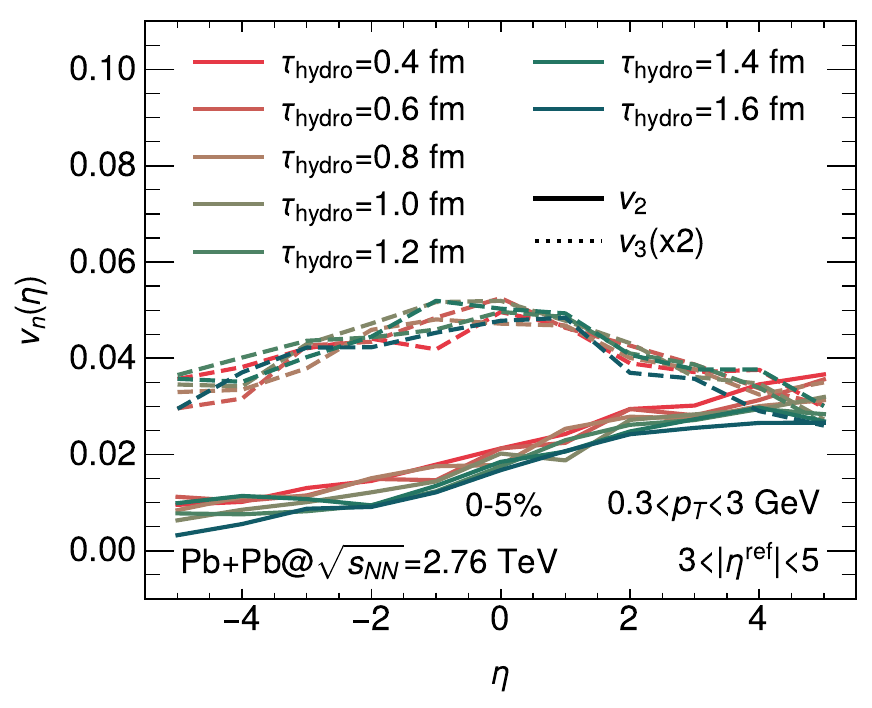}
    \caption{Sensitivity of the $p_T$ and $\eta$ dependence of anisotropic flow $v_2, v_3$ to hydro starting time $\tau_{\rm hydro}$.}
    \label{fig:vn}
\end{figure}

The anisotropic flow coefficients, which are among the most widely studied observables in heavy-ion collisions, are also examined for different durations of the pre-equilibrium stage in Fig.~\ref{fig:vn}. The upper panel shows the transverse momentum dependence of the second- and third-order flow coefficients at midrapidity, while the lower panel displays the rapidity dependence of the anisotropic flow for various hydrodynamization times $\tau_{\rm hydro}$. 
In this work, the $p_T$ and $\eta$ differential anisotropic flow ($v_2$ and $v_3$) are computed with the event plane method
as follows.
\begin{equation}
    v_n(p_T)= {1\over 2}
    \frac{\langle\cos n\left(\phi-{\Phi}_{n A}\right)\rangle
    + \langle\cos n\left(\phi-{\Phi}_{n B}\right)\rangle
    }
    {\sqrt{\cos n\left({\Phi}_{n A}-{\Phi}_{n B}\right)}}
\end{equation}
\begin{equation}
v_n(\eta)=
{\large
\begin{cases}
\frac{ \langle\cos n\left(\phi-{\Phi}_{n A}\right)\rangle } 
{\sqrt{\cos n\left({\Phi}_{n A}-{\Phi}_{n B}\right)}}, 
& \eta <0\\[0.1in]
{1\over 2}
\frac{ \langle\cos n\left(\phi-{\Phi}_{n A}\right)\rangle
+ \langle\cos n\left(\phi-{\Phi}_{n B}\right)\rangle }
{\sqrt{\cos n\left({\Phi}_{n A}-{\Phi}_{n B}\right)}}, 
& \eta=0\\[0.1in]
\frac{ \langle\cos n\left(\phi-{\Phi}_{n B}\right)\rangle } 
{\sqrt{\cos n\left({\Phi}_{n A}-{\Phi}_{n B}\right)}},
& \eta>0
\end{cases}    
}
\end{equation}
where 
$\langle \cdots \rangle$ denotes the average over all particles of interest,
$\phi$ is the azimuthal angle of particle of interest and $\Phi_{nA (B)}$ is the event plane of two reference subevents, chosen as $3<\eta_A^{\rm ref}<5$ and $-5<\eta_B^{\rm ref}<-3$ in the differential flow coefficients. In order to suppress short range non-flow correlations, the $\eta$ dependence of the anisotropic flow is evaluated in the following way: 
particles at $+\eta$ are correlated with reference particles from the backward region, 
while particles at $-\eta$ are correlated with reference particles from the forward region.
Since we only consider a single event, the differential observables as functions of $p_T$ and $\eta$ exhibit visible statistical fluctuations. Nevertheless, the curves remain very close to each other, both in their $p_T$ dependence and over the entire rapidity range, indicating only a weak sensitivity of anisotropic flow to $\tau_{\rm hydro}$ within our setup.

\section{Summary \& outlook}\label{sec:summary}
In this work, we developed the framework \kompost-3D to describe the (3+1)D pre-equilibrium dynamics of HICs. 
Based on non-equilibrium response functions calculated in kinetic theory in the relaxation-time approximation, we include the transverse and longitudinal structure of the response to initial energy perturbations on top of a longitudinal boost-invariant and transversely homogeneous background. 
We confirmed that the newly developed pre-equilibrium model, \kompost-3D, provides a smooth transition from 3D out-of-equilibrium initial conditions to (3+1)D near-equilibrium viscous hydrodynamics.
For example, the pre-equilibrium dynamics based on kinetic theory consistently capture the entropy production and the radial flow generation.
More importantly, the hydrodynamization can be achieved in most of the collision region, not only in the transverse plane but also in the longitudinal direction.
Subsequently, we embed the pre-equilibrium model into a (3+1)D framework to describe the full evolution of heavy-ion collisions, namely \Dipper+\kompost-3D+CLVisc+SMASH. 
We find that the above framework only shows a marginal dependence on the hydrodynamic initialization time $\tauhydro$. The longitudinal structure of $dN/d\eta$, $\langle p_T \rangle$ and $v_{2, 3}(\eta)$ shows little sensitivity to hydrodynamization time $\tauhydro$.

The development of a consistent theoretical framework for (3+1)D simulations of the space-time dynamics of heavy-ion collisions significantly improves the power of phenomenological studies, as a detailed exploration of the transverse and longitudinal dynamics could significantly improve the extraction of QCD transport coefficients from anisotropic flow measurements.  Our newly developed \kompost-3D framework reduces the uncertainty associated with the theoretical description of the early pre-equilibrium phase, e.g. by eliminating the dependence on the matching time to hydrodynamics, and can therefore be an important tool for future investigations.

In the future, there are still several improvements that can be done in this direction.
For example, a more realistic input of response functions from QCD kinetic theory is still needed to improve the description of pre-equilibrium dynamics.
In addition, it is also important to include the response to initial energy, momentum, and tensor perturbations simultaneously for a comprehensive propagation of information from initial conditions. 
Although these aspects are conceptually straightforward, they are computationally expensive. We also aim to explore the applicability of the \kompost ~framework by pushing the simulations to smaller system sizes and lower collision energies, where the duration of pre-equilibrium dynamics can become comparable to the system size.

\section*{Acknowledgments}
We would like to thank Travis Dore, Bao-chi Fu, Renata Krupczak, Aleksas Mazeliauskas, Stephan Ochsenfeld, Jean-Francois Paquet and Xiang-Yu Wu for productive discussions. 
This work is supported by the Deutsche Forschungsgemeinschaft (DFG, German Research Foundation) 
through the CRC-TR 211 ``Strong-interaction matter under extreme conditions'' Project No. 315477589--TRR 211. 
J.Z. is also supported in part by China Scholarship Council (CSC) under Grant No. 202306770009. 
X.D. is also supported by the Research Council of Finland, the Centre of Excellence in Quark Matter (project 364191).
Numerical simulations presented in this work were performed at the Paderborn Center for Parallel Computing ($\mathrm{PC}^2$) and at the National Energy Research Scientific Computing Center (NERSC), a U.S. Department of Energy Office of Science User Facility, supported by NERSC award NP-ERCAP0032835.

\bibliographystyle{apsrev4-1}

\bibliography{References}

\begin{thebibliography}{75}%
\makeatletter
\providecommand \@ifxundefined [1]{%
 \@ifx{#1\undefined}
}%
\providecommand \@ifnum [1]{%
 \ifnum #1\expandafter \@firstoftwo
 \else \expandafter \@secondoftwo
 \fi
}%
\providecommand \@ifx [1]{%
 \ifx #1\expandafter \@firstoftwo
 \else \expandafter \@secondoftwo
 \fi
}%
\providecommand \natexlab [1]{#1}%
\providecommand \enquote  [1]{``#1''}%
\providecommand \bibnamefont  [1]{#1}%
\providecommand \bibfnamefont [1]{#1}%
\providecommand \citenamefont [1]{#1}%
\providecommand \href@noop [0]{\@secondoftwo}%
\providecommand \href [0]{\begingroup \@sanitize@url \@href}%
\providecommand \@href[1]{\@@startlink{#1}\@@href}%
\providecommand \@@href[1]{\endgroup#1\@@endlink}%
\providecommand \@sanitize@url [0]{\catcode `\\12\catcode `\$12\catcode `\&12\catcode `\#12\catcode `\^12\catcode `\_12\catcode `\%12\relax}%
\providecommand \@@startlink[1]{}%
\providecommand \@@endlink[0]{}%
\providecommand \url  [0]{\begingroup\@sanitize@url \@url }%
\providecommand \@url [1]{\endgroup\@href {#1}{\urlprefix }}%
\providecommand \urlprefix  [0]{URL }%
\providecommand \Eprint [0]{\href }%
\providecommand \doibase [0]{http://dx.doi.org/}%
\providecommand \selectlanguage [0]{\@gobble}%
\providecommand \bibinfo  [0]{\@secondoftwo}%
\providecommand \bibfield  [0]{\@secondoftwo}%
\providecommand \translation [1]{[#1]}%
\providecommand \BibitemOpen [0]{}%
\providecommand \bibitemStop [0]{}%
\providecommand \bibitemNoStop [0]{.\EOS\space}%
\providecommand \EOS [0]{\spacefactor3000\relax}%
\providecommand \BibitemShut  [1]{\csname bibitem#1\endcsname}%
\let\auto@bib@innerbib\@empty
%</preamble>
\bibitem [{\citenamefont {Kurkela}\ \emph {et~al.}(2019{\natexlab{a}})\citenamefont {Kurkela}, \citenamefont {Mazeliauskas}, \citenamefont {Paquet}, \citenamefont {Schlichting},\ and\ \citenamefont {Teaney}}]{Kurkela:2018vqr}%
  \BibitemOpen
  \bibfield  {author} {\bibinfo {author} {\bibfnamefont {A.}~\bibnamefont {Kurkela}}, \bibinfo {author} {\bibfnamefont {A.}~\bibnamefont {Mazeliauskas}}, \bibinfo {author} {\bibfnamefont {J.-F.}\ \bibnamefont {Paquet}}, \bibinfo {author} {\bibfnamefont {S.}~\bibnamefont {Schlichting}}, \ and\ \bibinfo {author} {\bibfnamefont {D.}~\bibnamefont {Teaney}},\ }\href {\doibase 10.1103/PhysRevC.99.034910} {\bibfield  {journal} {\bibinfo  {journal} {Phys. Rev. C}\ }\textbf {\bibinfo {volume} {99}},\ \bibinfo {pages} {034910} (\bibinfo {year} {2019}{\natexlab{a}})},\ \Eprint {http://arxiv.org/abs/1805.00961} {arXiv:1805.00961 [hep-ph]} \BibitemShut {NoStop}%
\bibitem [{\citenamefont {Kurkela}\ \emph {et~al.}(2019{\natexlab{b}})\citenamefont {Kurkela}, \citenamefont {Mazeliauskas}, \citenamefont {Paquet}, \citenamefont {Schlichting},\ and\ \citenamefont {Teaney}}]{Kurkela:2018wud}%
  \BibitemOpen
  \bibfield  {author} {\bibinfo {author} {\bibfnamefont {A.}~\bibnamefont {Kurkela}}, \bibinfo {author} {\bibfnamefont {A.}~\bibnamefont {Mazeliauskas}}, \bibinfo {author} {\bibfnamefont {J.-F.}\ \bibnamefont {Paquet}}, \bibinfo {author} {\bibfnamefont {S.}~\bibnamefont {Schlichting}}, \ and\ \bibinfo {author} {\bibfnamefont {D.}~\bibnamefont {Teaney}},\ }\href {\doibase 10.1103/PhysRevLett.122.122302} {\bibfield  {journal} {\bibinfo  {journal} {Phys. Rev. Lett.}\ }\textbf {\bibinfo {volume} {122}},\ \bibinfo {pages} {122302} (\bibinfo {year} {2019}{\natexlab{b}})},\ \Eprint {http://arxiv.org/abs/1805.01604} {arXiv:1805.01604 [hep-ph]} \BibitemShut {NoStop}%
\bibitem [{\citenamefont {Muller}\ \emph {et~al.}(2012)\citenamefont {Muller}, \citenamefont {Schukraft},\ and\ \citenamefont {Wyslouch}}]{Muller:2012zq}%
  \BibitemOpen
  \bibfield  {author} {\bibinfo {author} {\bibfnamefont {B.}~\bibnamefont {Muller}}, \bibinfo {author} {\bibfnamefont {J.}~\bibnamefont {Schukraft}}, \ and\ \bibinfo {author} {\bibfnamefont {B.}~\bibnamefont {Wyslouch}},\ }\href {\doibase 10.1146/annurev-nucl-102711-094910} {\bibfield  {journal} {\bibinfo  {journal} {Ann. Rev. Nucl. Part. Sci.}\ }\textbf {\bibinfo {volume} {62}},\ \bibinfo {pages} {361} (\bibinfo {year} {2012})},\ \Eprint {http://arxiv.org/abs/1202.3233} {arXiv:1202.3233 [hep-ex]} \BibitemShut {NoStop}%
\bibitem [{\citenamefont {M\"uller}(2013)}]{Muller:2013dea}%
  \BibitemOpen
  \bibfield  {author} {\bibinfo {author} {\bibfnamefont {B.}~\bibnamefont {M\"uller}},\ }\href {\doibase 10.1088/0031-8949/2013/T158/014004} {\bibfield  {journal} {\bibinfo  {journal} {Phys. Scripta T}\ }\textbf {\bibinfo {volume} {158}},\ \bibinfo {pages} {014004} (\bibinfo {year} {2013})},\ \Eprint {http://arxiv.org/abs/1309.7616} {arXiv:1309.7616 [nucl-th]} \BibitemShut {NoStop}%
\bibitem [{\citenamefont {Moreland}\ \emph {et~al.}(2020)\citenamefont {Moreland}, \citenamefont {Bernhard},\ and\ \citenamefont {Bass}}]{Moreland:2018gsh}%
  \BibitemOpen
  \bibfield  {author} {\bibinfo {author} {\bibfnamefont {J.~S.}\ \bibnamefont {Moreland}}, \bibinfo {author} {\bibfnamefont {J.~E.}\ \bibnamefont {Bernhard}}, \ and\ \bibinfo {author} {\bibfnamefont {S.~A.}\ \bibnamefont {Bass}},\ }\href {\doibase 10.1103/PhysRevC.101.024911} {\bibfield  {journal} {\bibinfo  {journal} {Phys. Rev. C}\ }\textbf {\bibinfo {volume} {101}},\ \bibinfo {pages} {024911} (\bibinfo {year} {2020})},\ \Eprint {http://arxiv.org/abs/1808.02106} {arXiv:1808.02106 [nucl-th]} \BibitemShut {NoStop}%
\bibitem [{\citenamefont {Ke}\ \emph {et~al.}(2017)\citenamefont {Ke}, \citenamefont {Moreland}, \citenamefont {Bernhard},\ and\ \citenamefont {Bass}}]{Ke:2016jrd}%
  \BibitemOpen
  \bibfield  {author} {\bibinfo {author} {\bibfnamefont {W.}~\bibnamefont {Ke}}, \bibinfo {author} {\bibfnamefont {J.~S.}\ \bibnamefont {Moreland}}, \bibinfo {author} {\bibfnamefont {J.~E.}\ \bibnamefont {Bernhard}}, \ and\ \bibinfo {author} {\bibfnamefont {S.~A.}\ \bibnamefont {Bass}},\ }\href {\doibase 10.1103/PhysRevC.96.044912} {\bibfield  {journal} {\bibinfo  {journal} {Phys. Rev. C}\ }\textbf {\bibinfo {volume} {96}},\ \bibinfo {pages} {044912} (\bibinfo {year} {2017})},\ \Eprint {http://arxiv.org/abs/1610.08490} {arXiv:1610.08490 [nucl-th]} \BibitemShut {NoStop}%
\bibitem [{\citenamefont {Soeder}\ \emph {et~al.}(2023)\citenamefont {Soeder}, \citenamefont {Ke}, \citenamefont {Paquet},\ and\ \citenamefont {Bass}}]{Soeder:2023vdn}%
  \BibitemOpen
  \bibfield  {author} {\bibinfo {author} {\bibfnamefont {D.}~\bibnamefont {Soeder}}, \bibinfo {author} {\bibfnamefont {W.}~\bibnamefont {Ke}}, \bibinfo {author} {\bibfnamefont {J.~F.}\ \bibnamefont {Paquet}}, \ and\ \bibinfo {author} {\bibfnamefont {S.~A.}\ \bibnamefont {Bass}},\ }\href@noop {} {\  (\bibinfo {year} {2023})},\ \Eprint {http://arxiv.org/abs/2306.08665} {arXiv:2306.08665 [nucl-th]} \BibitemShut {NoStop}%
\bibitem [{\citenamefont {Schenke}\ \emph {et~al.}(2012{\natexlab{a}})\citenamefont {Schenke}, \citenamefont {Tribedy},\ and\ \citenamefont {Venugopalan}}]{Schenke:2012wb}%
  \BibitemOpen
  \bibfield  {author} {\bibinfo {author} {\bibfnamefont {B.}~\bibnamefont {Schenke}}, \bibinfo {author} {\bibfnamefont {P.}~\bibnamefont {Tribedy}}, \ and\ \bibinfo {author} {\bibfnamefont {R.}~\bibnamefont {Venugopalan}},\ }\href {\doibase 10.1103/PhysRevLett.108.252301} {\bibfield  {journal} {\bibinfo  {journal} {Phys. Rev. Lett.}\ }\textbf {\bibinfo {volume} {108}},\ \bibinfo {pages} {252301} (\bibinfo {year} {2012}{\natexlab{a}})},\ \Eprint {http://arxiv.org/abs/1202.6646} {arXiv:1202.6646 [nucl-th]} \BibitemShut {NoStop}%
\bibitem [{\citenamefont {Schenke}\ \emph {et~al.}(2012{\natexlab{b}})\citenamefont {Schenke}, \citenamefont {Tribedy},\ and\ \citenamefont {Venugopalan}}]{Schenke:2012hg}%
  \BibitemOpen
  \bibfield  {author} {\bibinfo {author} {\bibfnamefont {B.}~\bibnamefont {Schenke}}, \bibinfo {author} {\bibfnamefont {P.}~\bibnamefont {Tribedy}}, \ and\ \bibinfo {author} {\bibfnamefont {R.}~\bibnamefont {Venugopalan}},\ }\href {\doibase 10.1103/PhysRevC.86.034908} {\bibfield  {journal} {\bibinfo  {journal} {Phys. Rev. C}\ }\textbf {\bibinfo {volume} {86}},\ \bibinfo {pages} {034908} (\bibinfo {year} {2012}{\natexlab{b}})},\ \Eprint {http://arxiv.org/abs/1206.6805} {arXiv:1206.6805 [hep-ph]} \BibitemShut {NoStop}%
\bibitem [{\citenamefont {M\"antysaari}\ \emph {et~al.}(2017)\citenamefont {M\"antysaari}, \citenamefont {Schenke}, \citenamefont {Shen},\ and\ \citenamefont {Tribedy}}]{Mantysaari:2017cni}%
  \BibitemOpen
  \bibfield  {author} {\bibinfo {author} {\bibfnamefont {H.}~\bibnamefont {M\"antysaari}}, \bibinfo {author} {\bibfnamefont {B.}~\bibnamefont {Schenke}}, \bibinfo {author} {\bibfnamefont {C.}~\bibnamefont {Shen}}, \ and\ \bibinfo {author} {\bibfnamefont {P.}~\bibnamefont {Tribedy}},\ }\href {\doibase 10.1016/j.physletb.2017.07.038} {\bibfield  {journal} {\bibinfo  {journal} {Phys. Lett. B}\ }\textbf {\bibinfo {volume} {772}},\ \bibinfo {pages} {681} (\bibinfo {year} {2017})},\ \Eprint {http://arxiv.org/abs/1705.03177} {arXiv:1705.03177 [nucl-th]} \BibitemShut {NoStop}%
\bibitem [{\citenamefont {McDonald}\ \emph {et~al.}(2023)\citenamefont {McDonald}, \citenamefont {Jeon},\ and\ \citenamefont {Gale}}]{McDonald:2023qwc}%
  \BibitemOpen
  \bibfield  {author} {\bibinfo {author} {\bibfnamefont {S.}~\bibnamefont {McDonald}}, \bibinfo {author} {\bibfnamefont {S.}~\bibnamefont {Jeon}}, \ and\ \bibinfo {author} {\bibfnamefont {C.}~\bibnamefont {Gale}},\ }\href {\doibase 10.1103/PhysRevC.108.064910} {\bibfield  {journal} {\bibinfo  {journal} {Phys. Rev. C}\ }\textbf {\bibinfo {volume} {108}},\ \bibinfo {pages} {064910} (\bibinfo {year} {2023})},\ \Eprint {http://arxiv.org/abs/2306.04896} {arXiv:2306.04896 [hep-ph]} \BibitemShut {NoStop}%
\bibitem [{\citenamefont {Ipp}\ \emph {et~al.}(2024)\citenamefont {Ipp}, \citenamefont {Leuthner}, \citenamefont {M\"uller}, \citenamefont {Schlichting}, \citenamefont {Schmidt},\ and\ \citenamefont {Singh}}]{Ipp:2024ykh}%
  \BibitemOpen
  \bibfield  {author} {\bibinfo {author} {\bibfnamefont {A.}~\bibnamefont {Ipp}}, \bibinfo {author} {\bibfnamefont {M.}~\bibnamefont {Leuthner}}, \bibinfo {author} {\bibfnamefont {D.~I.}\ \bibnamefont {M\"uller}}, \bibinfo {author} {\bibfnamefont {S.}~\bibnamefont {Schlichting}}, \bibinfo {author} {\bibfnamefont {K.}~\bibnamefont {Schmidt}}, \ and\ \bibinfo {author} {\bibfnamefont {P.}~\bibnamefont {Singh}},\ }\href {\doibase 10.1103/PhysRevD.109.094040} {\bibfield  {journal} {\bibinfo  {journal} {Phys. Rev. D}\ }\textbf {\bibinfo {volume} {109}},\ \bibinfo {pages} {094040} (\bibinfo {year} {2024})},\ \Eprint {http://arxiv.org/abs/2401.10320} {arXiv:2401.10320 [hep-ph]} \BibitemShut {NoStop}%
\bibitem [{\citenamefont {Kuha}\ \emph {et~al.}(2025)\citenamefont {Kuha}, \citenamefont {Auvinen}, \citenamefont {Eskola}, \citenamefont {Hirvonen}, \citenamefont {Kanakubo},\ and\ \citenamefont {Niemi}}]{Kuha:2024kmq}%
  \BibitemOpen
  \bibfield  {author} {\bibinfo {author} {\bibfnamefont {M.}~\bibnamefont {Kuha}}, \bibinfo {author} {\bibfnamefont {J.}~\bibnamefont {Auvinen}}, \bibinfo {author} {\bibfnamefont {K.~J.}\ \bibnamefont {Eskola}}, \bibinfo {author} {\bibfnamefont {H.}~\bibnamefont {Hirvonen}}, \bibinfo {author} {\bibfnamefont {Y.}~\bibnamefont {Kanakubo}}, \ and\ \bibinfo {author} {\bibfnamefont {H.}~\bibnamefont {Niemi}},\ }\href {\doibase 10.1103/PhysRevC.111.054914} {\bibfield  {journal} {\bibinfo  {journal} {Phys. Rev. C}\ }\textbf {\bibinfo {volume} {111}},\ \bibinfo {pages} {054914} (\bibinfo {year} {2025})},\ \Eprint {http://arxiv.org/abs/2406.17592} {arXiv:2406.17592 [hep-ph]} \BibitemShut {NoStop}%
\bibitem [{\citenamefont {Hirvonen}\ \emph {et~al.}(2024)\citenamefont {Hirvonen}, \citenamefont {Kuha}, \citenamefont {Auvinen}, \citenamefont {Eskola}, \citenamefont {Kanakubo},\ and\ \citenamefont {Niemi}}]{Hirvonen:2024zne}%
  \BibitemOpen
  \bibfield  {author} {\bibinfo {author} {\bibfnamefont {H.}~\bibnamefont {Hirvonen}}, \bibinfo {author} {\bibfnamefont {M.}~\bibnamefont {Kuha}}, \bibinfo {author} {\bibfnamefont {J.}~\bibnamefont {Auvinen}}, \bibinfo {author} {\bibfnamefont {K.~J.}\ \bibnamefont {Eskola}}, \bibinfo {author} {\bibfnamefont {Y.}~\bibnamefont {Kanakubo}}, \ and\ \bibinfo {author} {\bibfnamefont {H.}~\bibnamefont {Niemi}},\ }\href {\doibase 10.1103/PhysRevC.110.034911} {\bibfield  {journal} {\bibinfo  {journal} {Phys. Rev. C}\ }\textbf {\bibinfo {volume} {110}},\ \bibinfo {pages} {034911} (\bibinfo {year} {2024})},\ \Eprint {http://arxiv.org/abs/2407.01338} {arXiv:2407.01338 [hep-ph]} \BibitemShut {NoStop}%
\bibitem [{\citenamefont {Shen}\ \emph {et~al.}(2016)\citenamefont {Shen}, \citenamefont {Qiu}, \citenamefont {Song}, \citenamefont {Bernhard}, \citenamefont {Bass},\ and\ \citenamefont {Heinz}}]{Shen:2014vra}%
  \BibitemOpen
  \bibfield  {author} {\bibinfo {author} {\bibfnamefont {C.}~\bibnamefont {Shen}}, \bibinfo {author} {\bibfnamefont {Z.}~\bibnamefont {Qiu}}, \bibinfo {author} {\bibfnamefont {H.}~\bibnamefont {Song}}, \bibinfo {author} {\bibfnamefont {J.}~\bibnamefont {Bernhard}}, \bibinfo {author} {\bibfnamefont {S.}~\bibnamefont {Bass}}, \ and\ \bibinfo {author} {\bibfnamefont {U.}~\bibnamefont {Heinz}},\ }\href {\doibase 10.1016/j.cpc.2015.08.039} {\bibfield  {journal} {\bibinfo  {journal} {Comput. Phys. Commun.}\ }\textbf {\bibinfo {volume} {199}},\ \bibinfo {pages} {61} (\bibinfo {year} {2016})},\ \Eprint {http://arxiv.org/abs/1409.8164} {arXiv:1409.8164 [nucl-th]} \BibitemShut {NoStop}%
%%CITATION = ARXIV:1409.8164;%%
\bibitem [{\citenamefont {Schenke}\ \emph {et~al.}(2010)\citenamefont {Schenke}, \citenamefont {Jeon},\ and\ \citenamefont {Gale}}]{Schenke:2010nt}%
  \BibitemOpen
  \bibfield  {author} {\bibinfo {author} {\bibfnamefont {B.}~\bibnamefont {Schenke}}, \bibinfo {author} {\bibfnamefont {S.}~\bibnamefont {Jeon}}, \ and\ \bibinfo {author} {\bibfnamefont {C.}~\bibnamefont {Gale}},\ }\href {\doibase 10.1103/PhysRevC.82.014903} {\bibfield  {journal} {\bibinfo  {journal} {Phys. Rev. C}\ }\textbf {\bibinfo {volume} {82}},\ \bibinfo {pages} {014903} (\bibinfo {year} {2010})},\ \Eprint {http://arxiv.org/abs/1004.1408} {arXiv:1004.1408 [hep-ph]} \BibitemShut {NoStop}%
\bibitem [{\citenamefont {Schenke}\ \emph {et~al.}(2011)\citenamefont {Schenke}, \citenamefont {Jeon},\ and\ \citenamefont {Gale}}]{Schenke:2010rr}%
  \BibitemOpen
  \bibfield  {author} {\bibinfo {author} {\bibfnamefont {B.}~\bibnamefont {Schenke}}, \bibinfo {author} {\bibfnamefont {S.}~\bibnamefont {Jeon}}, \ and\ \bibinfo {author} {\bibfnamefont {C.}~\bibnamefont {Gale}},\ }\href {\doibase 10.1103/PhysRevLett.106.042301} {\bibfield  {journal} {\bibinfo  {journal} {Phys. Rev. Lett.}\ }\textbf {\bibinfo {volume} {106}},\ \bibinfo {pages} {042301} (\bibinfo {year} {2011})},\ \Eprint {http://arxiv.org/abs/1009.3244} {arXiv:1009.3244 [hep-ph]} \BibitemShut {NoStop}%
\bibitem [{\citenamefont {Nijs}\ \emph {et~al.}(2021)\citenamefont {Nijs}, \citenamefont {van~der Schee}, \citenamefont {G{\"u}rsoy},\ and\ \citenamefont {Snellings}}]{Nijs:2020roc}%
  \BibitemOpen
  \bibfield  {author} {\bibinfo {author} {\bibfnamefont {G.}~\bibnamefont {Nijs}}, \bibinfo {author} {\bibfnamefont {W.}~\bibnamefont {van~der Schee}}, \bibinfo {author} {\bibfnamefont {U.}~\bibnamefont {G{\"u}rsoy}}, \ and\ \bibinfo {author} {\bibfnamefont {R.}~\bibnamefont {Snellings}},\ }\href {\doibase 10.1103/PhysRevC.103.054909} {\bibfield  {journal} {\bibinfo  {journal} {Phys. Rev. C}\ }\textbf {\bibinfo {volume} {103}},\ \bibinfo {pages} {054909} (\bibinfo {year} {2021})},\ \Eprint {http://arxiv.org/abs/2010.15134} {arXiv:2010.15134 [nucl-th]} \BibitemShut {NoStop}%
\bibitem [{\citenamefont {Bass}\ \emph {et~al.}(1998)\citenamefont {Bass} \emph {et~al.}}]{Bass:1998ca}%
  \BibitemOpen
  \bibfield  {author} {\bibinfo {author} {\bibfnamefont {S.~A.}\ \bibnamefont {Bass}} \emph {et~al.},\ }\href {\doibase 10.1016/S0146-6410(98)00058-1} {\bibfield  {journal} {\bibinfo  {journal} {Prog. Part. Nucl. Phys.}\ }\textbf {\bibinfo {volume} {41}},\ \bibinfo {pages} {255} (\bibinfo {year} {1998})},\ \Eprint {http://arxiv.org/abs/nucl-th/9803035} {arXiv:nucl-th/9803035} \BibitemShut {NoStop}%
%%CITATION = NUCL-TH/9803035;%%
\bibitem [{\citenamefont {Bleicher}\ \emph {et~al.}(1999)\citenamefont {Bleicher} \emph {et~al.}}]{Bleicher:1999xi}%
  \BibitemOpen
  \bibfield  {author} {\bibinfo {author} {\bibfnamefont {M.}~\bibnamefont {Bleicher}} \emph {et~al.},\ }\href {\doibase 10.1088/0954-3899/25/9/308} {\bibfield  {journal} {\bibinfo  {journal} {J. Phys. G}\ }\textbf {\bibinfo {volume} {25}},\ \bibinfo {pages} {1859} (\bibinfo {year} {1999})},\ \Eprint {http://arxiv.org/abs/hep-ph/9909407} {arXiv:hep-ph/9909407} \BibitemShut {NoStop}%
\bibitem [{\citenamefont {Wergieluk}\ \emph {et~al.}(2024)\citenamefont {Wergieluk}, \citenamefont {Weil}, \citenamefont {Tindall}, \citenamefont {Steinberg}, \citenamefont {Staudenmaier}, \citenamefont {Sorensen}, \citenamefont {Sciarra}, \citenamefont {Schäfer}, \citenamefont {Sattler}, \citenamefont {Ryu}, \citenamefont {Rothermel}, \citenamefont {Rose}, \citenamefont {Roch}, \citenamefont {Prinz}, \citenamefont {Petersen}, \citenamefont {Paulinyova}, \citenamefont {Pang}, \citenamefont {Oliinychenko}, \citenamefont {Mohs}, \citenamefont {Mitrovic}, \citenamefont {Mayer}, \citenamefont {Li}, \citenamefont {Kübler}, \citenamefont {Kretz}, \citenamefont {Kehrenberg}, \citenamefont {Inghirami}, \citenamefont {Hirayama}, \citenamefont {Hammelmann}, \citenamefont {Götz}, \citenamefont {Groebel}, \citenamefont {Goldschmidt}, \citenamefont {Geiger}, \citenamefont {Garcia-Montero}, \citenamefont {Elfner}, \citenamefont {Ehlert}, \citenamefont {Christensen}, \citenamefont {Bäuchle}, \citenamefont {Auvinen},\
  and\ \citenamefont {Attems}}]{wergieluk_2024_10707746}%
  \BibitemOpen
  \bibfield  {author} {\bibinfo {author} {\bibfnamefont {A.}~\bibnamefont {Wergieluk}}, \bibinfo {author} {\bibfnamefont {J.}~\bibnamefont {Weil}}, \bibinfo {author} {\bibfnamefont {J.}~\bibnamefont {Tindall}}, \bibinfo {author} {\bibfnamefont {V.}~\bibnamefont {Steinberg}}, \bibinfo {author} {\bibfnamefont {J.}~\bibnamefont {Staudenmaier}}, \bibinfo {author} {\bibfnamefont {A.}~\bibnamefont {Sorensen}}, \bibinfo {author} {\bibfnamefont {A.}~\bibnamefont {Sciarra}}, \bibinfo {author} {\bibfnamefont {A.}~\bibnamefont {Schäfer}}, \bibinfo {author} {\bibfnamefont {R.}~\bibnamefont {Sattler}}, \bibinfo {author} {\bibfnamefont {S.}~\bibnamefont {Ryu}}, \bibinfo {author} {\bibfnamefont {J.}~\bibnamefont {Rothermel}}, \bibinfo {author} {\bibfnamefont {J.-B.}\ \bibnamefont {Rose}}, \bibinfo {author} {\bibfnamefont {H.}~\bibnamefont {Roch}}, \bibinfo {author} {\bibfnamefont {L.}~\bibnamefont {Prinz}}, \bibinfo {author} {\bibfnamefont {H.}~\bibnamefont {Petersen}}, \bibinfo {author} {\bibfnamefont {Z.}~\bibnamefont
  {Paulinyova}}, \bibinfo {author} {\bibfnamefont {L.-G.}\ \bibnamefont {Pang}}, \bibinfo {author} {\bibfnamefont {D.}~\bibnamefont {Oliinychenko}}, \bibinfo {author} {\bibfnamefont {J.}~\bibnamefont {Mohs}}, \bibinfo {author} {\bibfnamefont {D.}~\bibnamefont {Mitrovic}}, \bibinfo {author} {\bibfnamefont {M.}~\bibnamefont {Mayer}}, \bibinfo {author} {\bibfnamefont {F.}~\bibnamefont {Li}}, \bibinfo {author} {\bibfnamefont {N.}~\bibnamefont {Kübler}}, \bibinfo {author} {\bibfnamefont {M.}~\bibnamefont {Kretz}}, \bibinfo {author} {\bibfnamefont {T.}~\bibnamefont {Kehrenberg}}, \bibinfo {author} {\bibfnamefont {G.}~\bibnamefont {Inghirami}}, \bibinfo {author} {\bibfnamefont {R.}~\bibnamefont {Hirayama}}, \bibinfo {author} {\bibfnamefont {J.}~\bibnamefont {Hammelmann}}, \bibinfo {author} {\bibfnamefont {N.}~\bibnamefont {Götz}}, \bibinfo {author} {\bibfnamefont {J.}~\bibnamefont {Groebel}}, \bibinfo {author} {\bibfnamefont {A.}~\bibnamefont {Goldschmidt}}, \bibinfo {author} {\bibfnamefont {L.}~\bibnamefont
  {Geiger}}, \bibinfo {author} {\bibfnamefont {O.}~\bibnamefont {Garcia-Montero}}, \bibinfo {author} {\bibfnamefont {H.}~\bibnamefont {Elfner}}, \bibinfo {author} {\bibfnamefont {N.}~\bibnamefont {Ehlert}}, \bibinfo {author} {\bibfnamefont {C.~H.}\ \bibnamefont {Christensen}}, \bibinfo {author} {\bibfnamefont {B.}~\bibnamefont {Bäuchle}}, \bibinfo {author} {\bibfnamefont {J.}~\bibnamefont {Auvinen}}, \ and\ \bibinfo {author} {\bibfnamefont {M.}~\bibnamefont {Attems}},\ }\href {\doibase 10.5281/zenodo.10707746} {\enquote {\bibinfo {title} {smash-transport/smash: Smash-3.1},}\ } (\bibinfo {year} {2024})\BibitemShut {NoStop}%
\bibitem [{\citenamefont {Weil}\ \emph {et~al.}(2016)\citenamefont {Weil} \emph {et~al.}}]{SMASH:2016zqf}%
  \BibitemOpen
  \bibfield  {author} {\bibinfo {author} {\bibfnamefont {J.}~\bibnamefont {Weil}} \emph {et~al.} (\bibinfo {collaboration} {SMASH}),\ }\href {\doibase 10.1103/PhysRevC.94.054905} {\bibfield  {journal} {\bibinfo  {journal} {Phys. Rev. C}\ }\textbf {\bibinfo {volume} {94}},\ \bibinfo {pages} {054905} (\bibinfo {year} {2016})},\ \Eprint {http://arxiv.org/abs/1606.06642} {arXiv:1606.06642 [nucl-th]} \BibitemShut {NoStop}%
\bibitem [{\citenamefont {Sjostrand}\ \emph {et~al.}(2006)\citenamefont {Sjostrand}, \citenamefont {Mrenna},\ and\ \citenamefont {Skands}}]{Sjostrand:2006za}%
  \BibitemOpen
  \bibfield  {author} {\bibinfo {author} {\bibfnamefont {T.}~\bibnamefont {Sjostrand}}, \bibinfo {author} {\bibfnamefont {S.}~\bibnamefont {Mrenna}}, \ and\ \bibinfo {author} {\bibfnamefont {P.~Z.}\ \bibnamefont {Skands}},\ }\href {\doibase 10.1088/1126-6708/2006/05/026} {\bibfield  {journal} {\bibinfo  {journal} {JHEP}\ }\textbf {\bibinfo {volume} {05}},\ \bibinfo {pages} {026} (\bibinfo {year} {2006})},\ \Eprint {http://arxiv.org/abs/hep-ph/0603175} {arXiv:hep-ph/0603175} \BibitemShut {NoStop}%
\bibitem [{\citenamefont {Sjostrand}\ \emph {et~al.}(2008)\citenamefont {Sjostrand}, \citenamefont {Mrenna},\ and\ \citenamefont {Skands}}]{Sjostrand:2007gs}%
  \BibitemOpen
  \bibfield  {author} {\bibinfo {author} {\bibfnamefont {T.}~\bibnamefont {Sjostrand}}, \bibinfo {author} {\bibfnamefont {S.}~\bibnamefont {Mrenna}}, \ and\ \bibinfo {author} {\bibfnamefont {P.~Z.}\ \bibnamefont {Skands}},\ }\href {\doibase 10.1016/j.cpc.2008.01.036} {\bibfield  {journal} {\bibinfo  {journal} {Comput. Phys. Commun.}\ }\textbf {\bibinfo {volume} {178}},\ \bibinfo {pages} {852} (\bibinfo {year} {2008})},\ \Eprint {http://arxiv.org/abs/0710.3820} {arXiv:0710.3820 [hep-ph]} \BibitemShut {NoStop}%
\bibitem [{\citenamefont {Miller}\ \emph {et~al.}(2007)\citenamefont {Miller}, \citenamefont {Reygers}, \citenamefont {Sanders},\ and\ \citenamefont {Steinberg}}]{Miller:2007ri}%
  \BibitemOpen
  \bibfield  {author} {\bibinfo {author} {\bibfnamefont {M.~L.}\ \bibnamefont {Miller}}, \bibinfo {author} {\bibfnamefont {K.}~\bibnamefont {Reygers}}, \bibinfo {author} {\bibfnamefont {S.~J.}\ \bibnamefont {Sanders}}, \ and\ \bibinfo {author} {\bibfnamefont {P.}~\bibnamefont {Steinberg}},\ }\href {\doibase 10.1146/annurev.nucl.57.090506.123020} {\bibfield  {journal} {\bibinfo  {journal} {Ann. Rev. Nucl. Part. Sci.}\ }\textbf {\bibinfo {volume} {57}},\ \bibinfo {pages} {205} (\bibinfo {year} {2007})},\ \Eprint {http://arxiv.org/abs/nucl-ex/0701025} {arXiv:nucl-ex/0701025 [nucl-ex]} \BibitemShut {NoStop}%
%%CITATION = NUCL-EX/0701025;%%
\bibitem [{\citenamefont {Moreland}\ \emph {et~al.}(2015)\citenamefont {Moreland}, \citenamefont {Bernhard},\ and\ \citenamefont {Bass}}]{Moreland:2014oya}%
  \BibitemOpen
  \bibfield  {author} {\bibinfo {author} {\bibfnamefont {J.~S.}\ \bibnamefont {Moreland}}, \bibinfo {author} {\bibfnamefont {J.~E.}\ \bibnamefont {Bernhard}}, \ and\ \bibinfo {author} {\bibfnamefont {S.~A.}\ \bibnamefont {Bass}},\ }\href {\doibase 10.1103/PhysRevC.92.011901} {\bibfield  {journal} {\bibinfo  {journal} {Phys. Rev. C}\ }\textbf {\bibinfo {volume} {92}},\ \bibinfo {pages} {011901} (\bibinfo {year} {2015})},\ \Eprint {http://arxiv.org/abs/1412.4708} {arXiv:1412.4708 [nucl-th]} \BibitemShut {NoStop}%
\bibitem [{\citenamefont {Blaizot}\ \emph {et~al.}(2014)\citenamefont {Blaizot}, \citenamefont {Broniowski},\ and\ \citenamefont {Ollitrault}}]{Blaizot:2014nia}%
  \BibitemOpen
  \bibfield  {author} {\bibinfo {author} {\bibfnamefont {J.-P.}\ \bibnamefont {Blaizot}}, \bibinfo {author} {\bibfnamefont {W.}~\bibnamefont {Broniowski}}, \ and\ \bibinfo {author} {\bibfnamefont {J.-Y.}\ \bibnamefont {Ollitrault}},\ }\href {\doibase 10.1016/j.physletb.2014.09.028} {\bibfield  {journal} {\bibinfo  {journal} {Phys. Lett. B}\ }\textbf {\bibinfo {volume} {738}},\ \bibinfo {pages} {166} (\bibinfo {year} {2014})},\ \Eprint {http://arxiv.org/abs/1405.3572} {arXiv:1405.3572 [nucl-th]} \BibitemShut {NoStop}%
\bibitem [{\citenamefont {Gr{\"o}nqvist}\ \emph {et~al.}(2016)\citenamefont {Gr{\"o}nqvist}, \citenamefont {Blaizot},\ and\ \citenamefont {Ollitrault}}]{Gronqvist:2016hym}%
  \BibitemOpen
  \bibfield  {author} {\bibinfo {author} {\bibfnamefont {H.}~\bibnamefont {Gr{\"o}nqvist}}, \bibinfo {author} {\bibfnamefont {J.-P.}\ \bibnamefont {Blaizot}}, \ and\ \bibinfo {author} {\bibfnamefont {J.-Y.}\ \bibnamefont {Ollitrault}},\ }\href {\doibase 10.1103/PhysRevC.94.034905} {\bibfield  {journal} {\bibinfo  {journal} {Phys. Rev. C}\ }\textbf {\bibinfo {volume} {94}},\ \bibinfo {pages} {034905} (\bibinfo {year} {2016})},\ \Eprint {http://arxiv.org/abs/1604.07230} {arXiv:1604.07230 [nucl-th]} \BibitemShut {NoStop}%
\bibitem [{\citenamefont {Broniowski}\ \emph {et~al.}(2009)\citenamefont {Broniowski}, \citenamefont {Florkowski}, \citenamefont {Chojnacki},\ and\ \citenamefont {Kisiel}}]{Broniowski:2008qk}%
  \BibitemOpen
  \bibfield  {author} {\bibinfo {author} {\bibfnamefont {W.}~\bibnamefont {Broniowski}}, \bibinfo {author} {\bibfnamefont {W.}~\bibnamefont {Florkowski}}, \bibinfo {author} {\bibfnamefont {M.}~\bibnamefont {Chojnacki}}, \ and\ \bibinfo {author} {\bibfnamefont {A.}~\bibnamefont {Kisiel}},\ }\href {\doibase 10.1103/PhysRevC.80.034902} {\bibfield  {journal} {\bibinfo  {journal} {Phys. Rev.}\ }\textbf {\bibinfo {volume} {C80}},\ \bibinfo {pages} {034902} (\bibinfo {year} {2009})},\ \Eprint {http://arxiv.org/abs/0812.3393} {arXiv:0812.3393 [nucl-th]} \BibitemShut {NoStop}%
%%CITATION = 0812.3393;%%
\bibitem [{\citenamefont {Liu}\ \emph {et~al.}(2015)\citenamefont {Liu}, \citenamefont {Shen},\ and\ \citenamefont {Heinz}}]{Liu:2015nwa}%
  \BibitemOpen
  \bibfield  {author} {\bibinfo {author} {\bibfnamefont {J.}~\bibnamefont {Liu}}, \bibinfo {author} {\bibfnamefont {C.}~\bibnamefont {Shen}}, \ and\ \bibinfo {author} {\bibfnamefont {U.}~\bibnamefont {Heinz}},\ }\href {\doibase 10.1103/PhysRevC.91.064906} {\bibfield  {journal} {\bibinfo  {journal} {Phys. Rev. C}\ }\textbf {\bibinfo {volume} {91}},\ \bibinfo {pages} {064906} (\bibinfo {year} {2015})},\ \bibinfo {note} {[Erratum: Phys.Rev.C 92, 049904 (2015)]},\ \Eprint {http://arxiv.org/abs/1504.02160} {arXiv:1504.02160 [nucl-th]} \BibitemShut {NoStop}%
\bibitem [{\citenamefont {Vredevoogd}\ and\ \citenamefont {Pratt}(2009)}]{Vredevoogd:2008id}%
  \BibitemOpen
  \bibfield  {author} {\bibinfo {author} {\bibfnamefont {J.}~\bibnamefont {Vredevoogd}}\ and\ \bibinfo {author} {\bibfnamefont {S.}~\bibnamefont {Pratt}},\ }\href {\doibase 10.1103/PhysRevC.79.044915} {\bibfield  {journal} {\bibinfo  {journal} {Phys. Rev. C}\ }\textbf {\bibinfo {volume} {79}},\ \bibinfo {pages} {044915} (\bibinfo {year} {2009})},\ \Eprint {http://arxiv.org/abs/0810.4325} {arXiv:0810.4325 [nucl-th]} \BibitemShut {NoStop}%
\bibitem [{\citenamefont {van~der Schee}\ \emph {et~al.}(2013)\citenamefont {van~der Schee}, \citenamefont {Romatschke},\ and\ \citenamefont {Pratt}}]{vanderSchee:2013pia}%
  \BibitemOpen
  \bibfield  {author} {\bibinfo {author} {\bibfnamefont {W.}~\bibnamefont {van~der Schee}}, \bibinfo {author} {\bibfnamefont {P.}~\bibnamefont {Romatschke}}, \ and\ \bibinfo {author} {\bibfnamefont {S.}~\bibnamefont {Pratt}},\ }\href {\doibase 10.1103/PhysRevLett.111.222302} {\bibfield  {journal} {\bibinfo  {journal} {Phys. Rev. Lett.}\ }\textbf {\bibinfo {volume} {111}},\ \bibinfo {pages} {222302} (\bibinfo {year} {2013})},\ \Eprint {http://arxiv.org/abs/1307.2539} {arXiv:1307.2539 [nucl-th]} \BibitemShut {NoStop}%
\bibitem [{\citenamefont {Romatschke}(2015)}]{Romatschke:2015gxa}%
  \BibitemOpen
  \bibfield  {author} {\bibinfo {author} {\bibfnamefont {P.}~\bibnamefont {Romatschke}},\ }\href {\doibase 10.1140/epjc/s10052-015-3509-3} {\bibfield  {journal} {\bibinfo  {journal} {Eur. Phys. J. C}\ }\textbf {\bibinfo {volume} {75}},\ \bibinfo {pages} {305} (\bibinfo {year} {2015})},\ \Eprint {http://arxiv.org/abs/1502.04745} {arXiv:1502.04745 [nucl-th]} \BibitemShut {NoStop}%
\bibitem [{\citenamefont {Wu}\ \emph {et~al.}(2024)\citenamefont {Wu}, \citenamefont {Du}, \citenamefont {Gale},\ and\ \citenamefont {Jeon}}]{Wu:2024pba}%
  \BibitemOpen
  \bibfield  {author} {\bibinfo {author} {\bibfnamefont {X.-Y.}\ \bibnamefont {Wu}}, \bibinfo {author} {\bibfnamefont {L.}~\bibnamefont {Du}}, \bibinfo {author} {\bibfnamefont {C.}~\bibnamefont {Gale}}, \ and\ \bibinfo {author} {\bibfnamefont {S.}~\bibnamefont {Jeon}},\ }\href {\doibase 10.1103/PhysRevC.110.054904} {\bibfield  {journal} {\bibinfo  {journal} {Phys. Rev. C}\ }\textbf {\bibinfo {volume} {110}},\ \bibinfo {pages} {054904} (\bibinfo {year} {2024})},\ \Eprint {http://arxiv.org/abs/2407.04156} {arXiv:2407.04156 [nucl-th]} \BibitemShut {NoStop}%
\bibitem [{\citenamefont {Garcia-Montero}\ \emph {et~al.}(2024{\natexlab{a}})\citenamefont {Garcia-Montero}, \citenamefont {Mazeliauskas}, \citenamefont {Plaschke},\ and\ \citenamefont {Schlichting}}]{Garcia-Montero:2023lrd}%
  \BibitemOpen
  \bibfield  {author} {\bibinfo {author} {\bibfnamefont {O.}~\bibnamefont {Garcia-Montero}}, \bibinfo {author} {\bibfnamefont {A.}~\bibnamefont {Mazeliauskas}}, \bibinfo {author} {\bibfnamefont {P.}~\bibnamefont {Plaschke}}, \ and\ \bibinfo {author} {\bibfnamefont {S.}~\bibnamefont {Schlichting}},\ }\href {\doibase 10.1007/JHEP03(2024)053} {\bibfield  {journal} {\bibinfo  {journal} {JHEP}\ }\textbf {\bibinfo {volume} {03}},\ \bibinfo {pages} {053} (\bibinfo {year} {2024}{\natexlab{a}})},\ \Eprint {http://arxiv.org/abs/2308.09747} {arXiv:2308.09747 [hep-ph]} \BibitemShut {NoStop}%
\bibitem [{\citenamefont {Garcia-Montero}\ \emph {et~al.}(2025{\natexlab{a}})\citenamefont {Garcia-Montero}, \citenamefont {Plaschke},\ and\ \citenamefont {Schlichting}}]{Garcia-Montero:2024lbl}%
  \BibitemOpen
  \bibfield  {author} {\bibinfo {author} {\bibfnamefont {O.}~\bibnamefont {Garcia-Montero}}, \bibinfo {author} {\bibfnamefont {P.}~\bibnamefont {Plaschke}}, \ and\ \bibinfo {author} {\bibfnamefont {S.}~\bibnamefont {Schlichting}},\ }\href {\doibase 10.1103/PhysRevD.111.034036} {\bibfield  {journal} {\bibinfo  {journal} {Phys. Rev. D}\ }\textbf {\bibinfo {volume} {111}},\ \bibinfo {pages} {034036} (\bibinfo {year} {2025}{\natexlab{a}})},\ \Eprint {http://arxiv.org/abs/2403.04846} {arXiv:2403.04846 [hep-ph]} \BibitemShut {NoStop}%
\bibitem [{\citenamefont {Schenke}\ \emph {et~al.}(2020)\citenamefont {Schenke}, \citenamefont {Shen},\ and\ \citenamefont {Teaney}}]{Schenke:2020uqq}%
  \BibitemOpen
  \bibfield  {author} {\bibinfo {author} {\bibfnamefont {B.}~\bibnamefont {Schenke}}, \bibinfo {author} {\bibfnamefont {C.}~\bibnamefont {Shen}}, \ and\ \bibinfo {author} {\bibfnamefont {D.}~\bibnamefont {Teaney}},\ }\href {\doibase 10.1103/PhysRevC.102.034905} {\bibfield  {journal} {\bibinfo  {journal} {Phys. Rev. C}\ }\textbf {\bibinfo {volume} {102}},\ \bibinfo {pages} {034905} (\bibinfo {year} {2020})},\ \Eprint {http://arxiv.org/abs/2004.00690} {arXiv:2004.00690 [nucl-th]} \BibitemShut {NoStop}%
\bibitem [{\citenamefont {Nunes~da Silva}\ \emph {et~al.}(2021)\citenamefont {Nunes~da Silva}, \citenamefont {Chinellato}, \citenamefont {Hippert}, \citenamefont {Serenone}, \citenamefont {Takahashi}, \citenamefont {Denicol}, \citenamefont {Luzum},\ and\ \citenamefont {Noronha}}]{NunesdaSilva:2020bfs}%
  \BibitemOpen
  \bibfield  {author} {\bibinfo {author} {\bibfnamefont {T.}~\bibnamefont {Nunes~da Silva}}, \bibinfo {author} {\bibfnamefont {D.}~\bibnamefont {Chinellato}}, \bibinfo {author} {\bibfnamefont {M.}~\bibnamefont {Hippert}}, \bibinfo {author} {\bibfnamefont {W.}~\bibnamefont {Serenone}}, \bibinfo {author} {\bibfnamefont {J.}~\bibnamefont {Takahashi}}, \bibinfo {author} {\bibfnamefont {G.~S.}\ \bibnamefont {Denicol}}, \bibinfo {author} {\bibfnamefont {M.}~\bibnamefont {Luzum}}, \ and\ \bibinfo {author} {\bibfnamefont {J.}~\bibnamefont {Noronha}},\ }\href {\doibase 10.1103/PhysRevC.103.054906} {\bibfield  {journal} {\bibinfo  {journal} {Phys. Rev. C}\ }\textbf {\bibinfo {volume} {103}},\ \bibinfo {pages} {054906} (\bibinfo {year} {2021})},\ \Eprint {http://arxiv.org/abs/2006.02324} {arXiv:2006.02324 [nucl-th]} \BibitemShut {NoStop}%
\bibitem [{\citenamefont {Kurkela}\ and\ \citenamefont {Zhu}(2015)}]{Kurkela:2015qoa}%
  \BibitemOpen
  \bibfield  {author} {\bibinfo {author} {\bibfnamefont {A.}~\bibnamefont {Kurkela}}\ and\ \bibinfo {author} {\bibfnamefont {Y.}~\bibnamefont {Zhu}},\ }\href {\doibase 10.1103/PhysRevLett.115.182301} {\bibfield  {journal} {\bibinfo  {journal} {Phys. Rev. Lett.}\ }\textbf {\bibinfo {volume} {115}},\ \bibinfo {pages} {182301} (\bibinfo {year} {2015})},\ \Eprint {http://arxiv.org/abs/1506.06647} {arXiv:1506.06647 [hep-ph]} \BibitemShut {NoStop}%
\bibitem [{\citenamefont {Giacalone}\ \emph {et~al.}(2019)\citenamefont {Giacalone}, \citenamefont {Mazeliauskas},\ and\ \citenamefont {Schlichting}}]{Giacalone:2019ldn}%
  \BibitemOpen
  \bibfield  {author} {\bibinfo {author} {\bibfnamefont {G.}~\bibnamefont {Giacalone}}, \bibinfo {author} {\bibfnamefont {A.}~\bibnamefont {Mazeliauskas}}, \ and\ \bibinfo {author} {\bibfnamefont {S.}~\bibnamefont {Schlichting}},\ }\href {\doibase 10.1103/PhysRevLett.123.262301} {\bibfield  {journal} {\bibinfo  {journal} {Phys. Rev. Lett.}\ }\textbf {\bibinfo {volume} {123}},\ \bibinfo {pages} {262301} (\bibinfo {year} {2019})},\ \Eprint {http://arxiv.org/abs/1908.02866} {arXiv:1908.02866 [hep-ph]} \BibitemShut {NoStop}%
\bibitem [{\citenamefont {Kurkela}(2016)}]{Kurkela:2016vts}%
  \BibitemOpen
  \bibfield  {author} {\bibinfo {author} {\bibfnamefont {A.}~\bibnamefont {Kurkela}},\ }\href {\doibase 10.1016/j.nuclphysa.2016.01.069} {\bibfield  {journal} {\bibinfo  {journal} {Nucl. Phys. A}\ }\textbf {\bibinfo {volume} {956}},\ \bibinfo {pages} {136} (\bibinfo {year} {2016})},\ \Eprint {http://arxiv.org/abs/1601.03283} {arXiv:1601.03283 [hep-ph]} \BibitemShut {NoStop}%
\bibitem [{\citenamefont {Kamata}\ \emph {et~al.}(2020)\citenamefont {Kamata}, \citenamefont {Martinez}, \citenamefont {Plaschke}, \citenamefont {Ochsenfeld},\ and\ \citenamefont {Schlichting}}]{Kamata:2020mka}%
  \BibitemOpen
  \bibfield  {author} {\bibinfo {author} {\bibfnamefont {S.}~\bibnamefont {Kamata}}, \bibinfo {author} {\bibfnamefont {M.}~\bibnamefont {Martinez}}, \bibinfo {author} {\bibfnamefont {P.}~\bibnamefont {Plaschke}}, \bibinfo {author} {\bibfnamefont {S.}~\bibnamefont {Ochsenfeld}}, \ and\ \bibinfo {author} {\bibfnamefont {S.}~\bibnamefont {Schlichting}},\ }\href {\doibase 10.1103/PhysRevD.102.056003} {\bibfield  {journal} {\bibinfo  {journal} {Phys. Rev. D}\ }\textbf {\bibinfo {volume} {102}},\ \bibinfo {pages} {056003} (\bibinfo {year} {2020})},\ \Eprint {http://arxiv.org/abs/2004.06751} {arXiv:2004.06751 [hep-ph]} \BibitemShut {NoStop}%
\bibitem [{\citenamefont {Keegan}\ \emph {et~al.}(2016)\citenamefont {Keegan}, \citenamefont {Kurkela}, \citenamefont {Mazeliauskas},\ and\ \citenamefont {Teaney}}]{Keegan:2016cpi}%
  \BibitemOpen
  \bibfield  {author} {\bibinfo {author} {\bibfnamefont {L.}~\bibnamefont {Keegan}}, \bibinfo {author} {\bibfnamefont {A.}~\bibnamefont {Kurkela}}, \bibinfo {author} {\bibfnamefont {A.}~\bibnamefont {Mazeliauskas}}, \ and\ \bibinfo {author} {\bibfnamefont {D.}~\bibnamefont {Teaney}},\ }\href {\doibase 10.1007/JHEP08(2016)171} {\bibfield  {journal} {\bibinfo  {journal} {JHEP}\ }\textbf {\bibinfo {volume} {08}},\ \bibinfo {pages} {171} (\bibinfo {year} {2016})},\ \Eprint {http://arxiv.org/abs/1605.04287} {arXiv:1605.04287 [hep-ph]} \BibitemShut {NoStop}%
\bibitem [{\citenamefont {York}\ and\ \citenamefont {Moore}(2009)}]{York:2008rr}%
  \BibitemOpen
  \bibfield  {author} {\bibinfo {author} {\bibfnamefont {M.~A.}\ \bibnamefont {York}}\ and\ \bibinfo {author} {\bibfnamefont {G.~D.}\ \bibnamefont {Moore}},\ }\href {\doibase 10.1103/PhysRevD.79.054011} {\bibfield  {journal} {\bibinfo  {journal} {Phys. Rev. D}\ }\textbf {\bibinfo {volume} {79}},\ \bibinfo {pages} {054011} (\bibinfo {year} {2009})},\ \Eprint {http://arxiv.org/abs/0811.0729} {arXiv:0811.0729 [hep-ph]} \BibitemShut {NoStop}%
\bibitem [{\citenamefont {Dash}\ \emph {et~al.}(2023)\citenamefont {Dash}, \citenamefont {Jaiswal}, \citenamefont {Bhadury},\ and\ \citenamefont {Jaiswal}}]{Dash:2023ppc}%
  \BibitemOpen
  \bibfield  {author} {\bibinfo {author} {\bibfnamefont {D.}~\bibnamefont {Dash}}, \bibinfo {author} {\bibfnamefont {S.}~\bibnamefont {Jaiswal}}, \bibinfo {author} {\bibfnamefont {S.}~\bibnamefont {Bhadury}}, \ and\ \bibinfo {author} {\bibfnamefont {A.}~\bibnamefont {Jaiswal}},\ }\href {\doibase 10.1103/PhysRevC.108.064913} {\bibfield  {journal} {\bibinfo  {journal} {Phys. Rev. C}\ }\textbf {\bibinfo {volume} {108}},\ \bibinfo {pages} {064913} (\bibinfo {year} {2023})},\ \Eprint {http://arxiv.org/abs/2307.06195} {arXiv:2307.06195 [nucl-th]} \BibitemShut {NoStop}%
\bibitem [{\citenamefont {Jaiswal}(2013)}]{Jaiswal:2013npa}%
  \BibitemOpen
  \bibfield  {author} {\bibinfo {author} {\bibfnamefont {A.}~\bibnamefont {Jaiswal}},\ }\href {\doibase 10.1103/PhysRevC.87.051901} {\bibfield  {journal} {\bibinfo  {journal} {Phys. Rev. C}\ }\textbf {\bibinfo {volume} {87}},\ \bibinfo {pages} {051901} (\bibinfo {year} {2013})},\ \Eprint {http://arxiv.org/abs/1302.6311} {arXiv:1302.6311 [nucl-th]} \BibitemShut {NoStop}%
\bibitem [{\citenamefont {Arnold}\ \emph {et~al.}(2003)\citenamefont {Arnold}, \citenamefont {Moore},\ and\ \citenamefont {Yaffe}}]{Arnold:2002zm}%
  \BibitemOpen
  \bibfield  {author} {\bibinfo {author} {\bibfnamefont {P.~B.}\ \bibnamefont {Arnold}}, \bibinfo {author} {\bibfnamefont {G.~D.}\ \bibnamefont {Moore}}, \ and\ \bibinfo {author} {\bibfnamefont {L.~G.}\ \bibnamefont {Yaffe}},\ }\href {\doibase 10.1088/1126-6708/2003/01/030} {\bibfield  {journal} {\bibinfo  {journal} {JHEP}\ }\textbf {\bibinfo {volume} {01}},\ \bibinfo {pages} {030} (\bibinfo {year} {2003})},\ \Eprint {http://arxiv.org/abs/hep-ph/0209353} {arXiv:hep-ph/0209353} \BibitemShut {NoStop}%
\bibitem [{\citenamefont {Du}\ and\ \citenamefont {Schlichting}(2021{\natexlab{a}})}]{Du:2020dvp}%
  \BibitemOpen
  \bibfield  {author} {\bibinfo {author} {\bibfnamefont {X.}~\bibnamefont {Du}}\ and\ \bibinfo {author} {\bibfnamefont {S.}~\bibnamefont {Schlichting}},\ }\href {\doibase 10.1103/PhysRevD.104.054011} {\bibfield  {journal} {\bibinfo  {journal} {Phys. Rev. D}\ }\textbf {\bibinfo {volume} {104}},\ \bibinfo {pages} {054011} (\bibinfo {year} {2021}{\natexlab{a}})},\ \Eprint {http://arxiv.org/abs/2012.09079} {arXiv:2012.09079 [hep-ph]} \BibitemShut {NoStop}%
\bibitem [{\citenamefont {Du}\ and\ \citenamefont {Schlichting}(2021{\natexlab{b}})}]{Du:2020zqg}%
  \BibitemOpen
  \bibfield  {author} {\bibinfo {author} {\bibfnamefont {X.}~\bibnamefont {Du}}\ and\ \bibinfo {author} {\bibfnamefont {S.}~\bibnamefont {Schlichting}},\ }\href {\doibase 10.1103/PhysRevLett.127.122301} {\bibfield  {journal} {\bibinfo  {journal} {Phys. Rev. Lett.}\ }\textbf {\bibinfo {volume} {127}},\ \bibinfo {pages} {122301} (\bibinfo {year} {2021}{\natexlab{b}})},\ \Eprint {http://arxiv.org/abs/2012.09068} {arXiv:2012.09068 [hep-ph]} \BibitemShut {NoStop}%
\bibitem [{\citenamefont {Dore}\ \emph {et~al.}(2025)\citenamefont {Dore}, \citenamefont {Du}, \citenamefont {Schlichting},\ and\ \citenamefont {Jie}}]{Dore:BI2025}%
  \BibitemOpen
  \bibfield  {author} {\bibinfo {author} {\bibfnamefont {T.}~\bibnamefont {Dore}}, \bibinfo {author} {\bibfnamefont {X.}~\bibnamefont {Du}}, \bibinfo {author} {\bibfnamefont {S.}~\bibnamefont {Schlichting}}, \ and\ \bibinfo {author} {\bibfnamefont {Z.}~\bibnamefont {Jie}},\ }\href@noop {} {\  (\bibinfo {year} {2025})},\ \Eprint {http://arxiv.org/abs/In preparation} {In preparation} \BibitemShut {NoStop}%
\bibitem [{\citenamefont {Du}\ \emph {et~al.}(2023)\citenamefont {Du}, \citenamefont {Ochsenfeld},\ and\ \citenamefont {Schlichting}}]{Du:2023bwi}%
  \BibitemOpen
  \bibfield  {author} {\bibinfo {author} {\bibfnamefont {X.}~\bibnamefont {Du}}, \bibinfo {author} {\bibfnamefont {S.}~\bibnamefont {Ochsenfeld}}, \ and\ \bibinfo {author} {\bibfnamefont {S.}~\bibnamefont {Schlichting}},\ }\href {\doibase 10.1016/j.physletb.2023.138161} {\bibfield  {journal} {\bibinfo  {journal} {Phys. Lett. B}\ }\textbf {\bibinfo {volume} {845}},\ \bibinfo {pages} {138161} (\bibinfo {year} {2023})},\ \Eprint {http://arxiv.org/abs/2306.09094} {arXiv:2306.09094 [hep-ph]} \BibitemShut {NoStop}%
\bibitem [{\citenamefont {Pang}\ \emph {et~al.}(2012)\citenamefont {Pang}, \citenamefont {Wang},\ and\ \citenamefont {Wang}}]{Pang:2012he}%
  \BibitemOpen
  \bibfield  {author} {\bibinfo {author} {\bibfnamefont {L.}~\bibnamefont {Pang}}, \bibinfo {author} {\bibfnamefont {Q.}~\bibnamefont {Wang}}, \ and\ \bibinfo {author} {\bibfnamefont {X.-N.}\ \bibnamefont {Wang}},\ }\href {\doibase 10.1103/PhysRevC.86.024911} {\bibfield  {journal} {\bibinfo  {journal} {Phys. Rev.}\ }\textbf {\bibinfo {volume} {C86}},\ \bibinfo {pages} {024911} (\bibinfo {year} {2012})},\ \Eprint {http://arxiv.org/abs/1205.5019} {arXiv:1205.5019 [nucl-th]} \BibitemShut {NoStop}%
%%CITATION = ARXIV:1205.5019;%%
\bibitem [{\citenamefont {Pang}\ \emph {et~al.}(2018)\citenamefont {Pang}, \citenamefont {Petersen},\ and\ \citenamefont {Wang}}]{Pang:2018zzo}%
  \BibitemOpen
  \bibfield  {author} {\bibinfo {author} {\bibfnamefont {L.-G.}\ \bibnamefont {Pang}}, \bibinfo {author} {\bibfnamefont {H.}~\bibnamefont {Petersen}}, \ and\ \bibinfo {author} {\bibfnamefont {X.-N.}\ \bibnamefont {Wang}},\ }\href {\doibase 10.1103/PhysRevC.97.064918} {\bibfield  {journal} {\bibinfo  {journal} {Phys. Rev. C}\ }\textbf {\bibinfo {volume} {97}},\ \bibinfo {pages} {064918} (\bibinfo {year} {2018})},\ \Eprint {http://arxiv.org/abs/1802.04449} {arXiv:1802.04449 [nucl-th]} \BibitemShut {NoStop}%
\bibitem [{\citenamefont {Wu}\ \emph {et~al.}(2022)\citenamefont {Wu}, \citenamefont {Qin}, \citenamefont {Pang},\ and\ \citenamefont {Wang}}]{Wu:2021fjf}%
  \BibitemOpen
  \bibfield  {author} {\bibinfo {author} {\bibfnamefont {X.-Y.}\ \bibnamefont {Wu}}, \bibinfo {author} {\bibfnamefont {G.-Y.}\ \bibnamefont {Qin}}, \bibinfo {author} {\bibfnamefont {L.-G.}\ \bibnamefont {Pang}}, \ and\ \bibinfo {author} {\bibfnamefont {X.-N.}\ \bibnamefont {Wang}},\ }\href {\doibase 10.1103/PhysRevC.105.034909} {\bibfield  {journal} {\bibinfo  {journal} {Phys. Rev. C}\ }\textbf {\bibinfo {volume} {105}},\ \bibinfo {pages} {034909} (\bibinfo {year} {2022})},\ \Eprint {http://arxiv.org/abs/2107.04949} {arXiv:2107.04949 [hep-ph]} \BibitemShut {NoStop}%
\bibitem [{\citenamefont {Garcia-Montero}\ \emph {et~al.}(2025{\natexlab{b}})\citenamefont {Garcia-Montero}, \citenamefont {Schlichting},\ and\ \citenamefont {Zhu}}]{Garcia-Montero:2025bpn}%
  \BibitemOpen
  \bibfield  {author} {\bibinfo {author} {\bibfnamefont {O.}~\bibnamefont {Garcia-Montero}}, \bibinfo {author} {\bibfnamefont {S.}~\bibnamefont {Schlichting}}, \ and\ \bibinfo {author} {\bibfnamefont {J.}~\bibnamefont {Zhu}},\ }\href@noop {} {\  (\bibinfo {year} {2025}{\natexlab{b}})},\ \Eprint {http://arxiv.org/abs/2501.14872} {arXiv:2501.14872 [nucl-th]} \BibitemShut {NoStop}%
\bibitem [{\citenamefont {Garcia-Montero}\ \emph {et~al.}(2024{\natexlab{b}})\citenamefont {Garcia-Montero}, \citenamefont {Elfner},\ and\ \citenamefont {Schlichting}}]{Garcia-Montero:2023gex}%
  \BibitemOpen
  \bibfield  {author} {\bibinfo {author} {\bibfnamefont {O.}~\bibnamefont {Garcia-Montero}}, \bibinfo {author} {\bibfnamefont {H.}~\bibnamefont {Elfner}}, \ and\ \bibinfo {author} {\bibfnamefont {S.}~\bibnamefont {Schlichting}},\ }\href {\doibase 10.1103/PhysRevC.109.044916} {\bibfield  {journal} {\bibinfo  {journal} {Phys. Rev. C}\ }\textbf {\bibinfo {volume} {109}},\ \bibinfo {pages} {044916} (\bibinfo {year} {2024}{\natexlab{b}})},\ \Eprint {http://arxiv.org/abs/2308.11713} {arXiv:2308.11713 [hep-ph]} \BibitemShut {NoStop}%
\bibitem [{\citenamefont {Blaizot}\ \emph {et~al.}(2004)\citenamefont {Blaizot}, \citenamefont {Gelis},\ and\ \citenamefont {Venugopalan}}]{Blaizot:2004wv}%
  \BibitemOpen
  \bibfield  {author} {\bibinfo {author} {\bibfnamefont {J.~P.}\ \bibnamefont {Blaizot}}, \bibinfo {author} {\bibfnamefont {F.}~\bibnamefont {Gelis}}, \ and\ \bibinfo {author} {\bibfnamefont {R.}~\bibnamefont {Venugopalan}},\ }\href {\doibase 10.1016/j.nuclphysa.2004.07.006} {\bibfield  {journal} {\bibinfo  {journal} {Nucl. Phys. A}\ }\textbf {\bibinfo {volume} {743}},\ \bibinfo {pages} {57} (\bibinfo {year} {2004})},\ \Eprint {http://arxiv.org/abs/hep-ph/0402257} {arXiv:hep-ph/0402257} \BibitemShut {NoStop}%
\bibitem [{\citenamefont {Gelis}\ and\ \citenamefont {Peshier}(2002{\natexlab{a}})}]{Gelis:2001da}%
  \BibitemOpen
  \bibfield  {author} {\bibinfo {author} {\bibfnamefont {F.}~\bibnamefont {Gelis}}\ and\ \bibinfo {author} {\bibfnamefont {A.}~\bibnamefont {Peshier}},\ }\href {\doibase 10.1016/S0375-9474(01)01264-7} {\bibfield  {journal} {\bibinfo  {journal} {Nucl. Phys. A}\ }\textbf {\bibinfo {volume} {697}},\ \bibinfo {pages} {879} (\bibinfo {year} {2002}{\natexlab{a}})},\ \Eprint {http://arxiv.org/abs/hep-ph/0107142} {arXiv:hep-ph/0107142} \BibitemShut {NoStop}%
\bibitem [{\citenamefont {Gelis}\ and\ \citenamefont {Peshier}(2002{\natexlab{b}})}]{Gelis:2001dh}%
  \BibitemOpen
  \bibfield  {author} {\bibinfo {author} {\bibfnamefont {F.}~\bibnamefont {Gelis}}\ and\ \bibinfo {author} {\bibfnamefont {A.}~\bibnamefont {Peshier}},\ }\href {\doibase 10.1016/S0375-9474(02)00752-2} {\bibfield  {journal} {\bibinfo  {journal} {Nucl. Phys. A}\ }\textbf {\bibinfo {volume} {707}},\ \bibinfo {pages} {175} (\bibinfo {year} {2002}{\natexlab{b}})},\ \Eprint {http://arxiv.org/abs/hep-ph/0111227} {arXiv:hep-ph/0111227} \BibitemShut {NoStop}%
\bibitem [{\citenamefont {Dumitru}\ and\ \citenamefont {McLerran}(2002)}]{Dumitru:2001ux}%
  \BibitemOpen
  \bibfield  {author} {\bibinfo {author} {\bibfnamefont {A.}~\bibnamefont {Dumitru}}\ and\ \bibinfo {author} {\bibfnamefont {L.~D.}\ \bibnamefont {McLerran}},\ }\href {\doibase 10.1016/S0375-9474(01)01301-X} {\bibfield  {journal} {\bibinfo  {journal} {Nucl. Phys. A}\ }\textbf {\bibinfo {volume} {700}},\ \bibinfo {pages} {492} (\bibinfo {year} {2002})},\ \Eprint {http://arxiv.org/abs/hep-ph/0105268} {arXiv:hep-ph/0105268} \BibitemShut {NoStop}%
\bibitem [{\citenamefont {Lappi}\ and\ \citenamefont {Schlichting}(2018)}]{Lappi:2017skr}%
  \BibitemOpen
  \bibfield  {author} {\bibinfo {author} {\bibfnamefont {T.}~\bibnamefont {Lappi}}\ and\ \bibinfo {author} {\bibfnamefont {S.}~\bibnamefont {Schlichting}},\ }\href {\doibase 10.1103/PhysRevD.97.034034} {\bibfield  {journal} {\bibinfo  {journal} {Phys. Rev. D}\ }\textbf {\bibinfo {volume} {97}},\ \bibinfo {pages} {034034} (\bibinfo {year} {2018})},\ \Eprint {http://arxiv.org/abs/1708.08625} {arXiv:1708.08625 [hep-ph]} \BibitemShut {NoStop}%
\bibitem [{\citenamefont {Dumitru}\ and\ \citenamefont {Jalilian-Marian}(2002)}]{Dumitru:2002qt}%
  \BibitemOpen
  \bibfield  {author} {\bibinfo {author} {\bibfnamefont {A.}~\bibnamefont {Dumitru}}\ and\ \bibinfo {author} {\bibfnamefont {J.}~\bibnamefont {Jalilian-Marian}},\ }\href {\doibase 10.1103/PhysRevLett.89.022301} {\bibfield  {journal} {\bibinfo  {journal} {Phys. Rev. Lett.}\ }\textbf {\bibinfo {volume} {89}},\ \bibinfo {pages} {022301} (\bibinfo {year} {2002})},\ \Eprint {http://arxiv.org/abs/hep-ph/0204028} {arXiv:hep-ph/0204028} \BibitemShut {NoStop}%
\bibitem [{\citenamefont {Dumitru}\ \emph {et~al.}(2006)\citenamefont {Dumitru}, \citenamefont {Hayashigaki},\ and\ \citenamefont {Jalilian-Marian}}]{Dumitru:2005gt}%
  \BibitemOpen
  \bibfield  {author} {\bibinfo {author} {\bibfnamefont {A.}~\bibnamefont {Dumitru}}, \bibinfo {author} {\bibfnamefont {A.}~\bibnamefont {Hayashigaki}}, \ and\ \bibinfo {author} {\bibfnamefont {J.}~\bibnamefont {Jalilian-Marian}},\ }\href {\doibase 10.1016/j.nuclphysa.2005.11.014} {\bibfield  {journal} {\bibinfo  {journal} {Nucl. Phys. A}\ }\textbf {\bibinfo {volume} {765}},\ \bibinfo {pages} {464} (\bibinfo {year} {2006})},\ \Eprint {http://arxiv.org/abs/hep-ph/0506308} {arXiv:hep-ph/0506308} \BibitemShut {NoStop}%
\bibitem [{\citenamefont {Kowalski}\ \emph {et~al.}(2006)\citenamefont {Kowalski}, \citenamefont {Motyka},\ and\ \citenamefont {Watt}}]{Kowalski:2006hc}%
  \BibitemOpen
  \bibfield  {author} {\bibinfo {author} {\bibfnamefont {H.}~\bibnamefont {Kowalski}}, \bibinfo {author} {\bibfnamefont {L.}~\bibnamefont {Motyka}}, \ and\ \bibinfo {author} {\bibfnamefont {G.}~\bibnamefont {Watt}},\ }\href {\doibase 10.1103/PhysRevD.74.074016} {\bibfield  {journal} {\bibinfo  {journal} {Phys. Rev. D}\ }\textbf {\bibinfo {volume} {74}},\ \bibinfo {pages} {074016} (\bibinfo {year} {2006})},\ \Eprint {http://arxiv.org/abs/hep-ph/0606272} {arXiv:hep-ph/0606272} \BibitemShut {NoStop}%
\bibitem [{\citenamefont {Kowalski}\ and\ \citenamefont {Teaney}(2003)}]{Kowalski:2003hm}%
  \BibitemOpen
  \bibfield  {author} {\bibinfo {author} {\bibfnamefont {H.}~\bibnamefont {Kowalski}}\ and\ \bibinfo {author} {\bibfnamefont {D.}~\bibnamefont {Teaney}},\ }\href {\doibase 10.1103/PhysRevD.68.114005} {\bibfield  {journal} {\bibinfo  {journal} {Phys. Rev. D}\ }\textbf {\bibinfo {volume} {68}},\ \bibinfo {pages} {114005} (\bibinfo {year} {2003})},\ \Eprint {http://arxiv.org/abs/hep-ph/0304189} {arXiv:hep-ph/0304189} \BibitemShut {NoStop}%
\bibitem [{\citenamefont {Denicol}\ \emph {et~al.}(2018)\citenamefont {Denicol}, \citenamefont {Gale}, \citenamefont {Jeon}, \citenamefont {Monnai}, \citenamefont {Schenke},\ and\ \citenamefont {Shen}}]{Denicol:2018wdp}%
  \BibitemOpen
  \bibfield  {author} {\bibinfo {author} {\bibfnamefont {G.~S.}\ \bibnamefont {Denicol}}, \bibinfo {author} {\bibfnamefont {C.}~\bibnamefont {Gale}}, \bibinfo {author} {\bibfnamefont {S.}~\bibnamefont {Jeon}}, \bibinfo {author} {\bibfnamefont {A.}~\bibnamefont {Monnai}}, \bibinfo {author} {\bibfnamefont {B.}~\bibnamefont {Schenke}}, \ and\ \bibinfo {author} {\bibfnamefont {C.}~\bibnamefont {Shen}},\ }\href {\doibase 10.1103/PhysRevC.98.034916} {\bibfield  {journal} {\bibinfo  {journal} {Phys. Rev. C}\ }\textbf {\bibinfo {volume} {98}},\ \bibinfo {pages} {034916} (\bibinfo {year} {2018})},\ \Eprint {http://arxiv.org/abs/1804.10557} {arXiv:1804.10557 [nucl-th]} \BibitemShut {NoStop}%
\bibitem [{\citenamefont {Denicol}\ \emph {et~al.}(2014)\citenamefont {Denicol}, \citenamefont {Jeon},\ and\ \citenamefont {Gale}}]{Denicol:2014vaa}%
  \BibitemOpen
  \bibfield  {author} {\bibinfo {author} {\bibfnamefont {G.~S.}\ \bibnamefont {Denicol}}, \bibinfo {author} {\bibfnamefont {S.}~\bibnamefont {Jeon}}, \ and\ \bibinfo {author} {\bibfnamefont {C.}~\bibnamefont {Gale}},\ }\href {\doibase 10.1103/PhysRevC.90.024912} {\bibfield  {journal} {\bibinfo  {journal} {Phys. Rev. C}\ }\textbf {\bibinfo {volume} {90}},\ \bibinfo {pages} {024912} (\bibinfo {year} {2014})},\ \Eprint {http://arxiv.org/abs/1403.0962} {arXiv:1403.0962 [nucl-th]} \BibitemShut {NoStop}%
\bibitem [{\citenamefont {Bazavov}\ \emph {et~al.}(2014)\citenamefont {Bazavov} \emph {et~al.}}]{HotQCD:2014kol}%
  \BibitemOpen
  \bibfield  {author} {\bibinfo {author} {\bibfnamefont {A.}~\bibnamefont {Bazavov}} \emph {et~al.} (\bibinfo {collaboration} {HotQCD}),\ }\href {\doibase 10.1103/PhysRevD.90.094503} {\bibfield  {journal} {\bibinfo  {journal} {Phys. Rev. D}\ }\textbf {\bibinfo {volume} {90}},\ \bibinfo {pages} {094503} (\bibinfo {year} {2014})},\ \Eprint {http://arxiv.org/abs/1407.6387} {arXiv:1407.6387 [hep-lat]} \BibitemShut {NoStop}%
\bibitem [{\citenamefont {Borsanyi}\ \emph {et~al.}(2016)\citenamefont {Borsanyi} \emph {et~al.}}]{Borsanyi:2016ksw}%
  \BibitemOpen
  \bibfield  {author} {\bibinfo {author} {\bibfnamefont {S.}~\bibnamefont {Borsanyi}} \emph {et~al.},\ }\href {\doibase 10.1038/nature20115} {\bibfield  {journal} {\bibinfo  {journal} {Nature}\ }\textbf {\bibinfo {volume} {539}},\ \bibinfo {pages} {69} (\bibinfo {year} {2016})},\ \Eprint {http://arxiv.org/abs/1606.07494} {arXiv:1606.07494 [hep-lat]} \BibitemShut {NoStop}%
\bibitem [{\citenamefont {Huovinen}\ and\ \citenamefont {Petersen}(2012)}]{Huovinen:2012is}%
  \BibitemOpen
  \bibfield  {author} {\bibinfo {author} {\bibfnamefont {P.}~\bibnamefont {Huovinen}}\ and\ \bibinfo {author} {\bibfnamefont {H.}~\bibnamefont {Petersen}},\ }\href {\doibase 10.1140/epja/i2012-12171-9} {\bibfield  {journal} {\bibinfo  {journal} {Eur. Phys. J. A}\ }\textbf {\bibinfo {volume} {48}},\ \bibinfo {pages} {171} (\bibinfo {year} {2012})},\ \Eprint {http://arxiv.org/abs/1206.3371} {arXiv:1206.3371 [nucl-th]} \BibitemShut {NoStop}%
\bibitem [{\citenamefont {Borghini}\ \emph {et~al.}(2024)\citenamefont {Borghini}, \citenamefont {Krupczak},\ and\ \citenamefont {Roch}}]{Borghini:2024kll}%
  \BibitemOpen
  \bibfield  {author} {\bibinfo {author} {\bibfnamefont {N.}~\bibnamefont {Borghini}}, \bibinfo {author} {\bibfnamefont {R.}~\bibnamefont {Krupczak}}, \ and\ \bibinfo {author} {\bibfnamefont {H.}~\bibnamefont {Roch}},\ }\href {\doibase 10.1140/epjc/s10052-024-13509-8} {\bibfield  {journal} {\bibinfo  {journal} {Eur. Phys. J. C}\ }\textbf {\bibinfo {volume} {84}},\ \bibinfo {pages} {1128} (\bibinfo {year} {2024})},\ \Eprint {http://arxiv.org/abs/2407.10634} {arXiv:2407.10634 [nucl-th]} \BibitemShut {NoStop}%
\bibitem [{\citenamefont {da~Silva}\ \emph {et~al.}(2023)\citenamefont {da~Silva}, \citenamefont {Chinellato}, \citenamefont {Giannini}, \citenamefont {Ferreira}, \citenamefont {Denicol}, \citenamefont {Hippert}, \citenamefont {Luzum}, \citenamefont {Noronha},\ and\ \citenamefont {Takahashi}}]{daSilva:2022xwu}%
  \BibitemOpen
  \bibfield  {author} {\bibinfo {author} {\bibfnamefont {T.~N.}\ \bibnamefont {da~Silva}}, \bibinfo {author} {\bibfnamefont {D.~D.}\ \bibnamefont {Chinellato}}, \bibinfo {author} {\bibfnamefont {A.~V.}\ \bibnamefont {Giannini}}, \bibinfo {author} {\bibfnamefont {M.~N.}\ \bibnamefont {Ferreira}}, \bibinfo {author} {\bibfnamefont {G.~S.}\ \bibnamefont {Denicol}}, \bibinfo {author} {\bibfnamefont {M.}~\bibnamefont {Hippert}}, \bibinfo {author} {\bibfnamefont {M.}~\bibnamefont {Luzum}}, \bibinfo {author} {\bibfnamefont {J.}~\bibnamefont {Noronha}}, \ and\ \bibinfo {author} {\bibfnamefont {J.}~\bibnamefont {Takahashi}} (\bibinfo {collaboration} {ExTrEMe}),\ }\href {\doibase 10.1103/PhysRevC.107.044901} {\bibfield  {journal} {\bibinfo  {journal} {Phys. Rev. C}\ }\textbf {\bibinfo {volume} {107}},\ \bibinfo {pages} {044901} (\bibinfo {year} {2023})},\ \Eprint {http://arxiv.org/abs/2211.10561} {arXiv:2211.10561 [nucl-th]} \BibitemShut {NoStop}%
\bibitem [{\citenamefont {Romatschke}(2018)}]{Romatschke:2017vte}%
  \BibitemOpen
  \bibfield  {author} {\bibinfo {author} {\bibfnamefont {P.}~\bibnamefont {Romatschke}},\ }\href {\doibase 10.1103/PhysRevLett.120.012301} {\bibfield  {journal} {\bibinfo  {journal} {Phys. Rev. Lett.}\ }\textbf {\bibinfo {volume} {120}},\ \bibinfo {pages} {012301} (\bibinfo {year} {2018})},\ \Eprint {http://arxiv.org/abs/1704.08699} {arXiv:1704.08699 [hep-th]} \BibitemShut {NoStop}%
\bibitem [{\citenamefont {Strickland}(2018)}]{Strickland:2018ayk}%
  \BibitemOpen
  \bibfield  {author} {\bibinfo {author} {\bibfnamefont {M.}~\bibnamefont {Strickland}},\ }\href {\doibase 10.1007/JHEP12(2018)128} {\bibfield  {journal} {\bibinfo  {journal} {JHEP}\ }\textbf {\bibinfo {volume} {12}},\ \bibinfo {pages} {128} (\bibinfo {year} {2018})},\ \Eprint {http://arxiv.org/abs/1809.01200} {arXiv:1809.01200 [nucl-th]} \BibitemShut {NoStop}%
\bibitem [{\citenamefont {Chattopadhyay}\ \emph {et~al.}(2022)\citenamefont {Chattopadhyay}, \citenamefont {Jaiswal}, \citenamefont {Du}, \citenamefont {Heinz},\ and\ \citenamefont {Pal}}]{Chattopadhyay:2021ive}%
  \BibitemOpen
  \bibfield  {author} {\bibinfo {author} {\bibfnamefont {C.}~\bibnamefont {Chattopadhyay}}, \bibinfo {author} {\bibfnamefont {S.}~\bibnamefont {Jaiswal}}, \bibinfo {author} {\bibfnamefont {L.}~\bibnamefont {Du}}, \bibinfo {author} {\bibfnamefont {U.}~\bibnamefont {Heinz}}, \ and\ \bibinfo {author} {\bibfnamefont {S.}~\bibnamefont {Pal}},\ }\href {\doibase 10.1016/j.physletb.2021.136820} {\bibfield  {journal} {\bibinfo  {journal} {Phys. Lett. B}\ }\textbf {\bibinfo {volume} {824}},\ \bibinfo {pages} {136820} (\bibinfo {year} {2022})},\ \Eprint {http://arxiv.org/abs/2107.05500} {arXiv:2107.05500 [nucl-th]} \BibitemShut {NoStop}%
\end{thebibliography}%

\appendix
\addcontentsline{toc}{section}{Appendix}

\section{Energy response functions in free-streaming limit}\label{free-streaming}
In this section, we will give the analytical results of 3D response functions under free-streaming limit. In this case, the Boltzmann equations are
\begin{align}
\partial_\tau f_{B G}\left(\tau, p_T,p_\eta\right)&=0,\\
\tau \partial_\tau \delta f_k(\tau, p_T, p_\eta)  &=\delta f_k(\tau, p_T, p_\eta)\nonumber\\
&\quad \left(\frac{i}{\tau p^\tau} k_\eta p_\eta-\frac{i \tau }{p^\tau}\mathbf{k}_{\mathbf{T}} \cdot \mathbf{p}_{\mathbf{T}}\right).
\end{align}
The solutions to them can be easily written down as
\begin{align}
    f_{BG}(\tau,p_T,p_\eta)&=f_{BG}(\tau_0,p_T,p_\eta),\\
    \delta f_k(\tau, p_T, p_\eta)&=\delta f_k(\tau_0, p_T, p_\eta) e^{-i k_\eta \left(
    \mathrm{arcsch} \frac{p_T \tau}{p_\eta} - \mathrm{arcsch} \frac{p_T \tau_0}{p_\eta} \right)}\nonumber\\
    &\qquad e^{-i\frac{\mathbf{p_T}\cdot \mathbf{k_T}}{p^2_T}
    \left(\sqrt{p_T^2\tau^2+p_\eta^2}-\sqrt{p_T^2 \tau_0^2+p_\eta^2}\right)}.
\end{align}
Following in the usual assumption in Ref.~\cite{Chattopadhyay:2021ive}, the initial distribution is taken as
\begin{align}
&f_{BG}(\tau_0)=\frac{2 A}{\lambda} \frac{Q_0}{|\mathbf{p}|} \frac{e^{-\frac{2}{3} \frac{\mathbf{p}^2}{Q_0^2}\left[1+\left(\xi^2-1\right) \cos ^2(\theta)\right]}}{\sqrt{1+\left(\xi^2-1\right) \cos ^2(\theta)}},\\
&\delta f_k(\tau_0)=-\frac{p}{3}[\partial_p f_{BG}(\tau_0)].
\end{align}
where $\cos\theta = p_\eta/(\tau_0 p)$ and $p^2=p_T^2+p_\eta^2/\tau_0^2$.

\begin{figure}[b]
    \centering
    \includegraphics[width=0.8\linewidth]{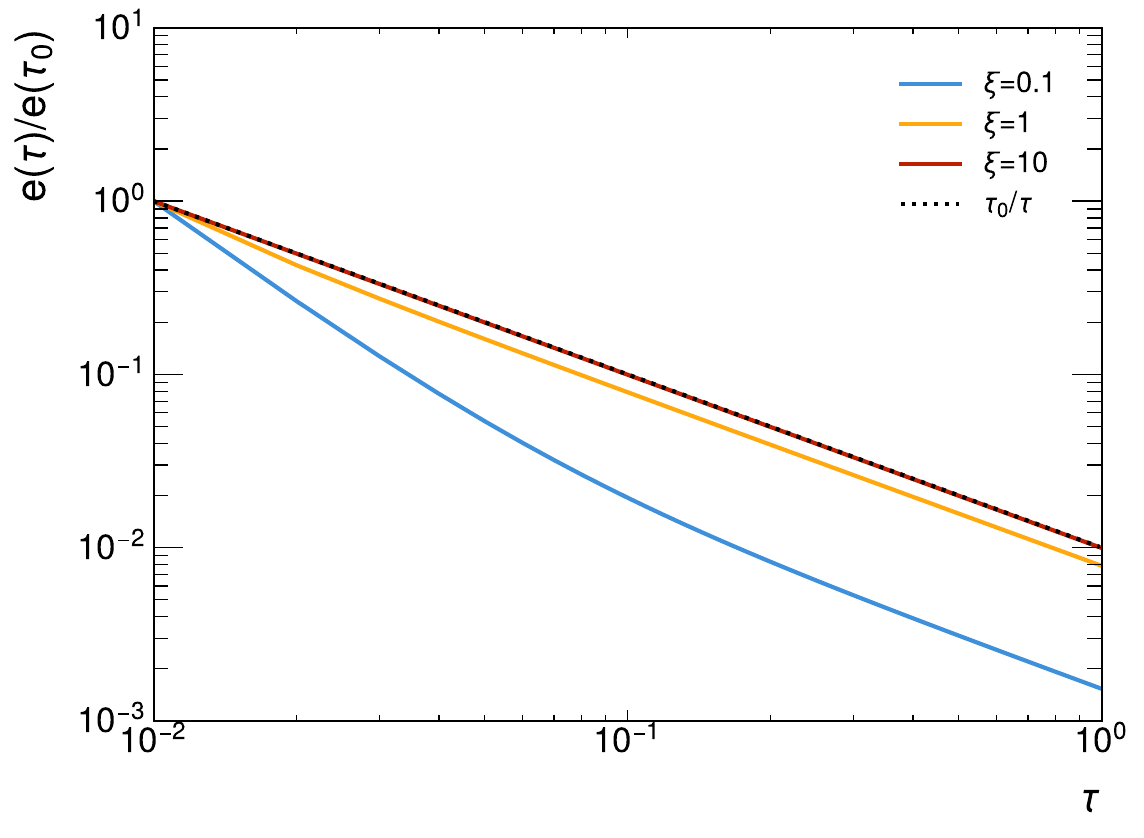}
    \caption{Evolution of the background energy density in the free-streaming limit.}
    \label{fig:fs-bg}
\end{figure}

Hence, the background ($\TBg^{\tau\tau}(\tau)$) evolution in free-streaming limit is
\begin{align}
    &\TBg^{\tau\tau}(\tau)=\frac{\nu_{eff}}{\tau}
    \int\frac{dp_\eta d^2p_T}{(2\pi)^3} p \cdot f_{BG}(\tau)\nonumber\\
    &=\frac{\nu_{eff}}{\tau}  \int\frac{dp_\eta d^2p_T}{(2\pi)^3} p \cdot 
    \left[\frac{2 A}{\lambda} \frac{Q_0}{|\mathbf{p}|} \frac{e^{-\frac{2}{3} \frac{\mathbf{p}^2}{Q_0^2}\left[1+\left(\xi^2-1\right) \cos ^2(\theta)\right]}}{\sqrt{1+\left(\xi^2-1\right) \cos ^2(\theta)}}\right]_{\tau=\tau_0}\nonumber\\
    &\propto     \int_{-1}^1 
    \frac{dx}{(1-x^2 + \xi^2 x^2 \tau^2/\tau_0^2)^2}
    %\equiv G(\xi,\tau,\tau_0)
    =\frac{1}{1+a}+\frac{\arctan \sqrt{a}}{\sqrt{a}},
\end{align}
where $x\equiv \cos\theta$ and $a\equiv\xi^2 \tau^2 / \tau_0^2 -1$.
We plot the background evolution in free-streaming limit in Fig.~\ref{fig:fs-bg}.
It is found that a highly anisotropic initial distribution (large $\xi$) tend to approach Bjorken solution $e(\tau) \sim 1/\tau$.
If we further assume $\tau/\tau_0 \gg 1$, we can obtain $\bar{T}^{\tau\tau}(\tau)\propto
\left(\frac{\tau_0}{\tau}\right)^2 \frac{1}{\xi^2}+\frac{\tau_0}{\tau} \frac{\pi}{2\xi}$, 
and $\TBg^{\tau\tau}(\tau_0)\propto \frac{1}{\xi^2}+\frac{\arctan \sqrt{\xi^2-1}}{\sqrt{\xi^2-1}}$.

Based on the definition of the perturbation,
\begin{align}
\delta T_{k}^{\mu \nu}(\tau)=\nu_{eff} \int \frac{d^3 \mathbf{p}}{(2 \pi)^3} \frac{p^\mu p^\nu}{p} \delta f_{k}(\tau, p_T, p_\eta),    
\end{align}
we express various responses to initial energy perturbation
$\delta T_k^{\tau\tau} (\tau_0) \propto
2\pi*(\frac{1}{\xi^2}+\frac{\arctan \sqrt{\xi^2-1}}{\sqrt{\xi^2-1}})$
as follows.
{\footnotesize
\begin{equation}
    \begin{aligned}
    \delta T_k^{\tau\tau} 
    &\propto
    {\frac{\tau_0}{\tau} 2\pi J_0\left(|\mathbf{k_T}|\tau\right)
        (1+k_\eta/\xi) e^{-k_\eta/\xi}},
    \\
    \delta T_k^{\tau i} i\frac{\mathbf{k_T}^i}{|\mathbf{k_T}|}
    &\propto
    {\frac{\tau_0}{\tau} 2\pi J_1\left(|\mathbf{k_T}|\tau\right)
    %\int dy \Re[P(y;k_\eta,\xi)]
    (1+k_\eta/\xi) e^{-k_\eta/\xi}},
    \\
    \delta T_k^{\tau\eta} i\frac{k_\eta}{\left|k_\eta\right|}
    &
    \propto
    {\frac{\tau_0^2}{\tau^3} 2\pi J_0\left(|\mathbf{k_T}|\tau \right)
    %\int dy \Im[P(y;k_\eta,\xi)]\cdot y
    \frac{k_\eta}{\left|k_\eta\right|}}
    \frac{k_\eta}{\xi^2} e^{-k_\eta/\xi},
    \\
    \delta T_k^{ij} (\delta^{i j} - \frac{\mathbf{k_T}^i \mathbf{k_T}^j}{|\mathbf{k_T}|^2})
    &\propto
    {\frac{\tau_0}{\tau} \pi (J_0+J_2)\left[|\mathbf{k_T}|\tau\right]
    %\int dy \Re[P(y;k_\eta,\xi)]
    (1+k_\eta/\xi) e^{-k_\eta/\xi}},
    \\
    \delta T_k^{ij} (2\frac{\mathbf{k_T}^i \mathbf{k_T}^j}{|\mathbf{k_T}|^2} - \delta^{i j})
    &\propto
    {\frac{\tau_0}{\tau} 2\pi (-1) J_2\left[|\mathbf{k_T}|\tau\right]
    (1+k_\eta/\xi) e^{-k_\eta/\xi}},
    \\
    \delta T_k^{i\eta} \frac{k_T^i k_\eta}{\left|\mathbf{k}_{\mathbf{T}}\right|\left|k_\eta\right|}
    &\propto
    {\frac{\tau_0^2}{\tau^3} 2\pi J_1\left(|\mathbf{k_T}|\tau\right) 
    \frac{k_\eta}{\left|k_\eta\right|}}
    \frac{-k_\eta}{\xi^2} e^{-k_\eta/\xi},
    \\
    \delta T_k^{\eta\eta}
    &\propto
    {\frac{\tau_0^3}{\tau^5}2\pi J_0\left[|\mathbf{k_T}|\tau \right]
    \frac{\xi-k_\eta}{\xi^3} e^{-k_\eta/\xi}}.
    \end{aligned}
\end{equation}
}where we assumed $\tau/\tau_0 \gg 1$, $\xi\gg 1$ and a common prefactor of perturbations is omitted here.

Utilizing the definition of response function,
\begin{equation}
    G_{\alpha \beta}^{\mu \nu}\left(\mathbf{k}_{\mathbf{T}}, k_\eta, \tau, \tau_0\right) =
    \frac{\delta T_k^{\mu \nu}(\tau)}{\TBg^{\tau \tau}(\tau)}
    /
    \frac{\delta T_k^{\alpha \beta}\left(\tau_0\right)}{\TBg^{\tau \tau}\left(\tau_0\right)},
\end{equation}
we derived various response functions in Fourier space. 

\begin{equation}
    \begin{aligned}
    & \tilde{G}_s^s(|\mathbf{k_T}|, |k_\eta|)
    \approx J_0\left(|\mathbf{k_T}|\tau\right) (1+k_\eta/\xi) e^{-k_\eta/\xi},
    \\
    &\tilde{G}_s^v(|\mathbf{k_T}|, |k_\eta|)
    \approx J_1\left(|\mathbf{k_T}|\tau\right) (1+k_\eta/\xi) e^{-k_\eta/\xi},
    \\
    &\tilde{G}_s^{s, \eta}\left(|\mathbf{k_T}|, |k_\eta|\right)
    \approx \frac{\tau_0}{\tau^2} J_0\left(|\mathbf{k_T}|\tau \right) 
    \frac{k_\eta}{\xi^2} e^{-k_\eta/\xi},
    \\
    & \tilde{G}_s^{t, \delta}(|\mathbf{k_T}|, |k_\eta|)
    \approx {1 \over 2} (J_0+J_2)\left[|\mathbf{k_T}|\tau\right]
    (1+k_\eta/\xi) e^{-k_\eta/\xi},
    \\
    &\tilde{G}_s^{t, k}(|\mathbf{k_T}|, |k_\eta|)
    \approx (-1) J_2\left[|\mathbf{k_T}|\tau\right] (1+k_\eta/\xi) e^{-k_\eta/\xi},
    \\
    &\tilde{G}_s^{v, \eta}\left(|\mathbf{k_T}|, |k_\eta|\right)
    \approx \frac{\tau_0}{\tau^2} J_1\left(|\mathbf{k_T}|\tau\right) 
    \frac{-k_\eta}{\xi^2} e^{-k_\eta/\xi},
    \\
    &\tilde{G}_s^\eta\left(|\mathbf{k_T}|, |k_\eta|\right)
    \approx \frac{\tau_0^2}{\tau^4} J_0\left[|\mathbf{k_T}|\tau \right] 
    \frac{\xi-k_\eta}{\xi^3} e^{-k_\eta/\xi}.
    \end{aligned}
\end{equation}

The response functions in coordinate space can be also obtained via the reverse Fourier transformation in Eq.~(\ref{eq:rev-fourier}).
    \begin{align}
    &
    G_s^s\left(|\mathbf{r}|, |\eta|\right) =
    G_s^v\left(|\mathbf{r}|, |\eta|\right) =
    G_s^{t,r}\left(|\mathbf{r}|, |\eta|\right) =
    {\delta(|\mathbf{r}|-\tau) \over \pi^2\tau}
    \frac{\xi }{(1+\eta^2 \xi^2)^2}, \nonumber
    \\
    &\tau G_s^{s, \eta}(|\mathbf{r}|, |\eta|)
    =-\tau G_s^{v, \eta}(|\mathbf{r}|, |\eta|)
    =\frac{\tau_0}{\tau} 
    {\delta(|\mathbf{r}|-\tau) \over \pi^2 \tau}
    \frac{|\eta|\xi}{(1+\eta^2\xi^2)^2}, \nonumber
    \\
    &\tau^2 G_s^\eta(|\mathbf{r}|, |\eta|)=
    \frac{\tau_0^2}{\tau^2} 
    {\delta(|\mathbf{r}|-\tau) \over \pi^2 \tau}
    \frac{\eta^2\xi}{(1+\eta^2 \xi^2)^2},
    \\
    &G_s^{t,\delta}\left(|\mathbf{r}|,|\eta|\right)=0. \nonumber
    \end{align}

\section{More images of response functions}\label{more-res}
\subsection{In Fourier space}
\begin{figure*}
    \centering
    \includegraphics[width=0.9\linewidth]{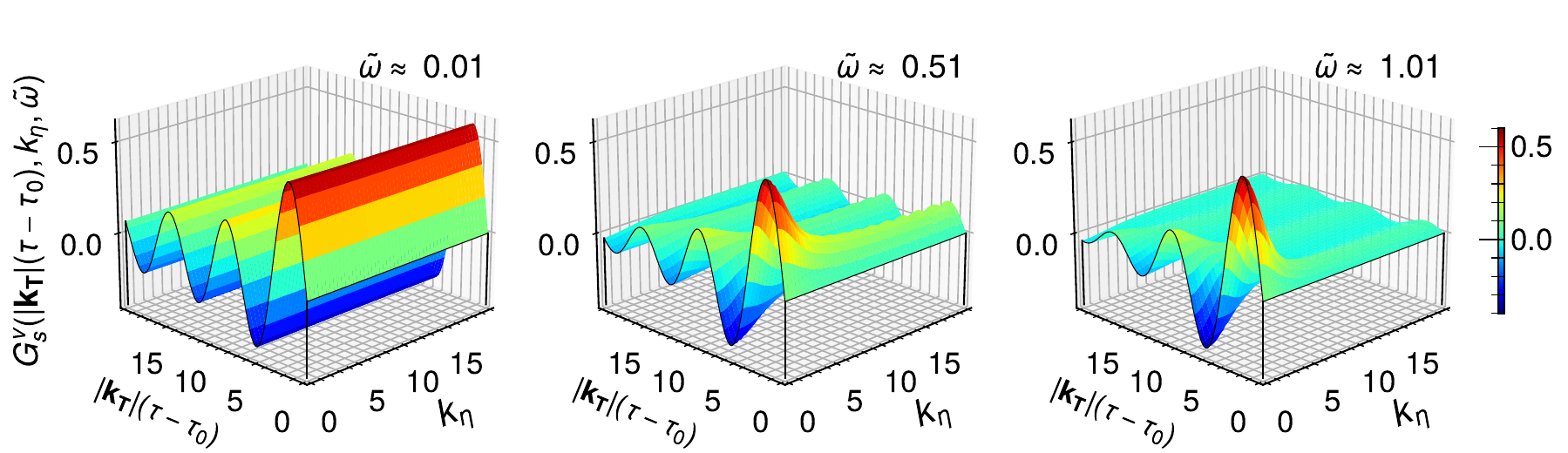}\\
    \includegraphics[width=0.9\linewidth]{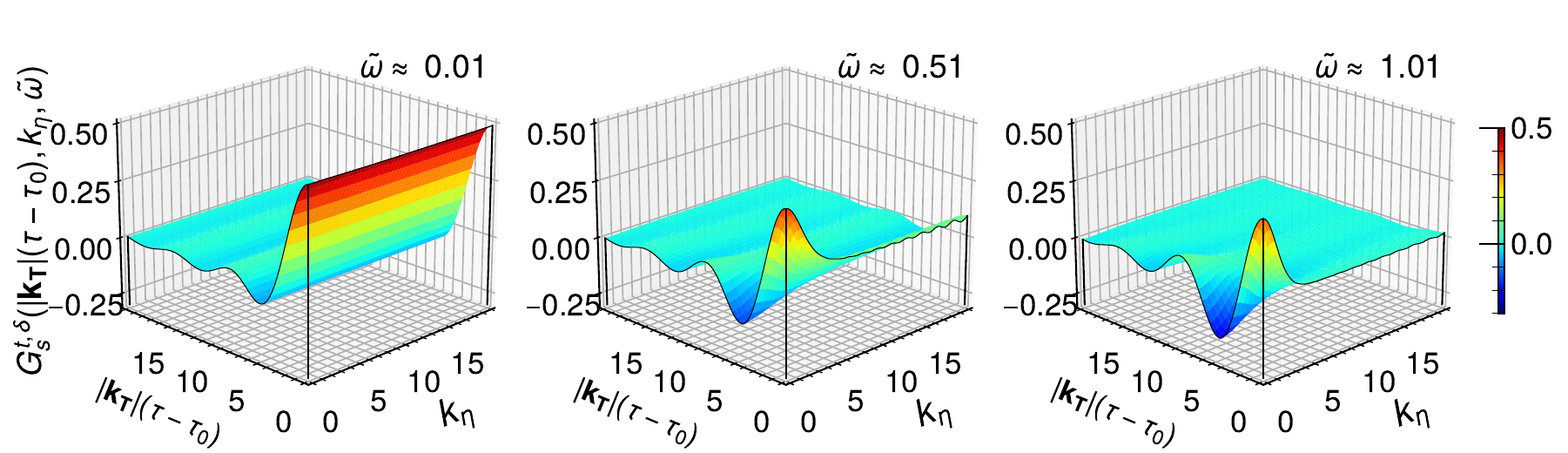}\\
    \includegraphics[width=0.9\linewidth]{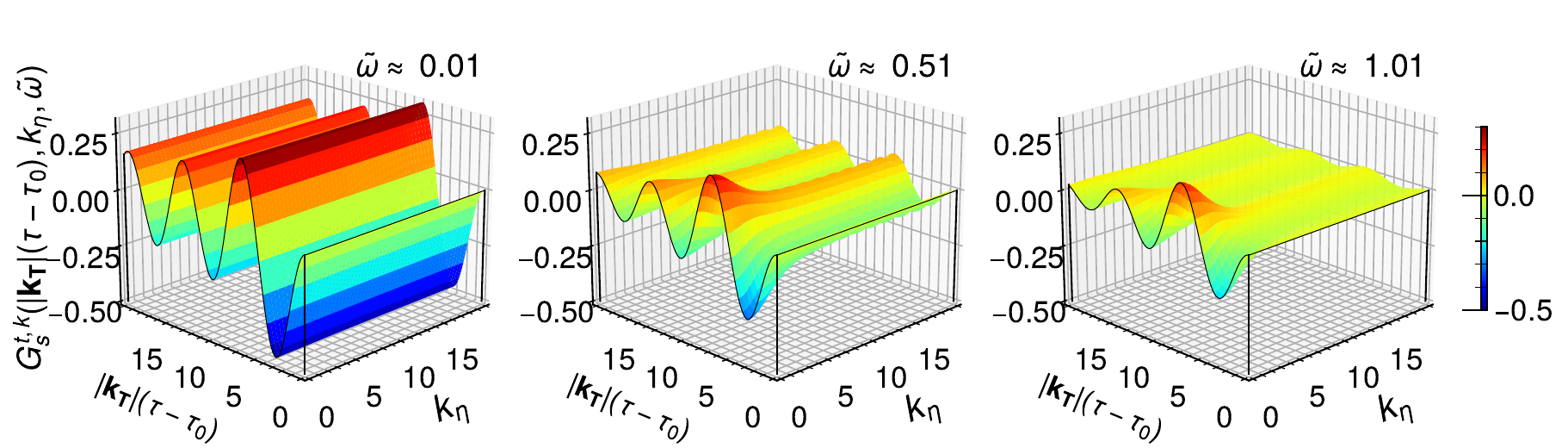}\\
    \includegraphics[width=0.9\linewidth]{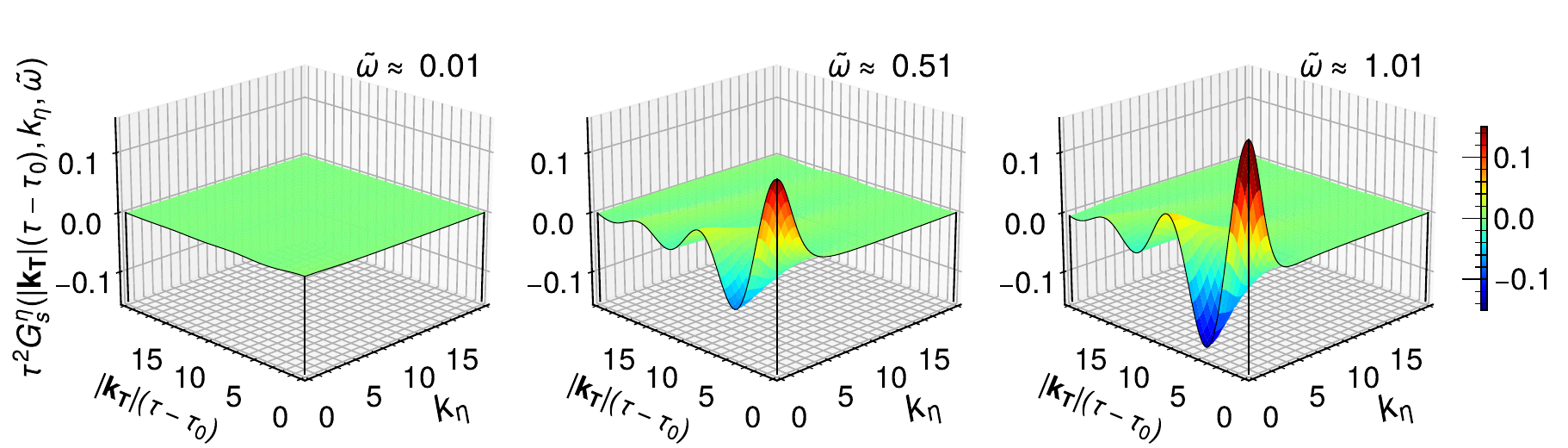}\\
    \includegraphics[width=0.9\linewidth]{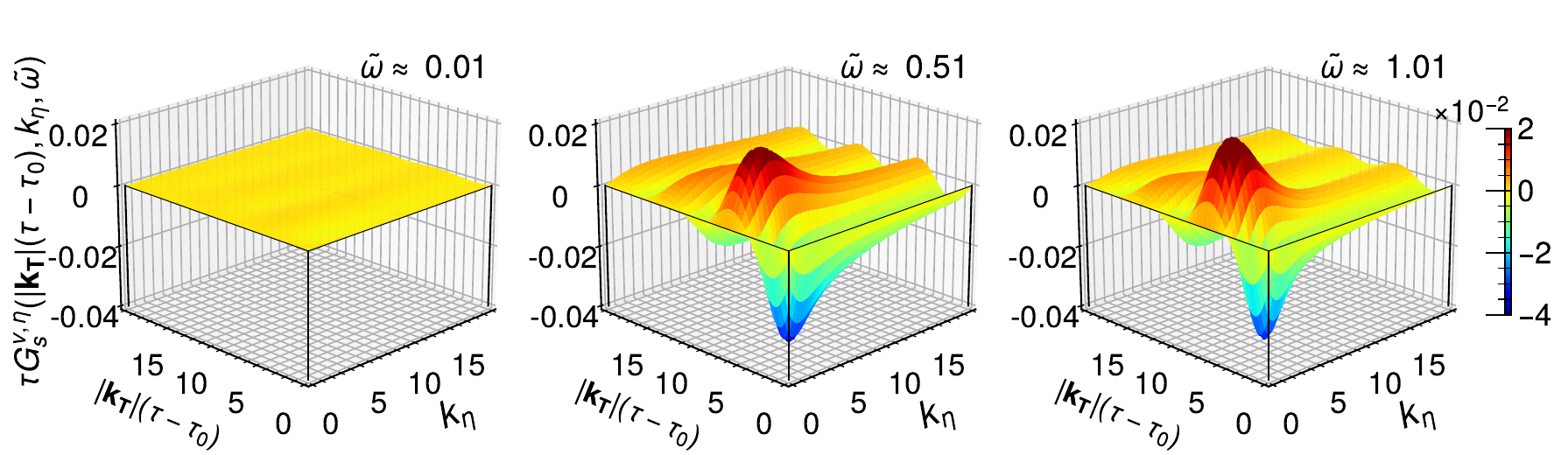}
    \caption{Evolution of transverse momentum response ($G_s^v(k\Delta\tau, k_\eta)$), pressure response ($G_s^{t,\delta}(k\Delta\tau, k_\eta)$ and $G_s^{\eta}(k\Delta\tau, k_\eta)$) and shear stress response ($G_s^{t,k}(k\Delta\tau, k_\eta)$ and $G_s^{v,\eta}(k\Delta\tau, k_\eta)$) to an initial energy perturbation as a function of $k\Delta\tau$ and $k_\eta$. Different panels correspond to different points in scaled time variable. 
    }
    \label{fig:more-Gs-k-RTA}
\end{figure*}

In Fig.~\ref{fig:more-Gs-k-RTA}, we show the evolution of the transverse momentum response ($G_s^v(k\Delta\tau, k_\eta)$), the pressure response ($G_s^{t,\delta}(k\Delta\tau, k_\eta)$ and $G_s^{\eta}(k\Delta\tau, k_\eta)$)
 and shear stress response ($G_s^{t,k}(k\Delta\tau, k_\eta)$ and $G_s^{v,\eta}(k\Delta\tau, k_\eta)$) to an initial energy perturbation as a function of $k\Delta\tau$ and $k_\eta$.

\subsection{In position space}
\begin{figure*}
    \centering
    \includegraphics[width=0.87\linewidth]{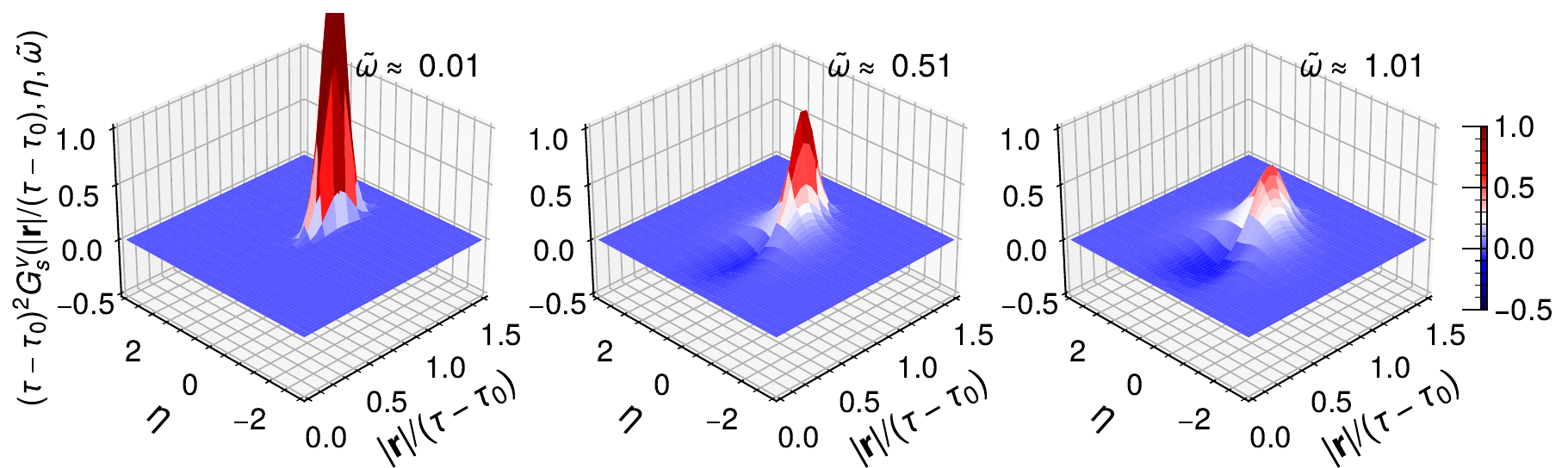}\\
    \includegraphics[width=0.87\linewidth]{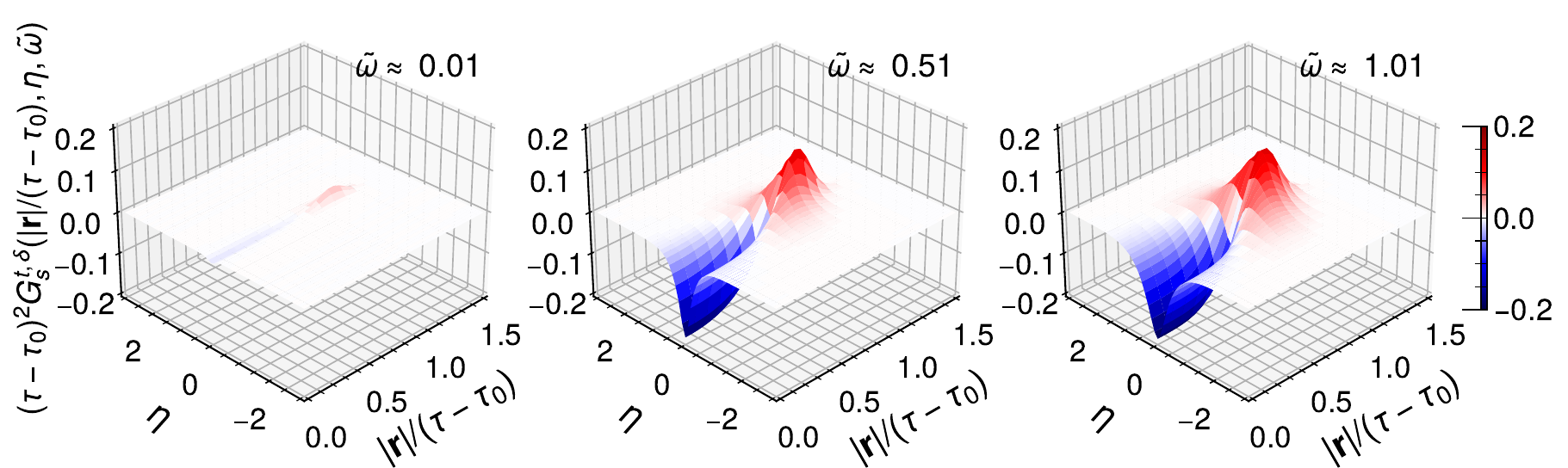}\\
    \includegraphics[width=0.87\linewidth]{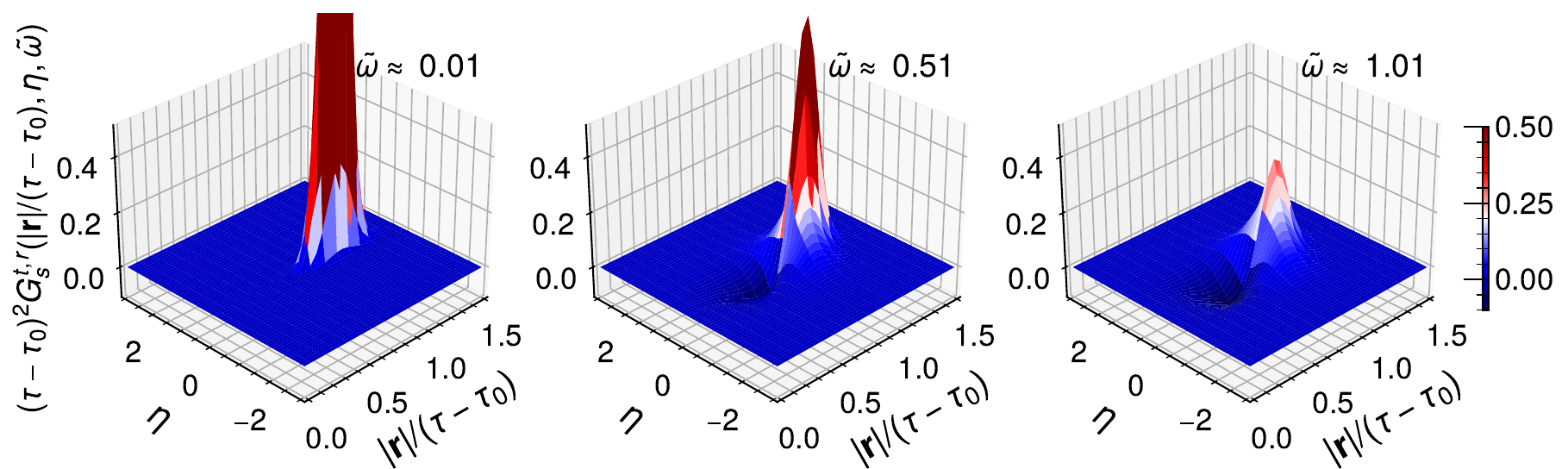}\\
    \includegraphics[width=0.87\linewidth]{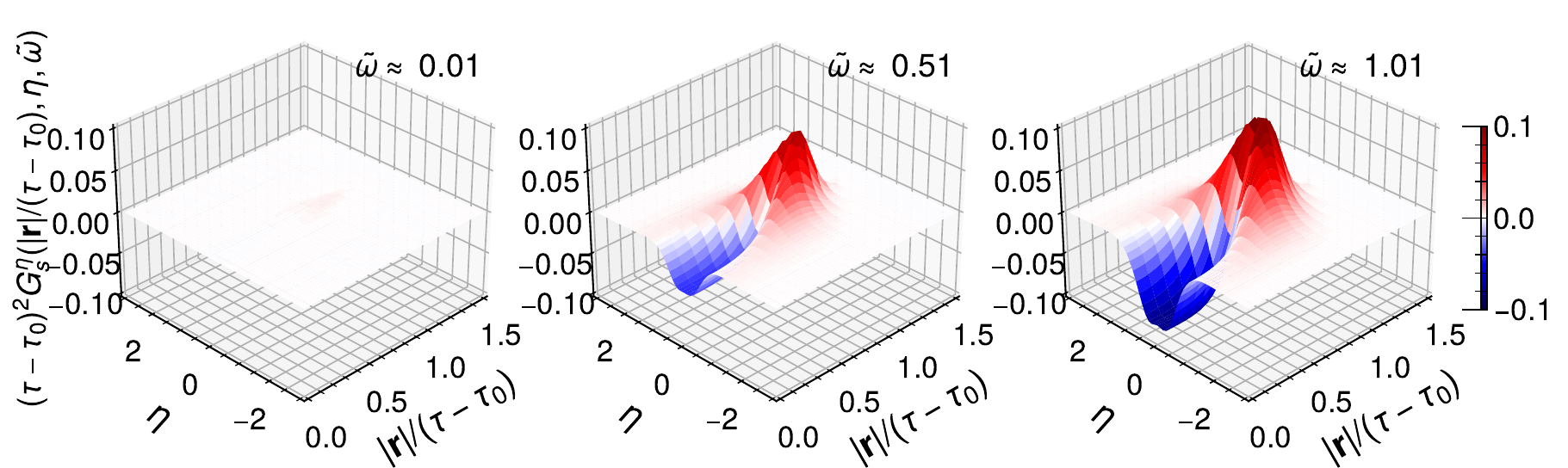}\\
    \includegraphics[width=0.87\linewidth]{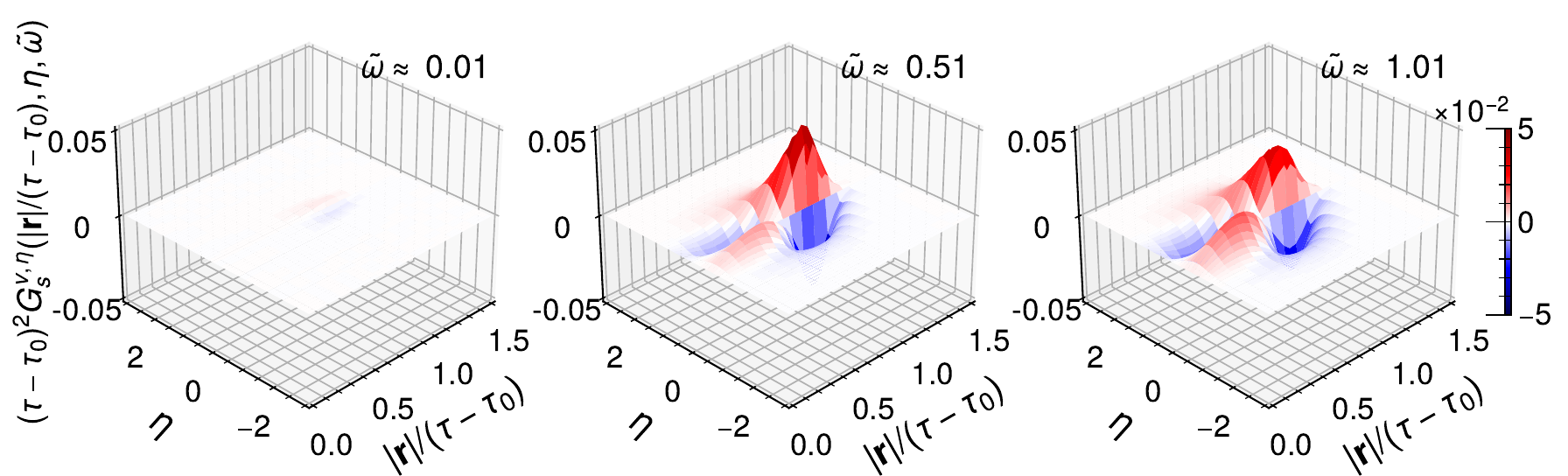}\\
    \caption{
    Evolution of the transverse momentum response ($G_s^v(r/\Delta\tau, \eta)$), the pressure response ($G_s^{t,\delta}(r/\Delta\tau, \eta)$ and $G_s^{\eta}(r/\Delta\tau, \eta)$) and shear stress response ($G_s^{t,k}(r/\Delta\tau, \eta)$ and $G_s^{v,\eta}(r/\Delta\tau, \eta)$) to an initial energy perturbation as a function of $r/\Delta\tau$ and $\eta$. Different panels correspond to different points in scaled time variable. 
    }
    \label{fig:more-Gs-r-RTA}
\end{figure*}

In Fig.~\ref{fig:more-Gs-r-RTA}, we show the evolution of the transverse momentum response ($G_s^v(r/\Delta\tau, \eta)$), the pressure response ($G_s^{t,\delta}(r/\Delta\tau, \eta)$ and $G_s^{\eta}(r/\Delta\tau, \eta)$)
 and shear stress response ($G_s^{t,k}(r/\Delta\tau, \eta)$ and $G_s^{v,\eta}(r/\Delta\tau, \eta)$) to an initial energy perturbation as a function of $r/\Delta\tau$ and $\eta$.

\section{Comparison of more components of shear stress tensor to their Navier-Stokes estimates}\label{more-shear}
\begin{figure*}
    \centering
    \includegraphics[width=0.63\linewidth]{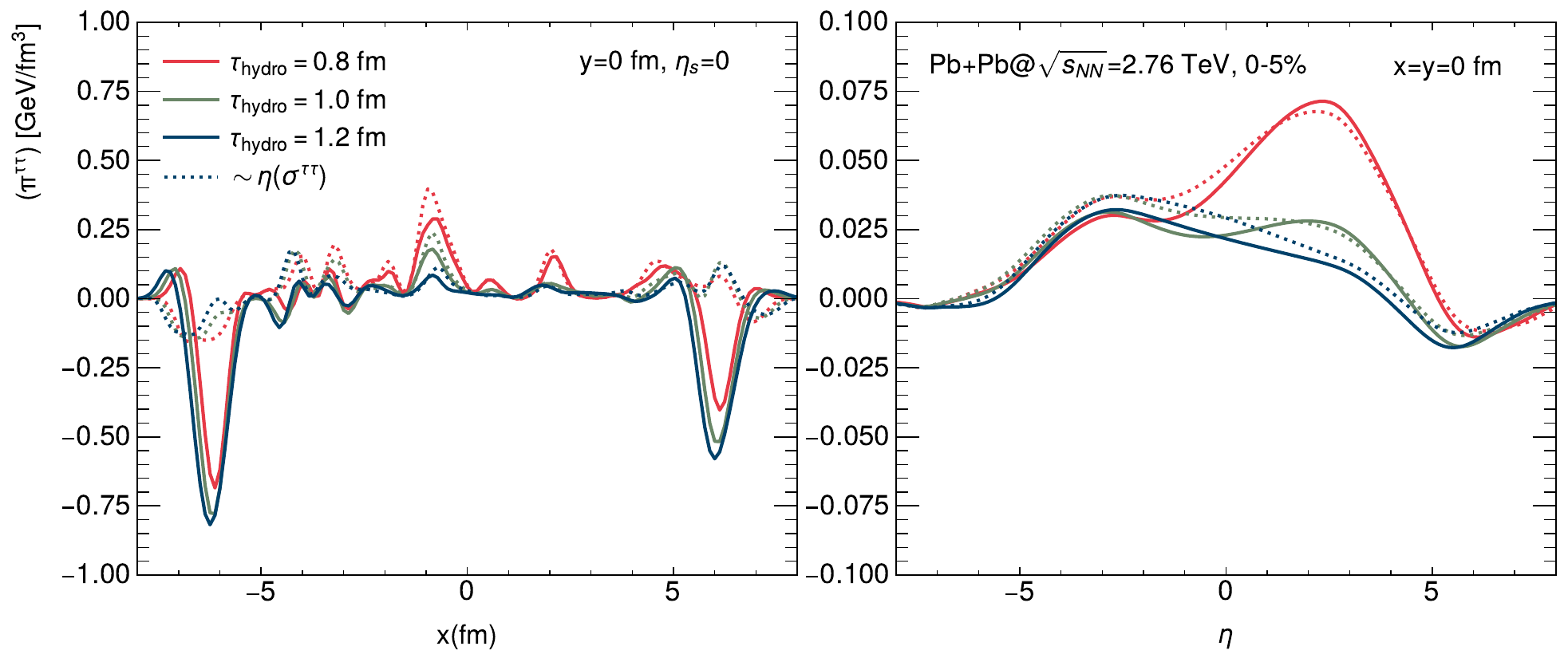}
    \includegraphics[width=0.63\linewidth]{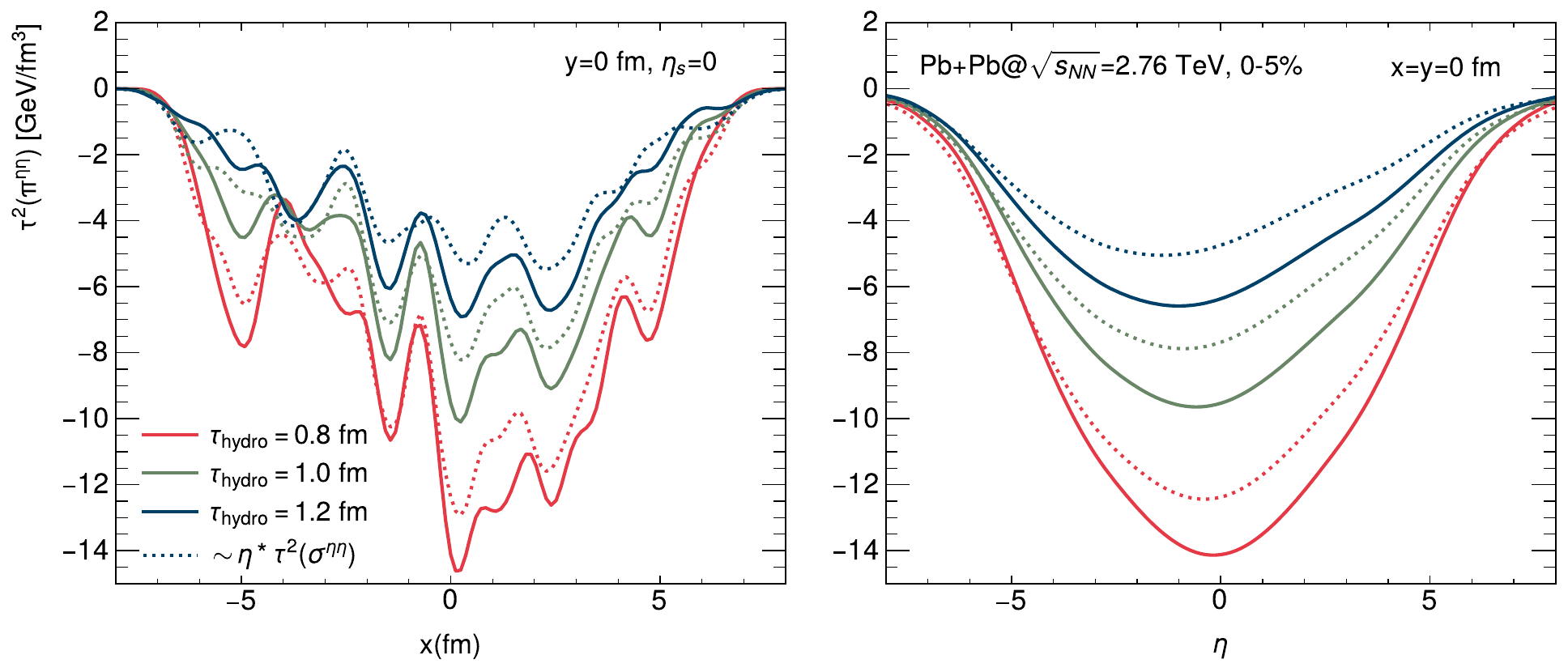}
    \includegraphics[width=0.63\linewidth]{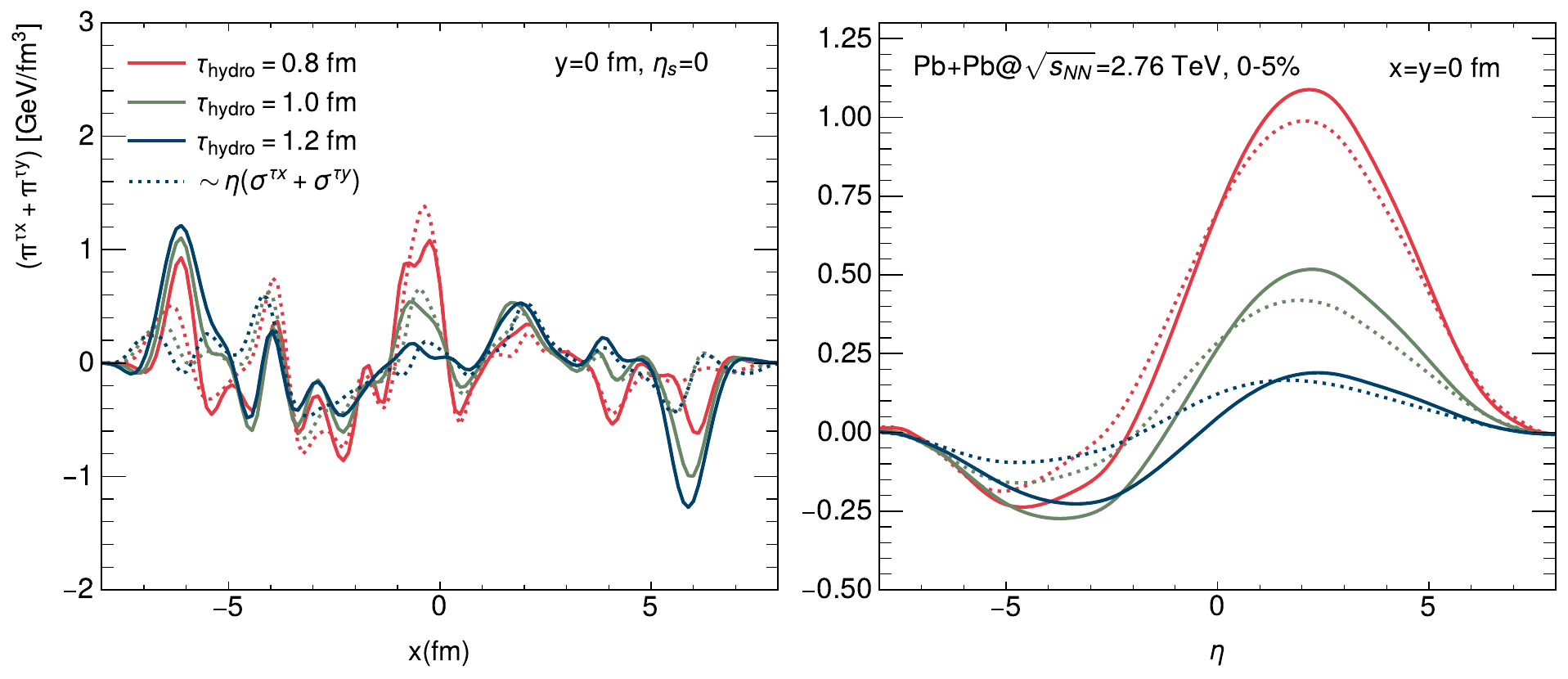}
    \includegraphics[width=0.63\linewidth]{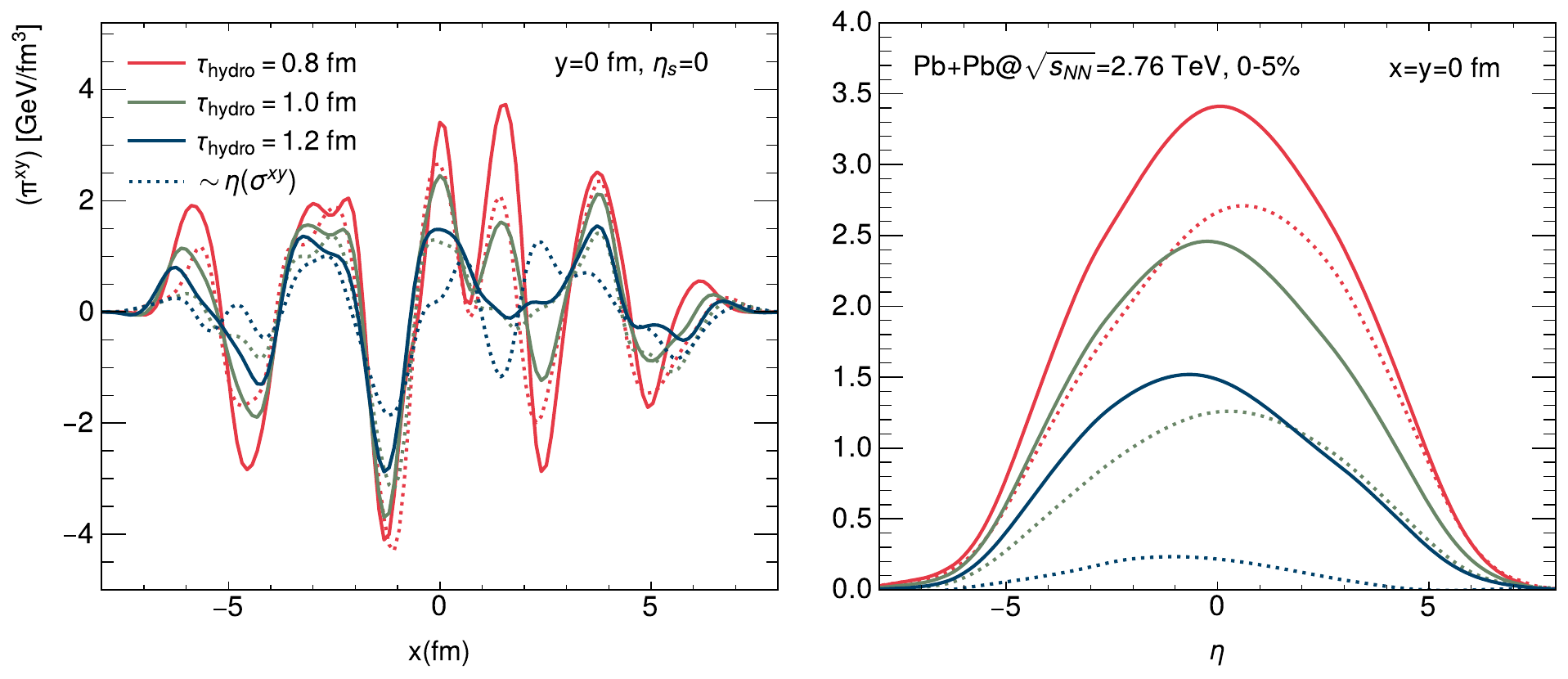}
    \includegraphics[width=0.63\linewidth]{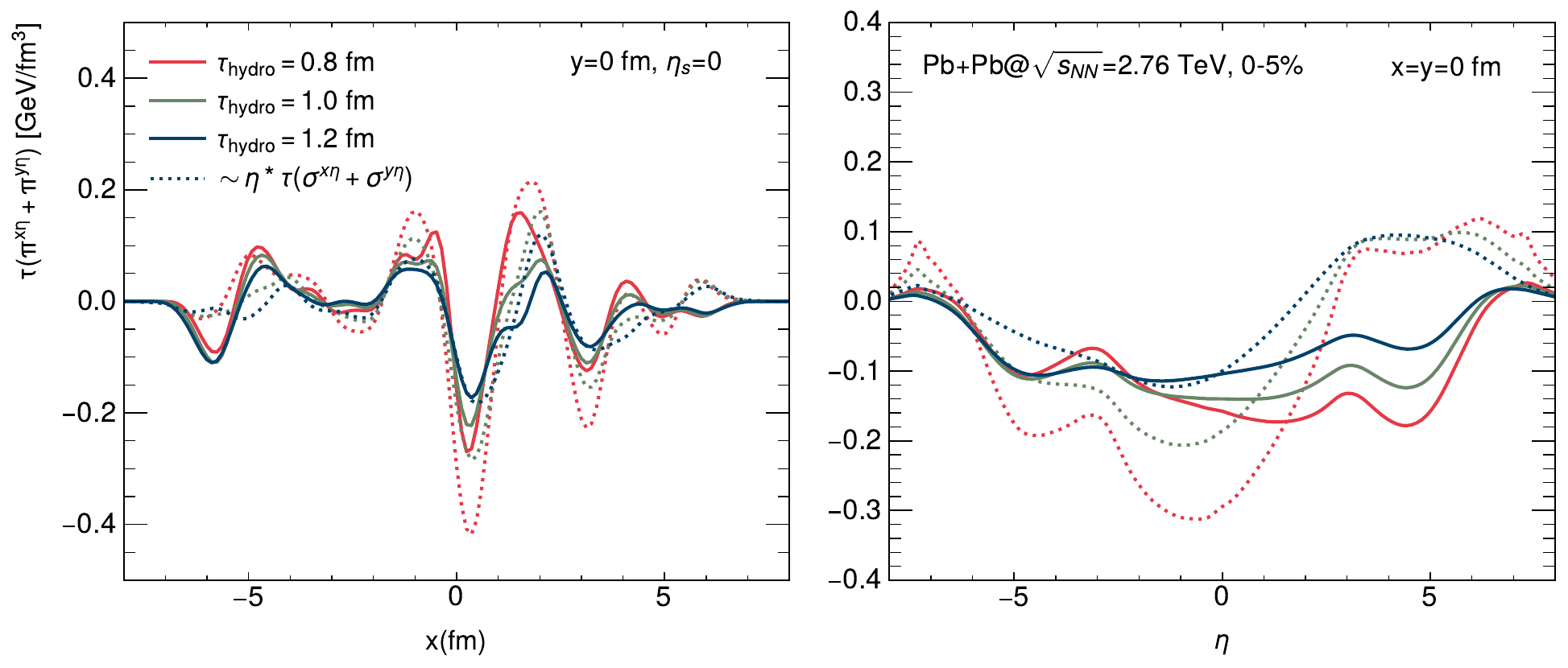}
    \caption{Hydrodynamization of more components of shear stress tensor}
    \label{fig:more-shear-hydrolize}
\end{figure*}

In Fig.~\ref{fig:more-shear-hydrolize},
we compare the remaining components of the out-of-equilibrium shear stress tensor $\pi^{\mu\nu}$ with their Navier-Stokes estimates at different hydrodynamics initialization times $\tau_{\rm hydro}$ = 0.4, 0.6, 0.8, 1.0, 1.2 fm. 
We find that most of them agree well with the hydrodynamic estimate over most of the collision region, both in the transverse plane and along the longitudinal direction, except the component related to longitudinal response.
However, the overall magnitude of these discrepancy is relatively small.
This indicates that it is safe to pass these components to subsequent hydrodynamic simulations.

\end{document}